\documentclass[12pt,preprint]{aastex}
\usepackage{natbib}
\newcommand{\be}{\begin{equation}}
\newcommand{\ee}{\end{equation}}

\newcommand{\nass}{\mbox{$N_{\rm ass}$}}
\newcommand{\nfalse}{\mbox{$N_{\rm false}$}}
\newcommand{\nfalsemc}{\mbox{$\langle \hat N_{\rm false} \rangle$}}

\slugcomment{Accepted for publication in Astrophysical Journal Supplement }

\shorttitle{Fermi LAT Second Catalog}
\shortauthors{Abdo et al.}

\begin{document}

\title{Fermi Large Area Telescope Second Source Catalog}


    
\author{
P.~L.~Nolan\altaffilmark{1,2}, 
A.~A.~Abdo\altaffilmark{3}, 
M.~Ackermann\altaffilmark{4}, 
M.~Ajello\altaffilmark{1}, 
A.~Allafort\altaffilmark{1}, 
E.~Antolini\altaffilmark{5,6}, 
W.~B.~Atwood\altaffilmark{7}, 
M.~Axelsson\altaffilmark{8,9,10}, 
L.~Baldini\altaffilmark{11}, 
J.~Ballet\altaffilmark{12,13}
% ...
}
\altaffiltext{1}{W. W. Hansen Experimental Physics Laboratory, Kavli Institute for Particle Astrophysics and Cosmology, Department of Physics and SLAC National Accelerator Laboratory, Stanford University, Stanford, CA 94305, USA}
\altaffiltext{2}{Deceased}
\altaffiltext{3}{Center for Earth Observing and Space Research, College of Science, George Mason University, Fairfax, VA 22030, resident at Naval Research Laboratory, Washington, DC 20375, USA}
\altaffiltext{4}{Deutsches Elektronen Synchrotron DESY, D-15738 Zeuthen, Germany}
\altaffiltext{5}{Istituto Nazionale di Fisica Nucleare, Sezione di Perugia, I-06123 Perugia, Italy}
\altaffiltext{6}{Dipartimento di Fisica, Universit\`a degli Studi di Perugia, I-06123 Perugia, Italy}
\altaffiltext{7}{Santa Cruz Institute for Particle Physics, Department of Physics and Department of Astronomy and Astrophysics, University of California at Santa Cruz, Santa Cruz, CA 95064, USA}
\altaffiltext{8}{Department of Astronomy, Stockholm University, SE-106 91 Stockholm, Sweden}
\altaffiltext{9}{The Oskar Klein Centre for Cosmoparticle Physics, AlbaNova, SE-106 91 Stockholm, Sweden}
\altaffiltext{10}{Department of Physics, Royal Institute of Technology (KTH), AlbaNova, SE-106 91 Stockholm, Sweden}
\altaffiltext{11}{Istituto Nazionale di Fisica Nucleare, Sezione di Pisa, I-56127 Pisa, Italy}
\altaffiltext{12}{Laboratoire AIM, CEA-IRFU/CNRS/Universit\'e Paris Diderot, Service d'Astrophysique, CEA Saclay, 91191 Gif sur Yvette, France}
\altaffiltext{13}{email: jean.ballet@cea.fr}
% ...

\begin{abstract}


We present the second catalog of high-energy $\gamma$-ray sources detected by the Large Area Telescope (LAT), the primary science instrument on the {\it Fermi Gamma-ray Space Telescope (Fermi)}, derived from data taken during the first 24 months of the science phase of the mission, which began on 2008 August 4.  Source detection is based on the average flux over the 24-month period.  The Second $Fermi$-LAT catalog (2FGL) includes source location regions, defined in terms of elliptical fits to the 95\% confidence regions and  spectral fits in terms of power-law, exponentially cutoff power-law, or log-normal forms.  Also included are flux measurements in 5 energy bands and light curves on monthly intervals for each source.  Twelve sources in the catalog are modeled as spatially extended.  We provide a detailed comparison of the results from this catalog with those from the first  $Fermi$-LAT catalog (1FGL).  Although the diffuse Galactic and isotropic models used in the 2FGL analysis are improved compared to the 1FGL catalog, we attach caution flags to 162 of the sources to indicate possible confusion with residual imperfections in the diffuse model.    The 2FGL catalog contains 1873 sources detected and characterized in the 100~MeV to 100~GeV range of which we consider 127 as being firmly identified and 1171 as being reliably associated with counterparts of known or likely $\gamma$-ray-producing source classes.

\end{abstract}


\section{Introduction}
\label{introduction}

This paper presents a catalog of high-energy $\gamma$-ray sources detected in the first two years of the {\it Fermi Gamma-ray Space Telescope} mission by the Large Area Telescope (LAT).  It is the successor to the LAT Bright Source List \citep{LAT09_BSL} and the First Fermi LAT \citep[1FGL, ][]{LAT10_1FGL} catalog, which were based on 3 months and 11 months of flight data, respectively.  The new catalog represents the deepest-ever catalog in the 100~MeV -- 100~GeV energy range and includes a number of analysis refinements.

Some important improvements compared to the 1FGL catalog are:

\begin{enumerate}
\item The 2FGL catalog is based on data from 24 months of observations. 
\item The data and Instrument Response Functions (IRFs) use the newer Pass 7 event selections, rather than the Pass 6 event selections used previously. 
\item This catalog employs a new, higher-resolution model of the diffuse Galactic and isotropic emissions.
\item Spatially extended sources and sources with spectra other than power laws are incorporated into the analysis. 
\item The source association process has been refined and expanded.  
\end{enumerate}

Owing to the nearly continuous all-sky survey observing mode and large field of view of the LAT, the catalog covers the entire sky with little observational bias.  The sensitivity is not uniform, due to the large range of brightness of the foreground diffuse Galactic $\gamma$-ray emission.  In addition, because the point-spread function (PSF) and effective area of the LAT depend on energy, the sensitivity limit depends markedly on the intrinsic source spectrum.   

As has been established with the 1FGL catalog, a number of source populations are known to be present in the data.  For individual sources, associations with objects in other astronomical catalogs are evaluated quantitatively.                                                    

In Section 2 we describe the LAT and the models for the diffuse  backgrounds, celestial and instrumental.  Section 3 describes how the catalog is constructed, with emphasis on what has changed since the analysis for the 1FGL catalog.  The 2FGL catalog itself is presented in Section 4, along with a comparison to the 1FGL catalog. We discuss associations and identifications in Section 5.  After the conclusions in Section 6  we provide appendices with technical details of the analysis and of the format of the electronic version of the 2FGL catalog.

\section{Instrument \& Background}
\label{lat_and_background}

\subsection{The data}
\label{DataDescription}

The LAT is a $\gamma$-ray detector designed to distinguish $\gamma$-rays in the energy range 20~MeV to more than 300~GeV from the intense background of energetic charged particles found in the 565~km~altitude orbit of the {\it Fermi} satellite.  For each $\gamma$-ray, the LAT measures its arrival time, direction, and energy.  The effective collecting area is $\sim$6500~cm$^2$ at 1~GeV (for the Pass~7 event selection used here; see below), the field of view is quite large ($>$2~sr), and the observing efficiency is very high, limited primarily by interruptions of data taking during passage of $Fermi$ through the South Atlantic Anomaly ($\sim$13\%) and trigger dead time fraction ($\sim$9\%).  The per-photon angular resolution is strongly dependent on energy; the 68\% containment radius is about $0\fdg8$ at 1~GeV (averaged over the acceptance of the LAT) and varies with energy approximately as $E^{-0.8}$, asymptoting at $\sim$$0\fdg2$ at high energies. The tracking section of the LAT has 36 layers of silicon strip detectors to record the tracks of charged particles, interleaved with 16 layers of tungsten foil (12 thin layers, 0.03 radiation length, at the top or {\it Front} of the instrument, followed by 4 thick layers, 0.18 radiation length, in the {\it Back} section) to promote $\gamma$-ray pair conversion.
Beneath the tracker is a calorimeter comprised of an 8-layer array of CsI crystals (1.08 radiation length per layer) to determine the 
$\gamma$-ray energy. The tracker is surrounded by segmented charged-particle anticoincidence detectors (plastic scintillators 
with photomultiplier tubes) to reject cosmic-ray background events.  More information about the LAT and the performance of the LAT is presented in  \citet{LAT09_instrument} and the in-flight calibration of the LAT is described in \citet{LAT09__calib} and  \citet{LAT11__calib}.

The data analyzed here for the 2FGL catalog were taken during the period 2008 August 4 (15:43 UTC) -- 2010 August 1 (01:17 UTC).  During most of this
time $Fermi$ was operated in sky-scanning survey mode (viewing direction rocking
 north and south of the zenith on alternate orbits). Time intervals flagged as `bad' (a very small fraction) were excluded.   Furthermore, a few minutes were excised around four bright GRBs (GRB~080916C: 243216749--243217979, GRB~090510: 263607771--263625987, GRB~090902B: 273582299--273586600, GRB~090926A: 275631598--275632048 in order to avoid having these bright transients distort the analysis of the more persistent catalog sources near these directions\footnote{These are Mission Elapsed Times, defined as seconds since 00:00:00 UTC on 2001 January 1.}).  We are preparing a separate catalog of LAT GRBs.

Previous analysis of the {\it Fermi} LAT data relied on criteria for selecting probable $\gamma$-ray events from all the instrument triggers as determined before launch or modified versions of these selections (called Pass~6\_V3 Diffuse\footnote{http://www.slac.stanford.edu/exp/glast/groups/canda/archive/pass6v3/lat\_Performance.htm}).  Experience with the data allowed us to develop an improved event selection process with lower instrumental background at energies above 10 GeV and higher effective area at energies below 200 MeV.   These Pass~7\_V6\footnote{http://www.slac.stanford.edu/exp/glast/groups/canda/archive/pass7v6/lat\_Performance.htm} (P7\_V6) Source class event selections are accompanied by a corresponding revised set of Instrument Response Functions \citep{LAT11__calib}, including an energy-dependent PSF calibrated using known celestial point sources.  The model for the diffuse gamma-ray background was fit using P7\_V6 Clean event selections and IRFs (see \S~\ref{DiffuseModel}).  The Clean event selection has lower residual background intensity than P7\_V6 Source at the cost of decreased effective area, a tradeoff that is worthwhile for studies of diffuse $\gamma$-ray emission.  The IRFs tabulate the effective area, PSF, and energy dispersions as functions of energy and inclination angle with respect to the LAT z-axis.  The IRFs are also tabulated as a function of the location of the $\gamma$-ray conversion in the LAT; $Front$ conversions occur in the top 12 tracking layers.  The tungsten foils are thinnest in this region and the PSF is narrower than for the $Back$ section, which has 4 layers of relatively thick conversion foils.  The 2FGL catalog is therefore derived from a new data set rather than simply an extension of the 1FGL data set.  

During the 1FGL time interval (up to 2009 July 4) the standard rocking angle for survey-mode observations was $35\degr$. During much of 2009 July and August it was set to $39\degr$. Then on 2009 September 2 the standard rocking angle was increased to $50\degr$ in order to lower the temperature of the spacecraft batteries and thus extend their lifetime. Time intervals during which the rocking angle of the LAT was greater than $52\degr$ were excluded.  The more-conservative 1FGL limit of $43\degr$ had to be raised to accommodate the larger standard rocking angle.  

For the 2FGL analysis we apply a more conservative cut on the zenith angles of the $\gamma$-rays, $100\degr$ instead of the $105\degr$ used for the 1FGL catalog.  This compensates for the increased contamination from atmospheric $\gamma$-rays from the earth's limb due to the larger rocking angle.  Another motivation for the tighter cut is that the new Pass 7 event selections used for the 2FGL analysis have much greater effective area at low energies than those used for the 1FGL analysis.  Because the point-spread function broadens with decreasing energy, a more conservative limit on zenith angle is warranted in any case.

\begin{figure}
\epsscale{.80}
\plotone{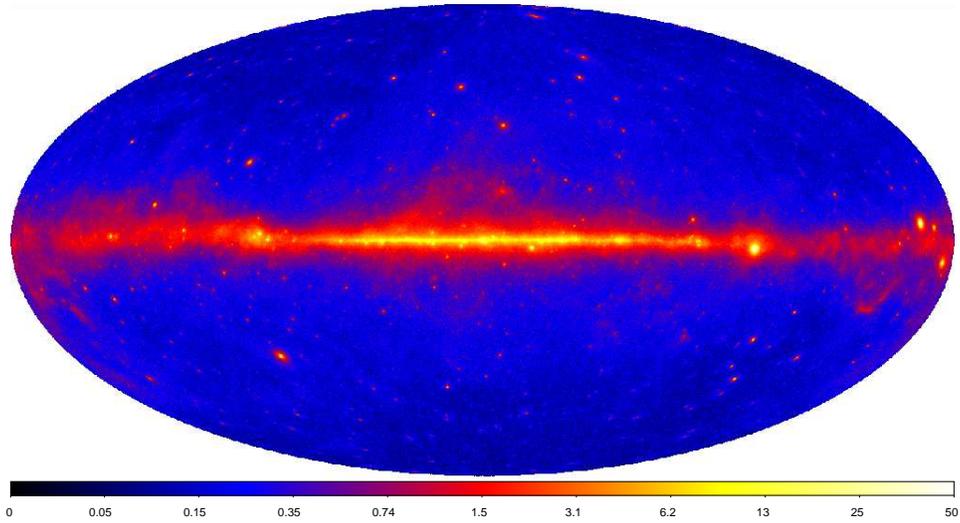}
\caption{Sky map of the energy flux derived from the LAT data for the time range analyzed in this paper, Aitoff projection in Galactic coordinates. The image shows $\gamma$-ray energy flux for energies between 100~MeV and 10~GeV, in units of 10$^{-7}$ erg cm$^{-2}$ s$^{-1}$ sr$^{-1}$.}
\label{fig:eflux_map}
\end{figure}

\begin{figure}
\epsscale{.80}
\plotone{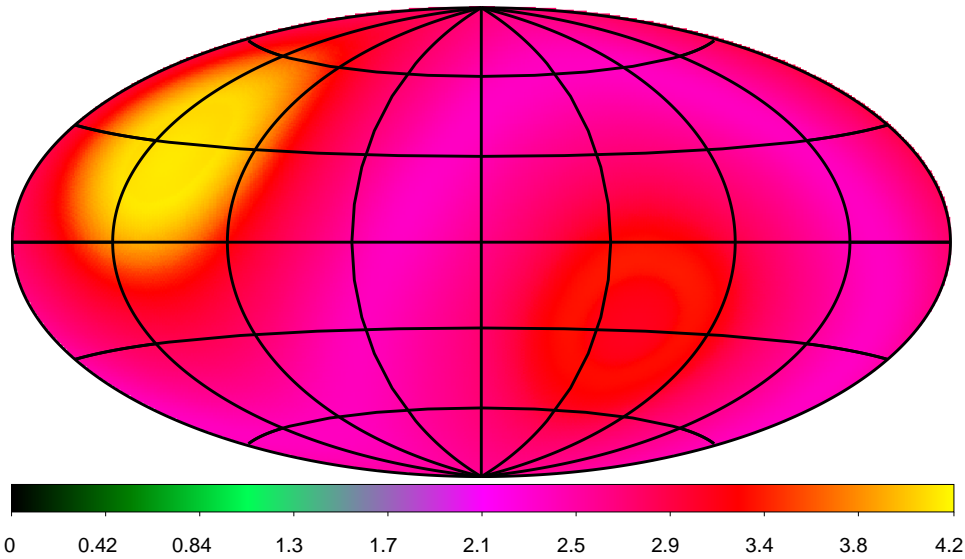}
\caption{Exposure of the LAT for the period September 2009 to July 2010 when the rocking angle was $50\degr$, Aitoff projection in Galactic coordinates. The units are equivalent on-axis exposure at 1 GeV in Ms.}
\label{fig:Expmap2ndYr}
\end{figure}

\begin{figure}
\epsscale{.80}
\plotone{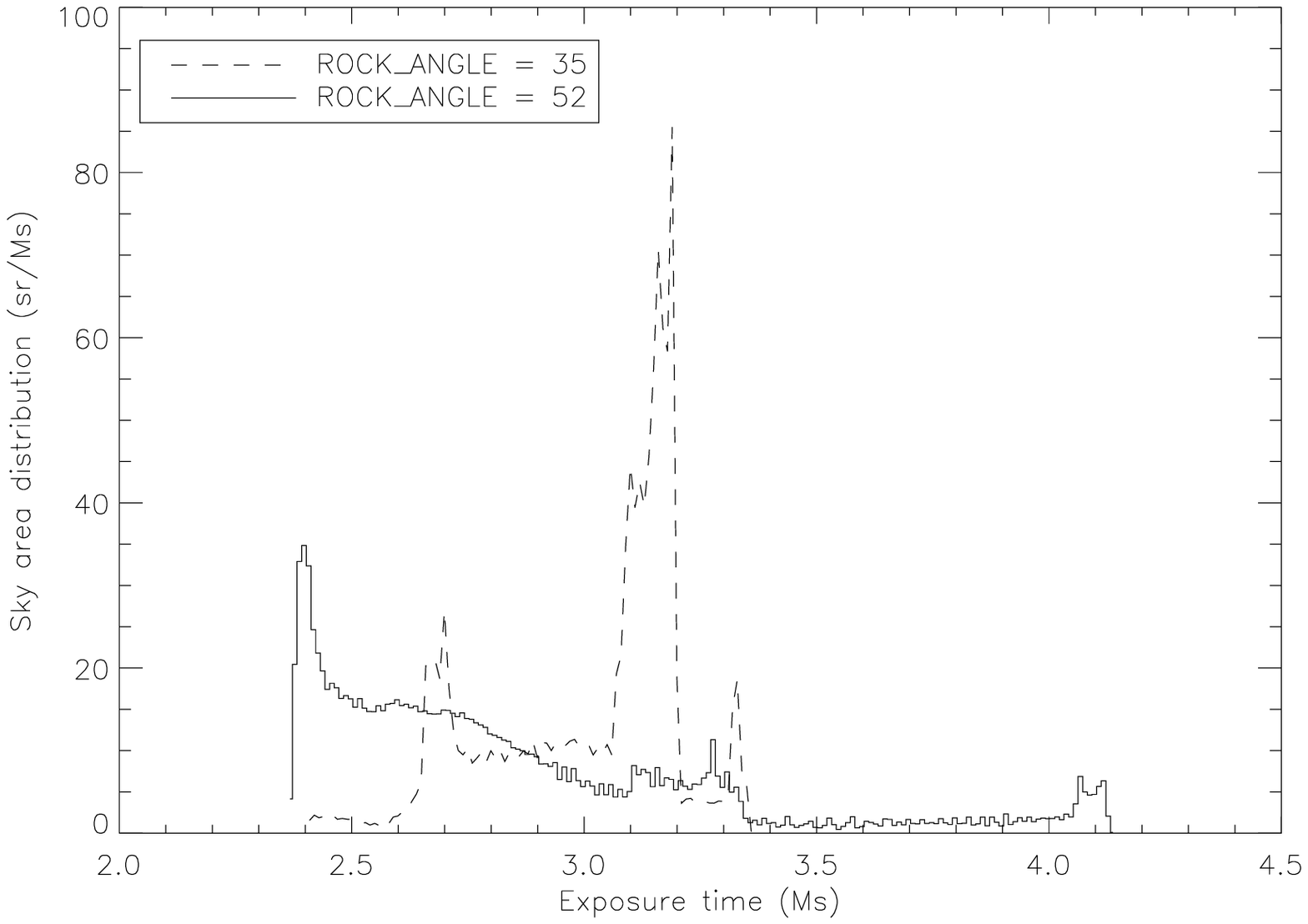}
\caption{Distribution of the equivalent on-axis exposure of the LAT at 1 GeV. The curves show the area of the sky exposed at that depth. The dashed curve is for the first 11 months (1FGL: 2008 August to 2009 June) when the rocking angle was $35\degr$, and the full curve is for the period when the rocking angle was $50\degr$ (2009 September to 2010 July, also 11 months).}
\label{fig:exposure_histogram}
\end{figure}

The energy flux map of Figure~\ref{fig:eflux_map} summarizes the data set used for this analysis.  The corresponding exposure is relatively uniform, owing to the large field-of-view and the rocking-scanning pattern of the sky survey. 
With the new rocking angle set to $50\degr$ the exposure is minimum at the celestial equator,  maximum at the North celestial pole and the contrast (maximum to minimum exposure ratio) is 1.75  (Fig.~\ref{fig:Expmap2ndYr}). The exposure with rocking angle $35\degr$ \citep[Fig. 2 of][]{LAT09_BSL} was least at the South celestial pole, with a contrast of 1.33. The North/South asymmetry is due to loss of exposure during passages of $Fermi$ through the South Atlantic Anomaly.
Figure~\ref{fig:exposure_histogram} shows that the original rocking scheme resulted in a very uniform exposure over the sky. The new rocking scheme is less uniform, although it still covers the entire sky to an adequate depth. 
The exposure map for 2FGL is about halfway between the $35\degr$ and $50\degr$ maps. It peaks
toward the North celestial pole and is rather uniform over the South celestial hemisphere, with a contrast of 1.37.
Note that the average etendue of the telescope is only slightly reduced, from 1.51 m$^2$ sr (at 1~GeV) in the first 11 months to 1.43 m$^2$ sr over the last 11 months. The reduction is due to the part of the field of view rejected by the newer zenith angle selection.

\subsection{Model for the Diffuse Gamma-Ray Background}
\label{DiffuseModel}

The $\gamma$-ray emission produced by the Galaxy originating from the interaction of  cosmic-ray electrons and protons with interstellar nucleons and photons is modeled with the same method as for the 1FGL catalog. We fit a linear combination of gas column densities, an Inverse Compton (IC) intensity map, and isotropic intensity to the LAT data using the P7\_V6 Clean data set. To account for the non-uniform cosmic-ray flux in the Galaxy, the gas column densities are distributed within galactocentric annuli. More details on the various radio and infrared surveys used to generate the maps for the different annuli are given at the Web site of the $Fermi$ Science Support Center\footnote{http://fermi.gsfc.nasa.gov/ssc/data/access/lat/BackgroundModels.html}.  Inverse Compton $\gamma$-rays from cosmic-ray electrons interacting on optical, infrared and CMB photons are modeled with GALPROP \citep{Strong2007}. In each energy band, the gas emissivities and IC normalization were left free to vary. 

For this study we have improved the modeling of the diffuse emission in several ways. With more than twice the $\gamma$-ray statistics we were better able to discriminate between the template maps described above and we were also able to increase the number of energy bins from 10 to 14, spanning 63~MeV to 40~GeV. Below 63~MeV, the combined effects of a low effective area and increased earth limb contamination owing to the increased breadth of the PSF prevent study of the diffuse emission. Above 40~GeV the statistics are too low to discriminate between the large number of templates that comprise the model. The quality of the determination of the linear coefficients (interpreted as the $\gamma$-ray emissivities for the gas) was also improved at high energies by using the P7\_V6 Clean data set, which has lower residual charged particle backgrounds at high energies than P6\_V3 Diffuse.  For energies below 63~MeV or above 40~GeV the diffuse emission model was derived by extrapolating the measured emissivities according to a fit of the emissivities in terms of bremsstrahlung and pion decay components.  

The spatial resolution of the model was improved from $0\fdg5$ to $0\fdg125$, which is the sampling of most of the CO survey \citep{Dame2001}.  The higher resolution in the fitting procedure helps discriminate H$_2$, H~{\sc I}, dark gas, and smoother distributions like inverse Compton.  For the actual fitting, for computational considerations we sampled the maps with $0\fdg25$ resolution to derive the emissivities and used the full resolution to reconstruct the model from the deduced emissivities. The final resolution of the model is then $0\fdg125$.  Given sufficient statistics this is crucial to discriminate point-like sources and molecular clouds at the PSF scale.

\begin{deluxetable}{lccccc}
\tabletypesize{\scriptsize}
\tablecaption{Additional Components in the Diffuse Emission Model
\label{tbl:patches}}
\tablewidth{0pt}
\tablehead{

\colhead{Designation}&
\colhead{Center}&
\colhead{Approx. dim.}&
\colhead{ $\Omega/ 4\pi$}&
\colhead{Fraction of total}&
\colhead{Fraction of intensity} \\

\colhead{}&
\colhead{$(l, b)$}&
\colhead{$(l \times b)$}&
\colhead{}&
\colhead{intensity}&
\colhead{within patch}
}

\startdata
First quadrant and inner   &  $25\degr, 0\degr$  &   $40\degr \times 30\degr$   &  1.9\% & 1.0\% & 13.4\% \\                                                          
Fourth quadrant                 &   $-$35, 9                  &     $40\times30$                      &   2.4 & 0.3 & 3.8 \\
Lobe North                          &     0, 25                      &     $50\times40$                      &   3.9 & 0.4 & 6.9 \\
Lobe South                         &      0,$-$30                &      $50\times40$                     &   3.7 & 0.4 & 14.1 \\
\enddata

\tablecomments{Description of the additional components added in the Galactic diffuse model. The centers and extents are in Galactic coordinates.  The extents are approximate because the shapes are irregular.  $\Omega$ is the solid angle.  To evaluate the fractional intensities, we integrated the intensity above 130 MeV for the First and Fourth quadrant patches and above 1.6 GeV for the Lobes patches.}

\end{deluxetable}

This procedure revealed regions with photon excesses not correlated with gas or templates defined by observations at other wavelengths. We found what appear to be two distinct origins for the excesses, depending on energy.  For both cases we introduce ad hoc `patches' in the diffuse emission model to account for their contributions.  The patches are regions of spatially uniform intensity whose shapes reproduce the shape of the excesses. The intensity of the emission associated with each patch is fitted for each energy band together with the other templates.  
The shapes of the patches were chosen to approximately encompass regions with an excess of  photons of the order of 20\% compared to the model outside the Galactic plane. 
Two of the regions have a hard spectrum and are lobe-shaped north and south from the Galactic center.  This emission was also observed and studied in detail by \cite{Su2010}.  Table~\ref{tbl:patches} summarizes the patches and their contributions to the model.  Images showing the locations and extents of the patches are available from the $Fermi$ Science Support Center at the URL cited above.  These regions do not correspond to fluctuations in the diffuse emission model.  We do not see large regions where the model exceeds the observed intensity, and we did not need to use `negative' patches. Four main regions were identified in the first and fourth quadrants, and north and south of the Galactic center. We added an extra inner patch to the first quadrant region where the intensity was greater than in the rest of the patch. The spectra of the patches were determined in the same way as for the other templates by extracting their intensities from fits in each energy bin.

At lower energies, below a few GeV, an excess of photons seems to be associated with the giant radio loop Loop~I.  
The North Polar Spur is clearly visible in the LAT data and can be roughly modeled with the 408~MHz radio map of \citet{Haslam1981} as well as a large rounded shape filling the Loop.
At low energies distinguishing between $\gamma$-rays originating from Loop~I and from larger distances is very difficult near the Galactic plane.  It is possible that the scaling of the model map for the Galactic inverse Compton emission as well as the fitted emissivities of inner Galaxy gas rings are artificially increased in the fitting procedure to account for $\gamma$-rays produced locally. While keeping the overall residual fairly flat, this may bias the diffuse emission spectrum and derived spectra and significances of faint sources in a large region of about 100$\degr$ wide in longitude and 30$\degr$ in latitude centered in the Galactic center. 
Independent of this effect, other regions are probably inadequately modeled, for example the Cygnus region, the Carina tangent, and  the Orion molecular cloud; see \S 3.9.

The spatial grid of the model now has a bin centered at latitude zero. Previously the Galactic ridge was split between two bins with the consequence of flattening the modeled ridge and possibly inducing the detection of spurious sources in the Galactic ridge.

We also created a template for the emission from the earth limb that is not completely removed from the P7\_V6 Source and Clean data sets at energies below 200~MeV.  These are $\gamma$-rays that are in the broad tails of the PSF and so pass the selection cut on zenith angle (see \S~\ref{DataDescription}). For the template we used the residuals in the 50--68~MeV energy range and assumed that the spatial shape is independent of energy.  The very soft spectrum was derived by adding this template to the model.  The template is specific to the data set analyzed here because the residual earth limb emission depends on the orientation of the LAT.

The isotropic component was derived for the P7\_V6 Source data set by fitting the data for the whole sky using the Galactic diffuse emission modeled as above.  By construction the isotropic component includes  the contribution of residual (misclassified) cosmic rays for the P7\_V6 Source event analysis class.  Treating the residual charged particles as effectively an isotropic component of the $\gamma$-ray sky brightness rests on the assumption that the acceptance for residual cosmic rays behaves similarly as for $\gamma$-rays; in particular we assume that the relative contributions of the $Front$ and $Back$ events to the isotropic intensity are according to their relative effective areas.  This approximation is necessary in the $gtlike$ analysis described in \S~\ref{catalog_significance}. The actual residual background rates for $Front$ and $Back$ events do not in fact scale precisely with the ($\gamma$-ray) effective areas, with the most notable difference being in the low energy range $<$400~MeV for which the background `leakage' in the $Back$ section of the tracker is appreciably greater than for the $Front$ section. This has the effect of decreasing the flux measurements at low energies (below $\sim$200~MeV) and hardening the spectra, with the greatest effects for low-significance, soft sources.  On average the spectral indices for power-law spectral fits are hardened by less than half of the typical uncertainty in the measured spectral index.  

The models for the Galactic diffuse emission and the isotropic background spectrum, along with more detailed descriptions of their derivation, are available from the $Fermi$ Science Support Center.

\section{Construction of the Catalog}
\label{catalog_main}

The procedure used to construct the 2FGL catalog has a number of improvements relative to what was done for the 1FGL catalog.  In this section we review the procedure, with an emphasis on what is being done differently.

As for the 1FGL catalog, the basic analysis steps are source detection, localization (position refinement), and significance estimation.  Once the final source list was determined, by applying a significance threshold, we evaluated the flux in 5 bands and the flux history (light curve of the integrated flux) for each source.

Also as for the 1FGL analysis, the source detection step was applied only to the data from the full 24-month time interval of the data set.  We did not search for transient sources that may have been bright for only a small fraction of the 2-year interval.  See \S~\ref{other_gev_detections} for a discussion of transient LAT sources reported in Astronomer's Telegrams.  Analysis of 2FGL catalog source variability is found in \S~\ref{catalog_variability}.

The 2FGL catalog is primarily a catalog of point (spatially unresolved) sources detected by the LAT in the 24-month interval.  As discussed below, the analysis and catalog also include a number of LAT sources that are known to be spatially extended.  These sources are defined specially in the analysis (see \S~\ref{catalog_extended}) but are considered members of the 2FGL catalog.

    \subsection{Detection and Localization}
\label{catalog_detection}

Detection of point sources involves iterating through three steps: (1) identification of potential point sources, denoted as `seeds', that have not already been selected in a previous iteration; (2) a full all-sky optimization of a model of the $\gamma$-ray sky (diffuse emission plus sources) including the new seeds to refine their estimated positions and evaluate their significances;  (3) creation of a `residual Test Statistic ($TS$) map'.  The $TS$ is evaluated as $TS = 2(\log\mathcal{L}({\rm source}) - \log\mathcal{L}({\rm no source}))$, where $\mathcal{L}$ represents the likelihood of the data given the model with or without a source present at a given position on the sky.  In each case the likelihood is assumed to have been maximized with respect to the adjustable parameters of the model \citep{mattox96}.

We performed this analysis using the {\it pointlike} analysis system, for which the data are partitioned by whether the conversion occurred in the $Front$ or $Back$ sections of the tracker and binned in energy with four bins per decade from 100~MeV to 316~GeV. For each such partition, or band, the $\gamma$-rays are partitioned according to their HEALPix \citep{Gorski2005} indices, with the $nside$ parameter chosen such that the angular size of the partition is small compared with the PSF for that energy and conversion position. Detailed simulations, analytic studies, and adjustments of the bin size have shown that this does not lose precision compared with a fully unbinned procedure.

We discuss each step of the iteration in turn.

\subsubsection{Determination of seeds}
We started with an initial model comprised of the 1FGL catalog of sources to which we added seeds from the wavelet-based methods, {\it mr\_filter} \citep{sp98} and {\it PGWave} \citep{dmm97, PGWAVE}, and a minimal spanning tree-based algorithm \citep{cmg08} as described in 1FGL. 
For the 2FGL catalog analysis, we also included in the model 12 spatially-extended sources that have been detected by the LAT; see \S~\ref{catalog_extended}.   In subsequent iterations, seeds may be added by examination of the residual $TS$ map, described below. Since source detection is an integral part of the iteration procedure, the efficiency of the initial seed-determination procedures is not critical.

\subsubsection{All-sky optimization}
\label{all_sky_optimization}
We define 1728 circular regions centered on points defined by a HEALPix tessellation with $nside=12$. All $\gamma$-ray data within a $5^{\circ}$ radius of each of the points are fit to a model including the diffuse components described in 
\S~\ref{DiffuseModel}
and all seeds within a radius of $10^{\circ}$. Each region was optimized independently. The parameters included the normalization of each diffuse component and the spectral parameters of the point sources lying inside the boundaries of the HEALPix pixel that defined the region.
Since neighboring regions are coupled, sharing data and sources, we repeated this step until the likelihoods were jointly optimized.  For some regions along the Galactic plane, convergence required up to 10 iterations. 

For point sources identified as pulsars by LAT phase analysis \citep{LAT10_PSRcat,LAT11_PSRcat}, the spectra were fit to a power law with an exponential cutoff; others were fitted to either a simple power law, or log-normal (also called log-parabola); the latter was used if it substantially improved the overall likelihood. These functions are described in \S~\ref{catalog_spectral_shapes}.
Each seed was characterized by two versions of the likelihood $TS$ \citep{mattox96}: one measuring the spectral-shape independent measure from independent fits of the fluxes in each energy band
($TS_{band}$), and another which is the result of a fit to the spectral model, ($TS_{model}$). 
The former always will be larger than the latter: the difference is 
used to decide to switch from a power law to a log parabola spectral shape. Seeds with $TS_{band}<10$ are eliminated from further analysis.  The rest are retained in the model for the {\it pointlike} optimization. After the optimization was complete, those with $TS_{model}>10$ were passed on to the {\it gtlike} step described below, with the {\it pointlike} fit as a starting point.

\subsubsection{Residual $TS$ map}
\label{residual_ts_map}
After the analysis in the previous step converged, we performed a special analysis of the full sky to search for missing point sources. A HEALPix tessellation with $nside=512$ is used to define 3.1M points on a $0\fdg1$ grid. For each point, we added a new point source with a power law spectrum and fixed spectral index 2.0, 
 to the model, and the likelihood was maximized as a function only of its flux.  The resulting array of values of $TS$ is plotted as a sky map.

\begin{figure}
\epsscale{0.9}
\plotone{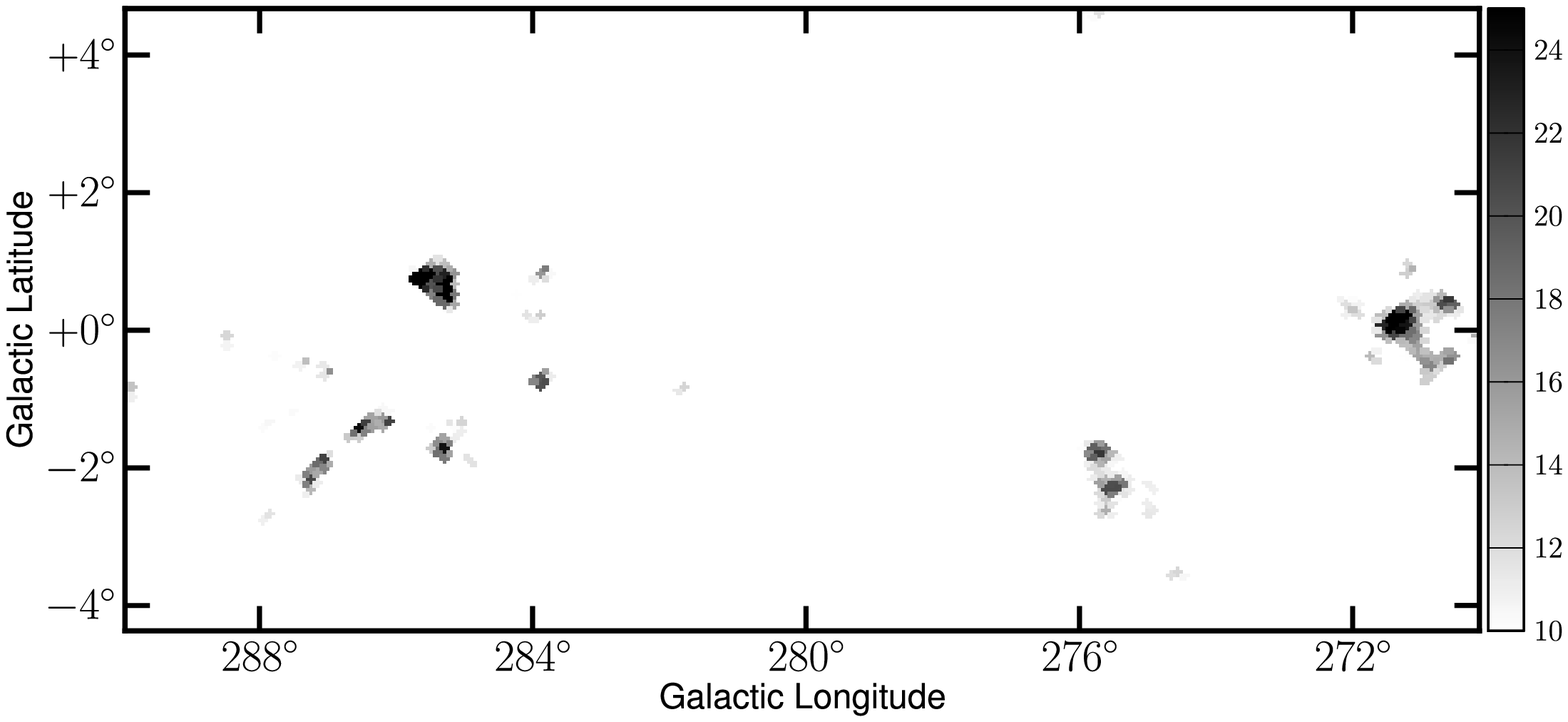}
\caption{A representative map of the $TS$ residuals along the Galactic ridge, corresponding to the final iteration. There are several clusters that could have generated seeds, but did not appear to be isolated point sources under visual inspection.} \label{fig:ts_residuals} 
\end{figure}

Clusters were defined by proximity: a cluster is the set of all pixels that occupy adjacent positions. The analysis generated a list of all clusters of such pixels with $TS>10$ on the map, used as seeds to be added  for the next iteration of the all-sky analysis. 
We estimated the position of a presumed source from the centroid of the pixels, weighted by $TS$; this position was refined later if the seed survived the full analysis.
Adding seeds from the map was done automatically for Galactic latitudes above $5\degr$; along the Galactic plane the data are not always well represented by either point sources or the model for diffuse Galactic emission, and we introduced new point sources only if they appeared to be well isolated under visual inspection. Figure \ref{fig:ts_residuals} shows the final such map for a region along the Galactic ridge.

In total, 3499 candidate sources were passed to the significance and thresholding step of the analysis. 

\subsubsection{Localization }
\label{catalog_localization}

The processing that created the residual $TS$ map used for source detection also performed local optimizations of the likelihood with respect to the position of each point source, using the spectral-shape independent definition of the likelihood, $TS_{band}$, described above, with the rest of the model fixed.
The positional uncertainty for each source was estimated by examining the shape of the log likelihood function, fitting the distribution to the expected quadratic form in the angular deviations from the best fit position. A measure of the quality of this fit is the mean square deviation of the log likelihood with respect to the fit on a circle of radius corresponding to two standard deviations.
For the catalog we tabulated the elliptical parameters including the fit position and the fit quality. As in the case of the 1FGL catalog, we made two empirical corrections based on comparison with the known locations of high-confidence associated sources: multiplied by a 1.1 scale factor, and added $0\fdg005$  in quadrature to the 95\% ellipse axes. This latter is comparable to the spacecraft alignment precision requirement of 10$\arcsec$.

We searched for systematic biases in source positions, using comparisons with counterpart positions (\S~\ref{source_assoc_main}). Two cases were considered: (1) sources near the Galactic plane, the positions of which might have been suspected to be biased by the strong gradient of the intensity of the Galactic diffuse emission, and (2) weak sources near much stronger ones. We did not find significant biases in either case.  In addition, in Appendix \ref{appendix_detection_localization} we show that the sizes of the localization regions for weak sources are consistent with expectations, as is the weak dependence on the source spectrum.

    \subsection{Significance and Thresholding}
\label{catalog_significance}

To evaluate the fluxes and spectral parameters, and also significances, for the catalog we use the standard LAT analysis tool $gtlike$ and associated LAT Science Tools\footnote{See http://fermi.gsfc.nasa.gov/ssc/data/analysis/documentation/Cicerone/} (version v9r23p0). The localization procedure (\S~\ref{catalog_localization}) provides spectra and significances as well, but we do not have as much experience with it so we prefer relying on the standard tools whenever possible.
This stage of the analysis is similar in principle to what was done for the 1FGL catalog \citep{LAT10_1FGL}. It splits the sky into Regions of Interest (RoI) in order to make the $\log \mathcal{L}$ (where $\mathcal{L}$ is the likelihood function) maximization tractable, varying typically half a dozen sources near the center of the RoI at the same time.  (There were 933 RoIs for 2FGL.) This requires an iterative scheme in order to inject the spectra of all sources in the outer parts of the RoI.
It uses the same energy range (100~MeV to 100~GeV) and adjusts the source spectra with positions fixed to the result of \S~\ref{catalog_localization}. The same parameters are used to refit the diffuse emission model (described in \S~\ref{DiffuseModel}) to each RoI: normalization and small corrective slope of the Galactic component and normalization of the isotropic component. We define the Test Statistic $TS=2\Delta \log \mathcal{L}$ for quantifying how significantly a source emerges from the background. The iteration scheme was also identical, as well as the threshold at $TS > 25$ applied to all sources, corresponding to a significance of just over 4 $\sigma$ evaluated from the $\chi^2$ distribution with 4 degrees of freedom \citep[position and spectral parameters, ][]{mattox96}.
We note that we require the predicted number of events from a source to be at least 10 over the full energy range, rejecting clusters of a few high-energy events without any low-energy counterpart. The same constraint was enforced for the 1FGL analysis.

The analysis does have a number of important differences with respect to 1FGL:
\begin{itemize}
\item The major change is that we switched from unbinned to binned likelihood (while still using {\it gtlike} or more precisely the {\it pyLikelihood} library in the Science Tools). The first reason for the change was to cap the computing time (which increases linearly with observing time in unbinned likelihood). The other important reason is that we discovered with simulations that the scale factors for the diffuse emission model terms returned by unbinned likelihood were significantly biased (overestimating the Galactic diffuse or isotropic diffuse intensity, whichever component was subdominant) whereas those returned by binned likelihood were not.
In order to treat the $Front$ and $Back$ events in the analysis according to their separate PSFs we added the $\log \mathcal{L}$ computed separately for {\it Front} and {\it Back} events. The energy binning was set to 10 bins per decade. RoIs are square for binned likelihood. We used the ARC projection with pixel size set to $0\fdg1$ for {\it Front} and $0\fdg2$ for {\it Back} events, in keeping with the high-energy PSF for each category. The sides of the RoIs were defined by adding $7\degr$ on each side to the diameter of the central part where all source parameters are free.
We note that the binned likelihood scheme is more conservative: in simulations comparable to the catalog depth (with or without sources) the significances of detections with unbinned likelihood tended to be around 1 $\sigma$ greater. This has important consequences for the number of sources in 2FGL (see \S~\ref{1fgl_comparison}).
\item We took advantage of the fact that the localization procedure (\S~\ref{catalog_localization}) also provides a spectral fit to all sources. We used it as the starting point for the procedure using {\it gtlike}, rather than starting with all sources set to 0.
\item We did not use exactly the result of the previous iteration to start the next one, but applied a damping factor $\delta$ (set to 0.1) to all parameters, defining the next starting point as $P_{n+1} = (1-\delta) P_n + \delta P_{n-1}$. It is a significant change because in all RoIs the number of sources (outside the core of the RoI) which are considered but frozen is much larger than that of free sources. The damping procedure avoids overshooting and improves convergence.
\item Many bright sources are fitted with curved spectra instead of power-law. This is described in \S~\ref{catalog_spectral_shapes}. In addition to providing more detailed descriptions of those bright sources, it also improves the reliability of the procedure for neighboring sources. The reason is that it greatly reduces the spectral residuals, which otherwise might have been picked up by neighboring sources. That kind of transfer can be an issue at low energy where the PSF is very broad and cross-talk between sources in the likelihood analysis is strong.
\item We introduce the Earth limb component obtained in \S~\ref{DiffuseModel}, without any adjustment or free parameter in the likelihood analysis.
\end{itemize}
Appendix~\ref{appendix_fit_quality} illustrates how well the full model (diffuse emission and individual sources) reproduces the $\gamma$-ray sky.

    \subsection{Spectral Shapes}
\label{catalog_spectral_shapes}

The 1FGL catalog considered only power-law (PL) spectra. This was simple and homogeneous, but not a good spectral representation of the bright sources, as could be easily seen from comparing the power-law fits with the fluxes in bands \citep[quantified by the Curvature\_Index column in][]{LAT10_1FGL}. As the exposure accumulated, the discrepancies grew statistically larger, to the point where it could affect the global fit in an RoI, altering the spectra of neighboring sources in order to get a better overall spectral fit.
For 2 years of data we had to allow for spectra that deviate from power laws. However increasing the number of free parameters means finding the true best fit is more difficult, so we chose spectral shapes with only one additional free parameter.

For the pulsars we chose exponentially cutoff power-laws (hereafter PLExpCutoff), which are a good representation of pulsar spectra in general \citep{LAT10_PSRcat}:
\begin{equation}
\frac{{\rm d}N}{{\rm d}E} = K \left (\frac{E}{E_0}\right )^{-\Gamma} \exp \left (-\frac{E-E_0}{E_c}\right )
\label{eq:expcutoff}
\end{equation}
This is just the product of power law and an exponential. The parameters are $K$, $\Gamma$ (as in the power law) and the cutoff energy $E_c$. $E_0$ is a reference energy that we are free to choose for each source. The value of $E_0$ started at 1~GeV but evolved separately for each source at each iteration as described below.
All the known $\gamma$-ray pulsars with significant LAT pulsations were fitted with the PLExpCutoff representation.

Other bright sources (mainly AGN) are also not very well represented by power-law spectra. Analysis of the bright blazars \citep{LAT10_AGN_spec} indicated that a broken power law was the best spectral representation. This however would add two free parameters and therefore was not stable enough for moderately bright sources. We adopted instead a log-normal representation (that we call LogParabola) which adds only one parameter while decreasing more smoothly at high energy than the PLExpCutoff form:
\begin{equation}
\frac{{\rm d}N}{{\rm d}E} = K \left (\frac{E}{E_0}\right )^{-\alpha -
\beta\log(E/E_0)}
\label{eq:logparabola}
\end{equation}

The parameters are $K$, $\alpha$ (spectral slope at $E_0$) and the curvature $\beta$, and $E_0$ is an arbitrary reference energy that evolves for each source along the iterations. Negative $\beta$ (spectra curved upwards) were allowed, although we did not get any.

In order to limit the number of free parameters, we did not fit every non-pulsar source as LogParabola, but only those in which the curvature was significant. In a procedure similar to that applied in the all-sky optimization for the source detection step (\S~\ref{all_sky_optimization}), we assessed that significance for a given source by
$TS_{curve} = 2 (\log \mathcal{L}({\rm LogParabola}) - \log \mathcal{L}$(power-law)),
where $\mathcal{L}$ represents the likelihood function, changing only the spectral representation of that source and refitting all free parameters in the RoI. Since power-law is a special case of LogParabola ($\beta$ = 0) and $\beta$ = 0 is inside the allowed interval we expect that $TS_{curve}$ is distributed as $\chi^2$ with one degree of freedom. We switched to LogParabola if $TS_{curve} > 16$, corresponding to 4 $\sigma$ significance for the curvature. All power-law sources were tested after each iteration, and we checked at the last iteration that $TS_{curve}$ for LogParabola sources was still $>$ 16 (if it was not, the source was downgraded to power law and the RoI was refit).
$TS_{curve}$ was computed for the LAT pulsars as well, but they were not downgraded to power-law if $TS_{curve} < 16$.

The extended sources (\S~\ref{catalog_extended}) were handled on a case by case basis and fitted with either PLExpCutoff, LogParabola or power-law.

The pivot energy $E_p$ (reported as \texttt{Pivot\_Energy}) was computed as the energy at which the relative uncertainty on the differential flux $K$ was minimal. This was done in the parabolic approximation using the covariance matrix between parameters. To improve the validity of the parabolic approximation, we changed the reference energy $E_0$ used for fitting to $E_p$ after each iteration (with the same damping procedure as in \S~\ref{catalog_significance}). This ensured that at the end $E_0$ was close enough to $E_p$.
The value of $\alpha$ (for LogParabola) depends on the reference energy, $\alpha(E_p) = \alpha(E_0) + 2 \beta \log(E_p/E_0)$. The uncertainties on $K$ and $\alpha$ at $E_p$ were derived from the covariance matrix on the actual fitted parameters (relative to $E_0$). The other parameters do not depend on the choice of $E_0$.

In the catalog the differential flux $K$ is reported as \texttt{Flux\_Density} at the reference energy $E_0 = E_p$ (where it is best determined).
The low energy spectral index $\Gamma$ (for PLExpCutoff) or the spectral slope $\alpha(E_p)$ (for LogParabola) are reported as Spectral\_Index. The cutoff energy $E_c$ is reported as \texttt{Cutoff}. The curvature $\beta$ is reported as \texttt{beta}.
For consistency with 1FGL and in order to allow statistical comparisons between the power-law sources and the curved ones, we also report the spectral index of the best power-law fit as \texttt{PowerLaw\_Index} for all sources.

The fitted curvatures $\beta$ sometimes tended to a large value, corresponding to very peaked spectra. There were cases (for example suspected millisecond pulsars) when this kind of spectrum could be real. However this occurred particularly in densely populated regions of the Galactic ridge, where the PSFs overlap and cross-talk between sources in the likelihood analysis is large at low energy. Even though one highly curved spectrum could lead to a better global fit for the RoI, it was not necessarily robust for that particular source, and in many cases we noted that the band fluxes (\S~\ref{catalog_flux_determination}) did not agree with the very curved fits.
In order to avoid extreme cases, we enforced the condition $\beta < 1$, corresponding to changing spectral slope by $2\log 10 = 4.6$ over one decade. Whenever $\beta$ reached 1 for a particular source, we fixed it to 1 and refitted in order to have a reasonable estimate of the errors on the other parameters. Sixty-four sources affected by this are flagged (see \S~3.10 for a description of the flags used in the catalog).
A similar difficulty occurred for 3 faint pulsars in which the low energy index $\Gamma$ tended to be very hard. We limited the values to $\Gamma > 0$ and refitted with $\Gamma$ fixed to 0 when it was reached. Those 3 pulsars were flagged in the same way.
Note that fixing one parameter tends to result in underestimating the errors on the photon and energy fluxes of those sources.
     \subsection{Extended Sources}
\label{catalog_extended}

In the analysis for the 1FGL catalog it became clear that a small number of sources were not properly modeled by a point source, leading to multiple detections being associated with the same source, e.g., the Large Magellanic Cloud (LMC). For the present analysis, twelve sources that have been shown to be extended in the LAT data were included as extended sources. The spatial templates were based on dedicated analysis of each source region, and have been normalized to contain the entire flux from the source ($>99\%$ of the flux for unlimited spatial distributions such as 2-D Gaussians, 2DG). The spectral form chosen for each source is the closest of those used in the catalog analysis (see \S~\ref{catalog_spectral_shapes}) to the spectrum determined by the dedicated analysis\footnote{The templates and spectral models will be made available through the $Fermi$ Science Support Center.  See Appendix~\ref{appendix_fits_format}.}.

The extended sources include seven supernova remnants (SNRs), two pulsar wind nebulae (PWNe), the LMC and the Small Magellanic Cloud (SMC), and the radio galaxy Centaurus A. Notes of interest for each source are provided below:

\begin{itemize}

\item {\bf SMC} -- (2DG, PLExpCutoff) We modeled the SMC using a two-dimensional (2-D) Gaussian function with a width $\sigma = 0\fdg9$. While this is the best-fitting simple geometric model, the morphology of the emission may be more complex \citep{LAT10_SMC}. 

\item {\bf LMC} -- (2$\times$2DG, PLExpCutoff) This complex region, which accounted for five point sources in the 1FGL catalog, has been modeled as a combination of two 2-D Gaussian profiles using the parameters specified in Table 3 of \citet{LAT10_LMC}. The first, with a width of $\sigma = 1\fdg2$, represents emission from the entire galaxy. The second, with a width of $\sigma = 0\fdg2$, corresponds to the $\gamma$-ray bright region near 30~Doradus. Although this model provides a reasonable first order description of the $\gamma$-ray emission seen from the LMC, it is clear that this composite geometric model is not sufficient to fully describe the complex morphology of the source \citep{LAT10_LMC}.
There are five sources in the 2FGL catalog that may be due to excess LMC emission after the fit, though two have blazar associations.

\item {\bf IC 443} -- (2DG, LogParabola) This SNR is modeled by a 2-D Gaussian profile with a width of $\sigma = 0\fdg26$. The log-parabola spectral form most closely matches the spectrum found for this source in the dedicated analysis \citep{LAT10_IC443}.

\item {\bf Vela~X} -- (Disk, PL) We modeled Vela~X using a simple disk with radius $r = 0\fdg88$ and a power law spectral form \citep{LAT10_VelaX}. Since the Vela pulsar is spatially coincident with the Vela~X PWN and significantly brighter, the detailed analysis was performed using the off-pulse events. For the catalog analysis it was necessary to fix the spectral parameters for the power law to the values determined by the off-pulse analysis.

\item {\bf Centaurus A} -- (map, PL) This large radio galaxy has $\gamma$-ray emitting lobes that extend $\sim10\degr$ across the sky. The template used for this source originated from the 22~GHz WMAP image, and excludes a $1\degr$ region around the core \citep{LAT10_CenAlobes}, which is modeled separately as a point source in the catalog. The lobes are clearly resolved in the LAT.

\item {\bf MSH 15$-$52} -- (Disk, PL) This PWN is spatially coincident with the bright $\gamma$-ray pulsar PSR~B1509$-$58. The PWN was detected above 1 GeV, while the pulsar was detected only below 1~GeV by the LAT. We were able to investigate the PWN emission using events from all pulsar phases by excluding data below 1~GeV. That analysis showed that a uniform disk with radius $r = 0\fdg249$ best fit the LAT data \citep{LAT10_PSR1509}. As with Vela~X, the power-law spectral parameters for this source were fixed during the catalog analysis. 

\item {\bf W28} -- (Disk, LogParabola) For W28, only the northern source at (R.A.,  Dec.) $=$ (270$\fdg$34, $-$23$\fdg$44) showed evidence for extension. We modeled this source using a disk with radius $r = 0\fdg39$, the best-fit spatial model found by detailed analysis \citep{LAT10_W28}. As with IC~443, a log-parabola spectral form fits the LAT data best.

\item {\bf W30} -- (Disk, LogParabola) The model for W30 uses a simple disk template centered at (R.A., Dec.) $=$ (271$\fdg$40, $-$21$\fdg$63) with a radius $r = 0\fdg37$. For the catalog analysis, a log-parabola spectral model best fits the source spectrum.

\item {\bf HESS J1825$-$137} -- (2DG, PL) This SNR is modeled with a 2-D Gaussian profile with a width of $\sigma = 0\fdg56$, which we found fit the source emission better than a disk. We tested a power-law spectrum both with and without an exponential cutoff and found that the data was best fit by a simple power-law \citep{LAT11_J1825}.

\item {\bf W44} -- (Ring, LogParabola) The template for the W44 SNR is an elliptical ring with axes $(a, b)_{\rm inner} = 0\fdg22,0\fdg14$, $(a, b)_{\rm outer} = 0\fdg30,0\fdg19$ and a position angle $\theta = 146\degr$ counterclockwise from north \citep{LAT10_W44}. Again, the best spectral model for the SNR is a log-parabola.

\item {\bf W51C} -- (Disk, LogParabola) W51C is well represented by an elliptical disk with axes $(a,b) = 0\fdg40,0\fdg25$ and a position angle $\theta = 0\degr$ \citep{LAT09_W51C}, using a log-parabola spectral form.  

\item {\bf Cygnus Loop} -- (Ring, PLExpCutoff) This relatively large SNR accounted for four sources in the 1FGL catalog. It is best represented by a ring located at  (R.A., Dec.) = (312$\fdg$75, 30$\fdg$85) with an outer radius of $r_{\rm outer} = 1\fdg6$ and an inner radius of $r_{\rm inner} = 0\fdg7$ \citep{LAT11_CygnusLoop}.

\end{itemize}

Table~\ref{tbl:extended} lists the source name, spatial template description, spectral form and the reference 
for the dedicated analysis, where available.  In the 2FGL catalog these sources are tabulated with the point sources, with the only distinction being that no position uncertainties are reported (see \S~3.1.4).

\begin{deluxetable}{lllll}
\tabletypesize{\scriptsize}
\tablecaption{Extended sources used in the 2FGL analysis
\label{tbl:extended}}
\tablewidth{0pt}
\tablehead{

\colhead{2FGL Name}&
\colhead{Extended Source}&
\colhead{Spatial Form}&
\colhead{Spectral Form}&
\colhead{Reference}
}

\startdata
2FGL J0059.0$-$7242e & SMC & 2D Gaussian & Exp Cutoff PL & \citet{LAT10_SMC} \\
2FGL J0526.6$-$6825e & LMC & 2D Gaussian\tablenotemark{a} & Exp Cutoff PL & \citet{LAT10_LMC} \\
2FGL J0617.2+2234e & IC 443 & 2D Gaussian & Log Parabola & \citet{LAT10_IC443} \\
2FGL J0833.1$-$4511e & Vela X & Disk & Power Law & \citet{LAT10_VelaX} \\
2FGL J1324.0$-$4330e & Centaurus A (lobes) & Contour Map & Power Law & \citet{LAT10_CenAlobes} \\
2FGL J1514.0$-$5915e & MSH 15$-$52 & Disk & Power Law & \citet{LAT10_PSR1509} \\
2FGL J1801.3$-$2326e & W28 & Disk & Log Parabola & \citet{LAT10_W28} \\
2FGL J1805.6$-$2136e & W30 & Disk & Log Parabola & \nodata \\
2FGL J1824.5$-$1351e & HESS J1825$-$137 & 2D Gaussian & Power Law & \citet{LAT11_J1825} \\
2FGL J1855.9+0121e & W44 & Ring & Log Parabola & \citet{LAT10_W44} \\
2FGL J1923.2+1408e & W51C & Disk & Log Parabola & \citet{LAT09_W51C} \\
2FGL J2051.0+3040e & Cygnus Loop & Ring & Exp Cutoff PL & \citet{LAT11_CygnusLoop} \\
\enddata

\tablenotetext{a}{To fit the LMC we used a combination of two 2D Gaussian spatial templates.}

\tablecomments{Twelve 2FGL sources that have been modeled as extended sources. More detail regarding the parameters used in the analysis can be found in the text. The publications describing the detailed analysis for W30 is still in preparation.}

\end{deluxetable}

    \subsection{Flux Determination}
\label{catalog_flux_determination}

The source photon fluxes are reported in the 2FGL catalog in the same five energy bands (100 to 300~MeV; 300~MeV to 1~GeV; 1 to 3~GeV; 3 to 10~GeV; 10 to 100~GeV) as in 1FGL. The fluxes were obtained by freezing the spectral index to that obtained in the fit over the full range and adjusting the normalization in each spectral band. For the curved spectra (\S~\ref{catalog_spectral_shapes}) the spectral index in a band was set to the local spectral slope at the logarithmic mid-point of the band $\sqrt{E_n E_{n+1}}$, restricted to be in the interval [0,5]. 
We used binned likelihood in all bands, but contrary to \S~\ref{catalog_significance} we did not distinguish $Front$ and $Back$ events. The pixel sizes in each band were $0\fdg3$, $0\fdg2$, $0\fdg15$, $0\fdg1$, $0\fdg1$ decreasing in size with energy as the PSF improves.

The procedure for reporting either a measurement or an upper limit is the same as for the 1FGL catalog.
For bands where the source was too weak to be detected, those
with Test Statistic in the band $TS_i < 10$ or relative uncertainty on the flux $\Delta F_i / F_i > 0.5$, 2 $\sigma$ upper limits were
calculated, $F_i^{UL}$. Two methods were used, the \textit{profile}
and \textit{Bayesian} methods. In the first \citep{rolke05}, which is used when
$1<TS<10$, the profile likelihood function, $\log\mathcal{L}(F_i)$,
is assumed to be distributed as $\chi^2/2$ and the upper limit
corresponds to the point where $\log\mathcal{L}(F_i)$ decreases by
2 from its maximum value. In the Bayesian method \citep{helene83}, which is used
when $TS<1$, the limit is found by integrating $\mathcal{L}(F_i)$ from 0 up to
the flux that encompasses 95\% of the posterior probability.
With the probability chosen in
this way the upper limits calculated with each method are similar for
sources with $TS=1$.
The 2 $\sigma$ upper limit is then reported in the flux column and the uncertainty is set to 0.

\begin{figure}
\epsscale{.80}
\plotone{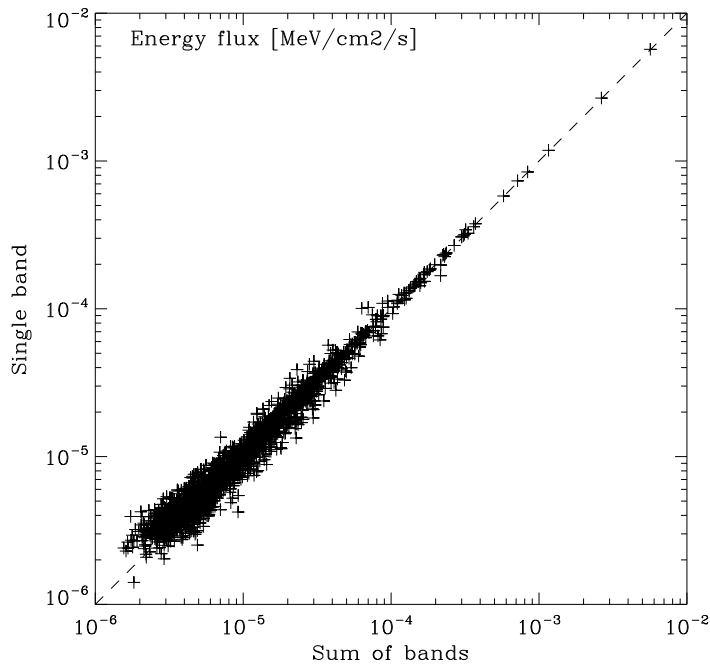}
\caption{Comparison of estimates of the energy flux from 100~MeV to 100~GeV $S_{25}$ from the sum of bands (abscissa) and the fit to the full band (ordinate). No obvious bias can be observed.}
\label{fig:eflux_correlplot}
\end{figure}

In the 1FGL catalog the photon flux between 1 and 100~GeV and the energy flux between 100~MeV and 100~GeV 
\citep[$F_{35}$ and $S_{25}$ in Table \ref{tab:desc},][]{LAT10_1FGL} 
were estimated from the sum of band fluxes because the result of the fit over the full band was biased by the power-law approximation and was inconsistent with the sum of band fluxes for the bright sources. 
In the 2FGL catalog analysis the curved spectral shapes are precise enough to overcome that limitation (Fig.~\ref{fig:eflux_correlplot}). The main advantage of the full spectral fit is that it is statistically more precise because it incorporates the (reasonable) constraint that the spectral shape should be smoothly varying with energy.
Even using the newer data set (with larger effective area at low energy), the relative uncertainties in the lower energy bands tend to be very large. The relative uncertainty on the full photon flux between 100~MeV and 100~GeV ($F_{25}$, dominated by low energy) is much larger than that on $F_{35}$ or $S_{25}$ (23\% vs 15\% and 14\% respectively for a $TS$ = 100 source with spectral index 2.2) and strongly depends on spectral index (whereas that on $F_{35}$ does not). So we do not report the photon flux over the full band in 2FGL. We report $F_{35}$ and $S_{25}$, as in 1FGL, but estimated from the fit over the full band.
For comparison, the relative uncertainties on estimates of $F_{35}$ and $S_{25}$ from the sum of bands (as in 1FGL) are 20\% for the same typical source.
The procedure for reporting upper limits described above applies to $F_{35}$ and $S_{25}$ as well.  Five sources (4 very hard and 1 very soft) have relative uncertainty on $F_{35}$ larger than 0.5. The faintest of those 5 also has relative uncertainty on $S_{25}$ larger than 0.5.

\begin{figure}
\epsscale{.80}
\plotone{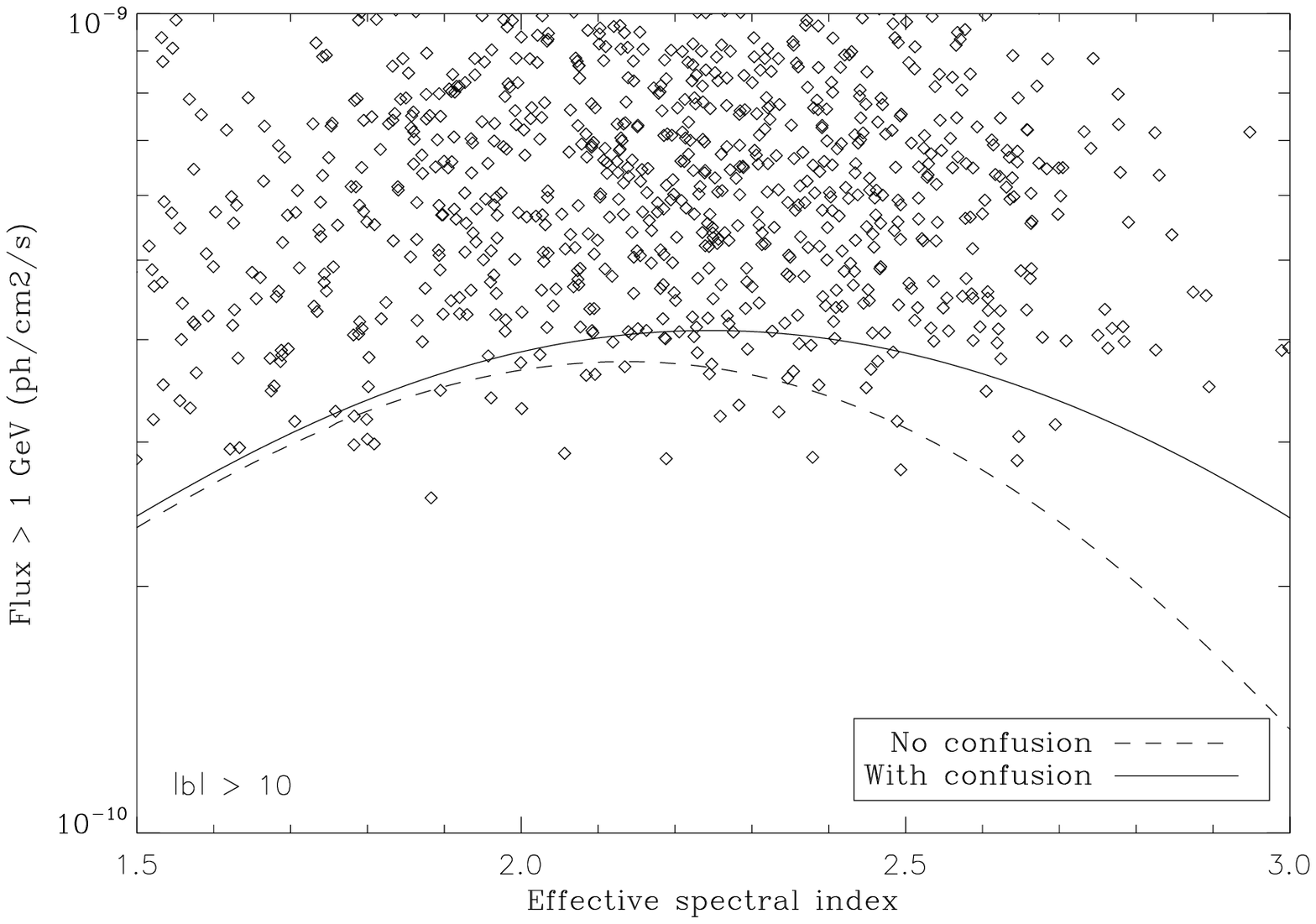}
\caption{Distribution of sources in 2FGL excluding the Galactic plane in the spectral index - photon flux plane. The spectral index is the effective PowerLaw\_Index (power-law fit even for curved sources). The photon flux is between 1 and 100~GeV ($F_{35}$). The low flux threshold is quite sharp around $4 \times 10^{-10}$ ph cm$^{-2}$ s$^{-1}$. The full line shows the expected threshold following App. A of \citet{LAT10_1FGL} accounting for the average confusion, and the dashed line for an isolated source.}
\label{fig:detThreshGeV}
\end{figure}

\begin{figure}
\epsscale{.80}
\plotone{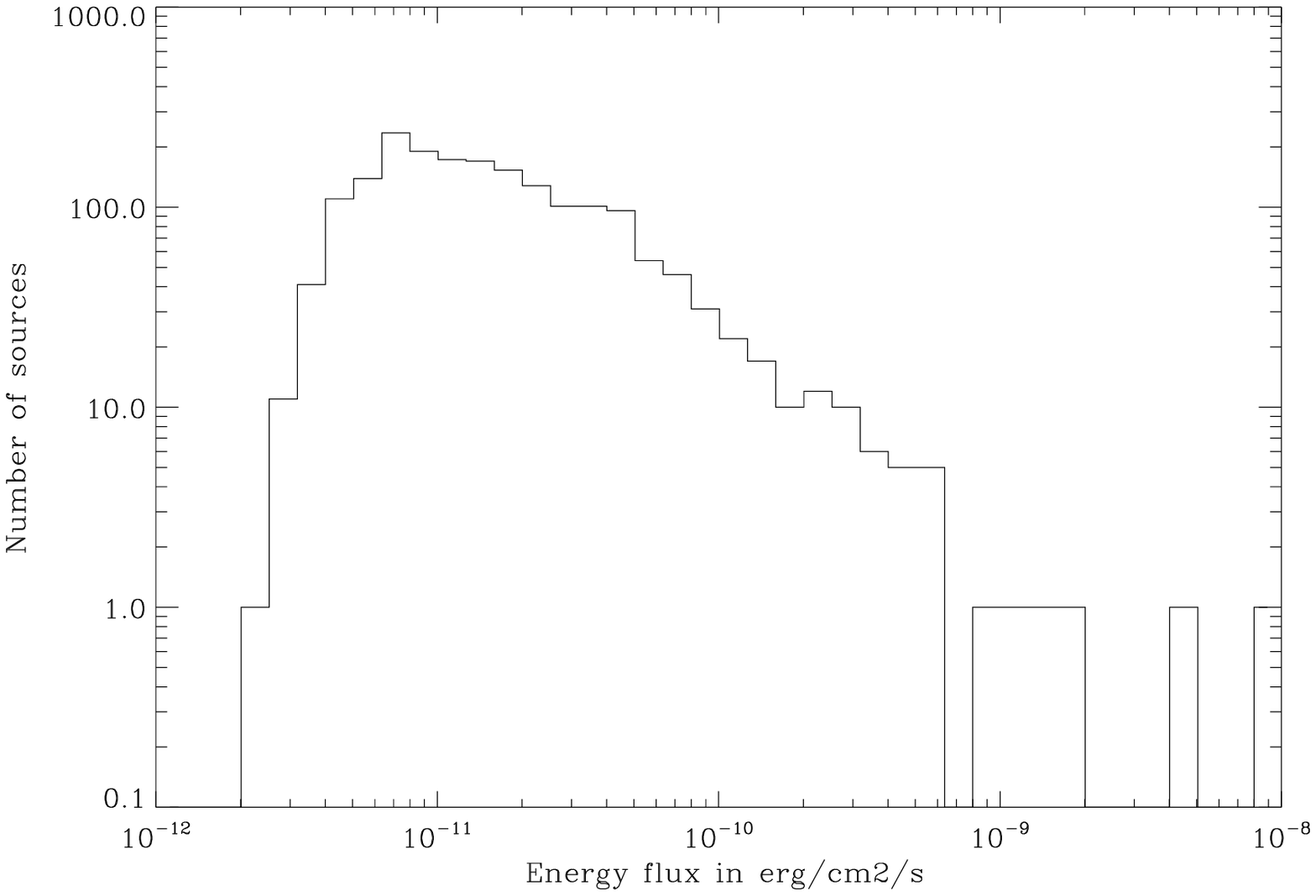}
\caption{Distribution of all sources in 2FGL with respect to log(Energy flux). The low flux threshold is quite sharp around $5 \times 10^{-12}$ erg cm$^{-2}$ s$^{-1}$, indicating that the $TS$ cut that is applied is not too far from a cut on the energy flux $S_{25}$ over the full band (100~MeV to 100~GeV).}
\label{fig:EFluxHisto}
\end{figure}

We show the photon and energy flux distributions for the 2FGL sources in two different ways in Figures \ref{fig:detThreshGeV} and \ref{fig:EFluxHisto}. Figure \ref{fig:EFluxHisto} shows that the range of energy fluxes among the 2FGL sources is greater than 3 decades. Figure~20 of \citet{LAT10_1FGL} was the same plot as Figure~\ref{fig:detThreshGeV} but on the photon flux between 100~MeV and 100~GeV. The detection threshold on the photon flux over the full band depends sensitively on the spectral index of the source. Building a flux-limited sample on that quantity required raising the minimum flux to the detection threshold for soft sources and resulted in discarding most of the hard sources. The photon flux above 1~GeV (or the energy flux), which we show in these figures, is more appropriate to build a flux-limited sample because it discards few sources.

\begin{figure}
\epsscale{.80}
\plotone{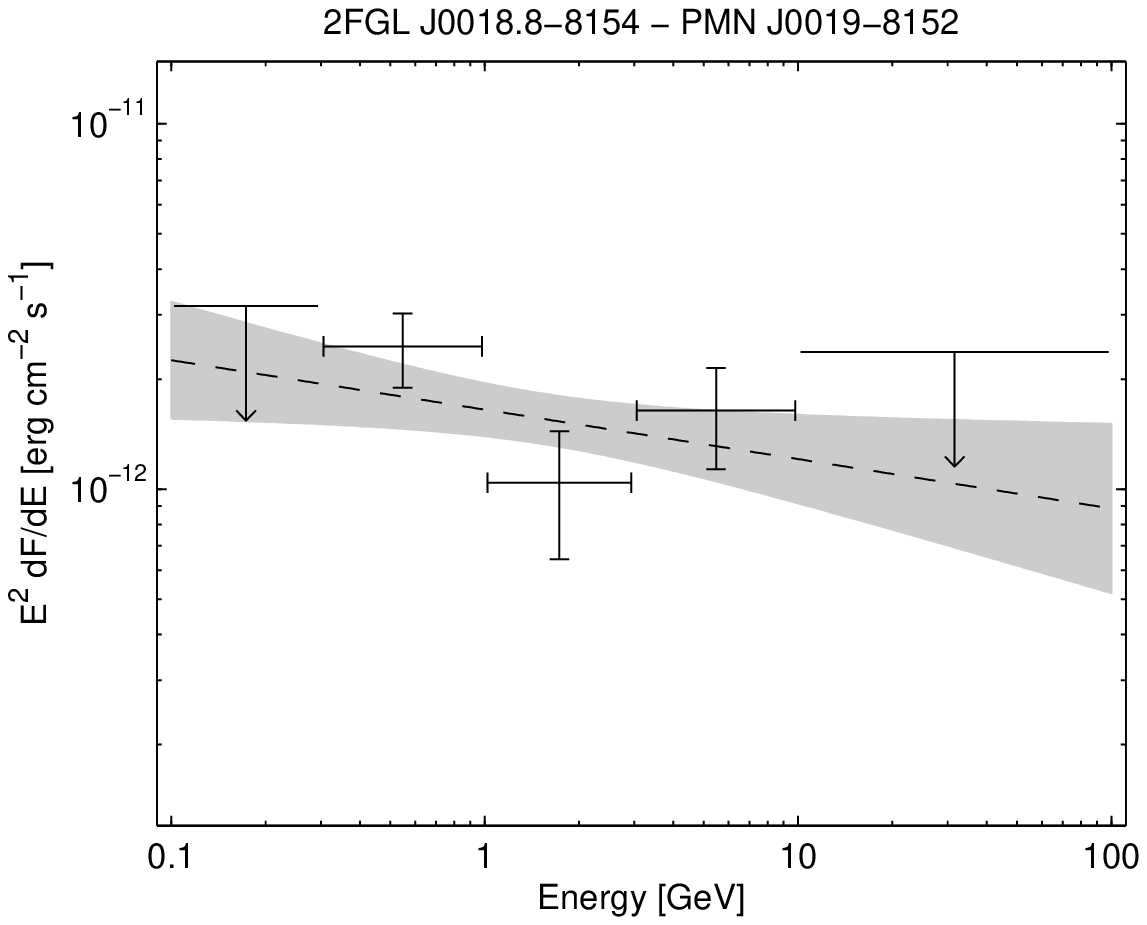}
\caption{Spectrum of a faint AGN, as an example of a power-law spectrum. The fit over the full band (dashed line) is overlaid over the five band fluxes converted to $\nu F_\nu$ units. The grey shaded area (butterfly) shows the formal 1 $\sigma$ statistical error on log(differential flux) as a function of energy, obtained using the covariance matrix involving the parameters of that particular source. The upper limits (here the lowest-energy and highest-energy bands) are 2 $\sigma$.}
\label{fig:PowerLawspec}
\end{figure}

\begin{figure}
\epsscale{.80}
\plotone{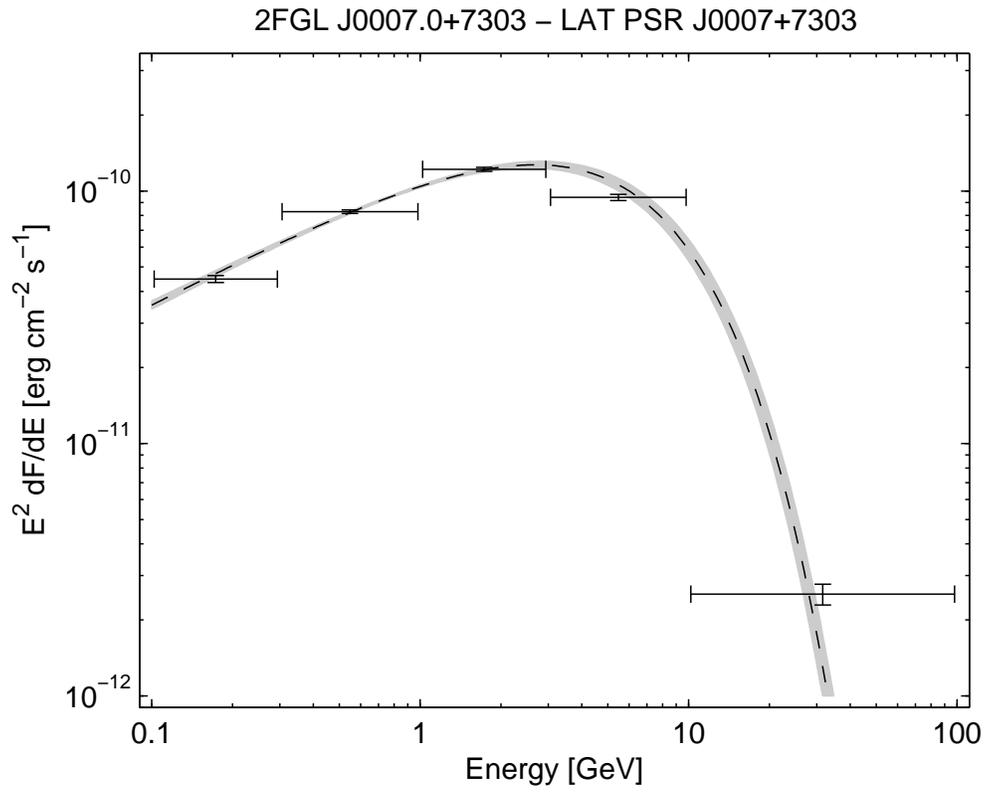}
\caption{Spectrum of the pulsar in CTA1, as an example of an exponentially cutoff spectrum. See Figure~\ref{fig:PowerLawspec} for details.}
\label{fig:PLExpCutoffspec}
\end{figure}

\begin{figure}
\epsscale{.80}
\plotone{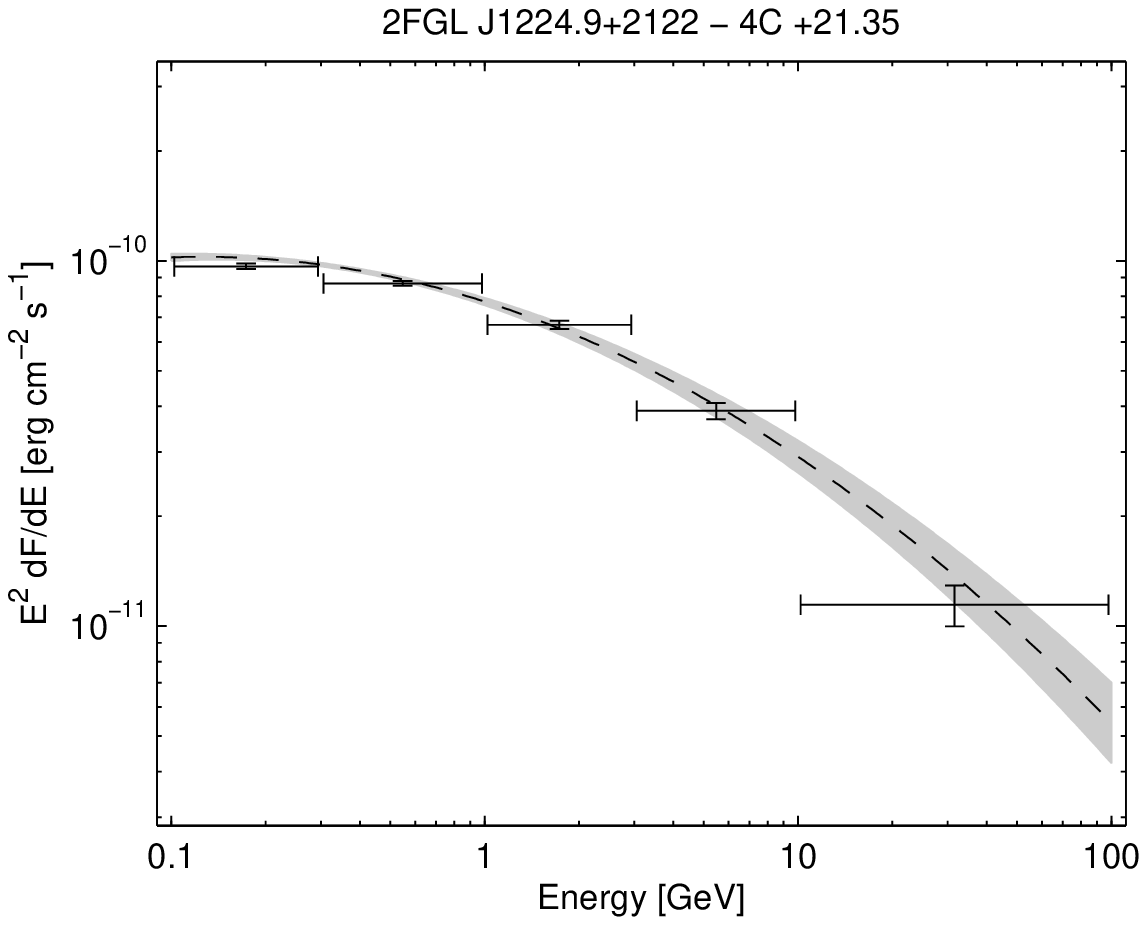}
\caption{Spectrum of the bright AGN 4C +21.35, as an example of a LogParabola spectrum. See Figure~\ref{fig:PowerLawspec} for details.}
\label{fig:LogParabolaspec}
\end{figure}

Figures~\ref{fig:PowerLawspec}, \ref{fig:PLExpCutoffspec}, and \ref{fig:LogParabolaspec} show examples of the band fluxes, with the best fit over the full range overlaid.
From this kind of plot one may build a spectral fit quality indicator similar to the \texttt{Curvature\_Index} of 1FGL.
\begin{equation}
\label{eq:fitquality}
C_{\rm syst} = \sum_i \frac{(F_i - F_i^{\rm fit})^2}{\sigma_i^2 + (f_i^{\rm rel} F_i^{\rm fit})^2}
\end{equation}
where $i$ runs over all bands and $F_i^{\rm fit}$ is the flux predicted
in that band from the spectral fit to the full band. $f_i^{\rm rel}$ reflects the systematic uncertainty on effective area (\S~\ref{catalog_limitations}). They were set to 0.1, 0.05, 0.05, 0.08, 0.1 in our five bands.
Since, in 2FGL, curvature is accounted for in the spectral shape, the interpretation of that quantity is now whether the proposed spectral shape agrees well with the band fluxes or not. We did not report that in the table, but we set a flag (Flag 10 of Table \ref{tab:flags}) whenever $C_{\rm syst} > 16.3$, corresponding to a probability of $10^{-3}$ assuming a $\chi^2$ distribution with 3 degrees of freedom ($5 - 2$, since the majority of sources are fitted with power-law spectra which have 2 free parameters). Thirty-three sources are flagged in this way, including the two brightest pulsars (Geminga and Vela) whose spectrum does not decrease as fast as a simple PLExpCutoff.

A few percent error in the effective area calibration as a function of energy may result in an incorrect report of significant curvature for very bright sources.
There is no obvious rigorous way to enter systematic uncertainties in the $TS_{curve}$ calculation (\S~\ref{catalog_spectral_shapes}).
In order to do that approximately, we note that $TS_{curve}$ is an improved estimator of how much the spectrum deviates from a power-law. The analog of $TS_{curve}$ in 1FGL was $C^{\rm PL}_{\rm nosyst}$, applying Eq.~\ref{eq:fitquality} to the power-law fit with no $f_i^{\rm rel}$ term ($TS_{curve}$ is a purely statistical quantity). We can compare $C^{\rm PL}_{\rm nosyst}$ with the same quantity $C^{\rm PL}_{\rm syst}$ obtained with the $f_i^{\rm rel}$ term (\texttt{Curvature\_Index} of 1FGL). Their ratio is a measure of how much the systematic uncertainties reduced \texttt{Curvature\_Index}.
We can then apply that same ratio to $TS_{curve}$ and we report in the catalog \texttt{Signif\_Curve} = $\sqrt{TS_{curve} \; C^{\rm PL}_{\rm syst} / C^{\rm PL}_{\rm nosyst}}$, converting to $\sigma$ units.

We consider that sources with \texttt{Signif\_Curve} $>$ 4 are significantly curved. The consequence of introducing the systematic uncertainties is that 40 sources in the catalog have a LogParabola spectrum because $TS_{curve} > 16$ (\S~\ref{catalog_spectral_shapes}) even though \texttt{Signif\_Curve} $<$ 4. We do not claim that the curvature is real for those sources, even though it is statistically significant.

    \subsection{Variability}
\label{catalog_variability}

Temporal variability is relatively common in $\gamma$-ray sources and
provides a powerful tool to associate them definitively with objects
known at other wavelengths and to study the physical processes
powering them. We present a light curve for each source in the catalog,
produced by dividing the data into approximately monthly time bins
and applying the likelihood analysis procedure to each. The details of
the light curve analysis and how the results are presented are
summarized below:
\begin{itemize}
\item There are 24 time bins, starting at the beginning of the data set, approximately
  $54682.66$~MJD (\S~\ref{DataDescription}). The first 23 bins have durations of $30.37$\,days; the
  final has a duration of $27.88$\,days. The first 11 time bins
  correspond exactly to those of 1FGL.
\item The parameters describing the spectral shapes of the sources in
  the RoI are fixed in the light curve calculation. Only the
  normalizations of the source of interest, the diffuse backgrounds,
  and bright and nearby catalog sources (see
  section~\ref{catalog_significance}) are allowed to vary.
  We use binned likelihood, but do not distinguish $Front$ and $Back$ events. The pixel size is set to $0\fdg2$.
\item For each time bin, the photon flux over the full energy range
  (100~MeV to 100~GeV), $F_i$, its error, $\Delta F_i$ and the
  detection significance, $TS_i$, are presented in the catalog. With
  the spectral shape of each source frozen in the light curve analysis,
  the relative uncertainty on $F_{25}$ is the same as that of $F_{35}$
  and $S_{25}$, and it is reasonable to present the photon flux over
  the full energy range in this case.
\item For time bins where the source is too weak to be detected, those
  with $TS_i<10$ or $\Delta F_i/F_i>0.5$, 95\% upper limits $F_i^{UL}$
  are calculated following the same method as in
  \S~\ref{catalog_flux_determination}.
  A fraction of those have flux exactly equal to 0,
  because the Poisson likelihood framework that we use does not accept
  negative flux values.
\item In the case of an upper limit, the best-fit flux value is given
  in the catalog, and the error is replaced by $0.5(F_i^{UL}-F_i)$. 
  This allows bands with upper limits to be treated consistently with
  the other bands while preserving enough information to extract the 
  upper limits.  The FITS version of the catalog\footnote{The FITS version of the catalog is available through the $Fermi$ Science Support Center.  See Appendix~\ref{appendix_fits_format}.} has a flag column to indicate when an entry in a flux 
  history is an upper limit.  Please note that this is a different convention to that used to report flux upper limits for the energy bands (\S~3.5). See Appendix C for more information.
\item A total of 340 sources have only upper limits on monthly
  timescales. These sources have an average integrated significance over the full 2-year data set of 5.3 $\sigma$.  At the
  opposite extreme, 94 sources are detected significantly in every one
  of the time periods.
\end{itemize}

\begin{figure}[p]
\centering
\includegraphics[width=0.99\textwidth]{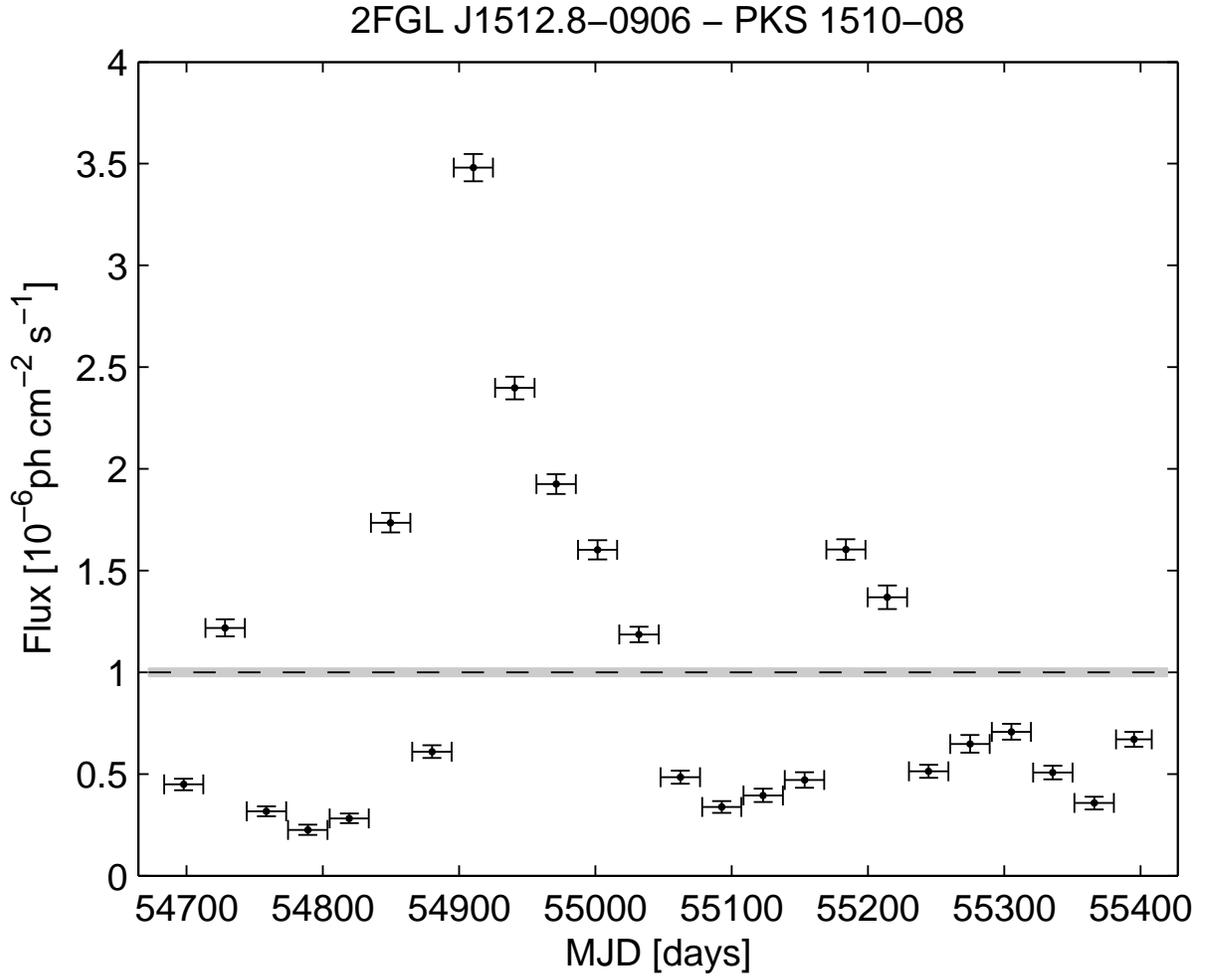}
\caption{Light curve for the bright blazar PKS~1510$-$089
in the full energy range (100~MeV to 100~GeV). The dashed
  line depicts the average flux from the analysis of the full 24-month
  dataset.}
\label{fig:lc_pks1510}
\end{figure}

To test for variability in each source we construct a variability
index from the value of the likelihood in the null hypothesis, that the
source flux is constant across the full 2-year period, and the value
under the alternate hypothesis where the flux in each bin is
optimized:
\begin{equation}
\label{EQ::TSVAR}
TS_{var} = 2\left[\log\mathcal{L}(\{F_i\})-\log\mathcal{L}(F_{Const})\right] = 
2\sum_i\left[\log\mathcal{L}_i(F_i)-\log\mathcal{L}_i(F_{Const})\right] 
= 2\sum_iV_i^2
\end{equation}
where the log likelihood for the full time period,
$\log\mathcal{L}(\{F_i\})$, can be expressed as a sum of terms for the
individual time bands, $\log\mathcal{L}_i$. If the null hypothesis is
correct $TS_{var}$ is distributed as $\chi^2$ with 23 degrees of
freedom, and a value of $TS_{var}>41.6$ is used to identify variable
sources at a 99\% confidence level. For most sources the value for
$F_{Const}$ is close to the value derived from the likelihood analysis
of the full time period, although strong variability in nearby
background sources can cause to them to differ in some cases. The
light curve for PKS~1510$-$089, a bright blazar, is shown in
Figure~\ref{fig:lc_pks1510}. This source is easily flagged as variable,
with $TS_{var}=6406$.

Upper limits calculated through the profile method are handled
naturally in the variability index procedure described above, but
those calculated using the Bayesian method would have to included in
an ad hoc manner. Instead, when calculating the variability index, the
results of the profile method are used for all upper limits.

\begin{figure}[p]
\centering
\includegraphics[width=0.99\textwidth]{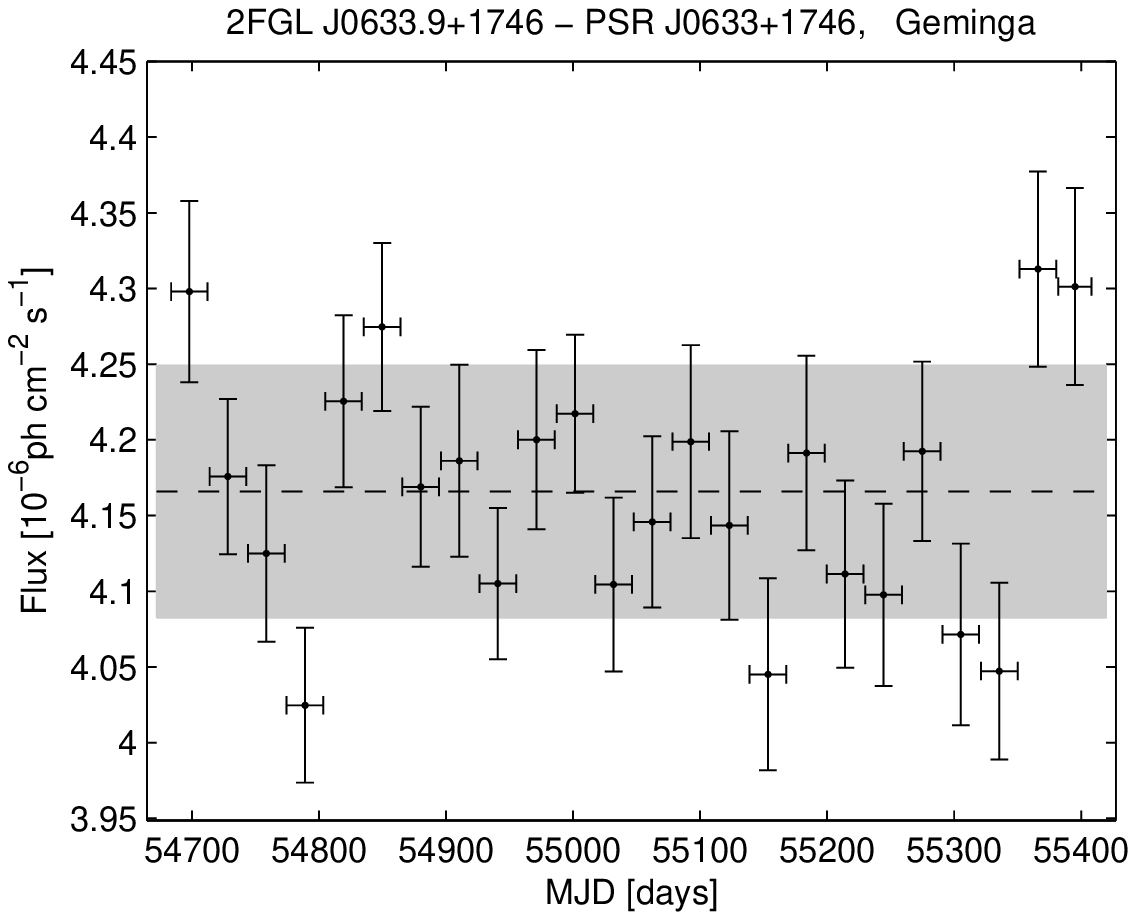}
\caption{Light curve for the bright pulsar Geminga
(100~MeV to 100~GeV). The gray band
depicts the size of the 2\% systematic correction applied to the
calculation of the variability index. The error bars on the flux
points show the statistical errors only.}
\label{fig:lc_geminga}
\end{figure}

As in 1FGL, the brightest pulsars detected by the LAT are flagged as
being variable with this procedure. This apparent variability is
caused by systematic errors in the calculation of the source exposure,
resulting from small inaccuracies in the dependence of the IRFs on the
source viewing angle, coupled with changes in the observing profile as
the orbit of the spacecraft precesses. We introduce a correction
factor to account for these errors, and fix the size of this
correction such that the bright pulsars are steady. Specifically, we scale each
$V_i^2$ in the summation of $TS_{var}$ by a factor 
which combines the error on the flux each time bin in quadrature with a fixed fraction of the overall flux,
\begin{displaymath}
TS_{var} = 2
\sum_i \frac{\Delta F_i^2}{\Delta F_i^2 + f^2 F_{Const}^2} V_i^2.
\end{displaymath}
A value of $f=0.02$, i.e. a 2\% systematic correction factor, was
found sufficient such that only PSR~J1741$-$2054 remains (marginally)
above threshold among the LAT pulsars, excluding the Crab which was
recently discovered to have a highly variable nebular component at LAT
energies \citep{AGILE11_CrabFlares,LAT11_CrabFlares}. This is smaller
than the 3\% correction required in 1FGL, the improvement resulting
from the higher-fidelity IRFs used in this work. This systematic error
component is included in the flux errors reported in the catalog FITS
file. Figure~\ref{fig:lc_geminga} shows the light curve for the pulsar
Geminga (PSR~J0633$+$1746), one of the brightest non-variable sources
in 2FGL.

\begin{figure}[p]
\centering
\includegraphics[width=0.99\textwidth]{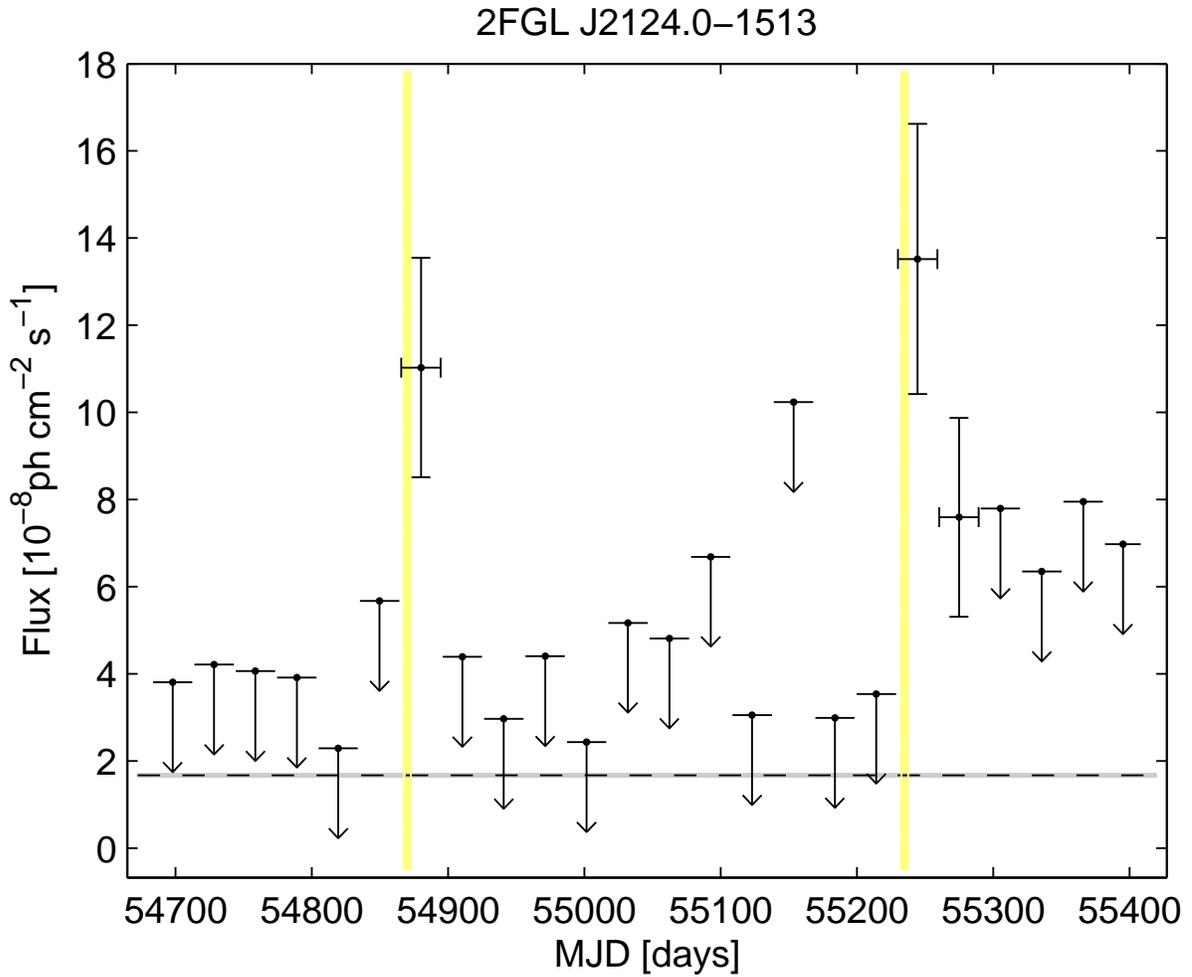}
\caption{Light curve for the unassociated source 2FGL~J2124.0$-$1513
  (100~MeV to 100~GeV).
  Time periods in which the sun is closer than $2\fdg5$ to the
  source are marked with yellow vertical bands. In this case, a large
  fraction of the detection significance is accumulated during these
  periods, and the source is flagged as suspicious in the catalog.}
\label{fig:lc_j2124}
\end{figure}

The Sun is a bright, extended source of $\gamma$ rays, both from cosmic-ray interactions in its outer atmosphere and from IC scattering of cosmic-ray electrons on the solar radiation field, which produces an extended $\gamma$-ray halo around the Sun \citep{LAT11_Sun}.  For sources close to the ecliptic, solar conjunctions can lead to
significant enhancements of the flux detected during the time periods
when the Sun is closer than approximately $2\fdg5$ to the
source. Sources for which a large fraction of the total detection
significance comes during such periods are flagged as suspicious in
the catalog. The light curve for such a source, 2FGL~J2124.0$-$1513, is
shown in Figure~\ref{fig:lc_j2124}. The Moon is comparably bright to the Sun in $\gamma$ rays \citep{LAT11_Moon} and lunar conjunctions also
potentially affect the fluxes measured from LAT sources, but the
higher apparent speed of the Moon and
parallax from the motion of the spacecraft mean that such close
conjunctions are brief, and the precession of the orbit of the Moon means that for any given source conjunctions are less frequent than the $\sim$28 d period of the orbit.  In addition, the $\gamma$-ray emission of the Moon does not include a bright, extended IC component \citep{LAT11_Moon}. Hence, we do not attempt to identify sources which may be affected by the Moon nor to flag time periods where lunar conjunctions
occur.

Light curves for all 2FGL sources are available from the Fermi Science
Support Center.

    \subsection{Limitations and Systematic Uncertainties}
\label{catalog_limitations}

\begin{figure}
\epsscale{.80}
\plotone{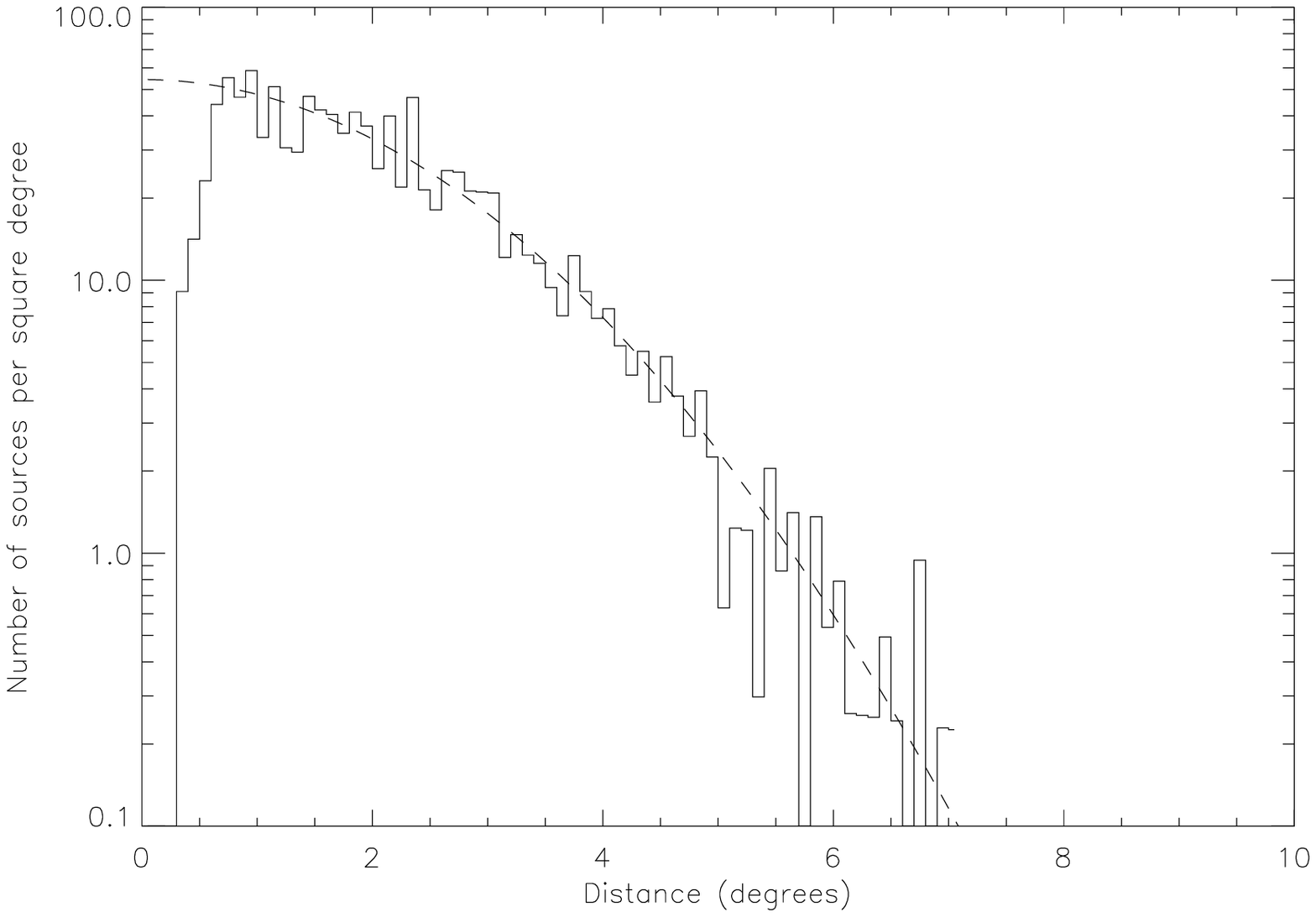}
\caption{Distribution of the distances $D_n$ to the nearest neighbors
of all detected sources at $|b| > 10\degr$. The number of entries is divided
by $2 \pi D_n \, \Delta D_n$ in which $\Delta D_n$ is the distance bin,
in order to eliminate the 2-dimensional geometry.
The overlaid curve is the expected Gaussian distribution
for a uniform distribution of sources with no confusion.}
\label{fig:confusion}
\end{figure}

A limitation for the catalog analysis is source confusion.  (The related issue of systematics for localization is discussed
in \S~\ref{catalog_localization}.)
Confusion is of course strong in the inner Galaxy, where the source density is very high, 
but it is also a significant issue elsewhere.
The average distance between sources outside the Galactic plane
is $2\fdg8$ (it was $3\degr$ in 1FGL), 
to be compared with a per photon containment radius
$r_{68} = 0\fdg8$ at 1~GeV where the sensitivity is best.
The ratio between these numbers is not large enough
that confusion can be neglected.
As for the 1FGL catalog analysis \citep{LAT10_1FGL} we study source confusion by evaluating the distribution
of distances between each source and its nearest neighbor ($D_n$)
in the area of the sky where the source density is approximately uniform, i.e.,
outside the Galactic plane. 
This is shown in Figure~\ref{fig:confusion}, to be compared with Figure~9 of \citet{LAT10_1FGL} which also details the expected distribution.
The histogram still falls off toward $D_n=0$, but follows the expected distribution down to $1\degr$ or so instead of $1\fdg5$ in 1FGL.
We estimate that some 43 sources within $1\degr$ of another one were missed because of confusion (to be compared with the 1319 sources observed at $|b| > 10\degr$). This means that the fraction of missed sources decreased from 7.7\% in the 1FGL analysis to 3.3\% for 2FGL.
This attests to the progress made in the detection process (\S~\ref{catalog_detection}).

An important issue for the evaluation of spectra is the systematic
uncertainties of the effective area of the instrument.
Compared to the 1FGL instrument response  functions (P6\_V3), the current P7\_V6 response functions have somewhat reduced 
systematic uncertainties. 
 The current estimate of the remaining systematic
uncertainty is 10\% at 100~MeV,  decreasing to 5\% at 560~MeV and increasing to 10\%
at 10~GeV and above \citep{LAT11__calib}. This
uncertainty applies uniformly to all sources.
Our relative errors (comparing one source to another
or the same source as a function of time) are much smaller,
as indicated in \S~\ref{catalog_variability}.

The model of diffuse emission is the other important source
of uncertainties. Contrary to the effective area, it does not affect all sources
equally: its effects are smaller outside the Galactic plane 
where the diffuse emission is fainter and varying on larger angular scales.
It is also less of a concern in the high energy bands ($>$ 3~GeV) where
the core of the PSF is narrow enough that the sources dominate the background
under the PSF.
But it is a serious concern inside the Galactic plane
in the low-energy bands ($<$ 1~GeV) and particularly
inside the Galactic ridge ($|l| < 60\degr$) where the diffuse emission
is strongest and very structured, following the molecular cloud distribution.
It is not easy to assess precisely how large the uncertainties are, because they relate to uncertainties in the distributions of interstellar gas, the interstellar radiation field, and cosmic rays, which depend in detail on position on the sky.
We discuss the Galactic ridge more specifically in \S~\ref{catalog_ism}.

For an automatic assessment we have tried re-extracting the source locations and fluxes assuming the same diffuse model that we used for 1FGL, and also the same event selection as in 1FGL but with improved calibration (P6\_v11).
The results show that the systematic uncertainty more or less follows the statistical one (i.e., it is larger for fainter sources in relative terms) and is of the same order.
More precisely, the dispersions of flux and spectral index are 0.8 $\sigma$ at $|b| > 10\degr$, and 1.3 $\sigma$ at $|b| < 10\degr$.
We have not increased the errors accordingly because this older model does not fit the data as well as the newer one.
From that point of view we may expect this estimate to be an upper limit.
On the other hand both models rely on nearly the same set of H~{\sc I} and CO maps of the gas in the interstellar medium, which we know are an imperfect representation of the mass.
That is, potentially large systematic uncertainties are not accounted for by the comparison.
So we present the figures as qualitative estimates.
We also use the same comparison to flag outliers as suspect (\S~\ref{catalog_analysis_flags}).

Finally, we note that handling $Front$ and $Back$ events separately for the significance and spectral shape computation (\S~\ref{catalog_significance}) introduces another approximation. Because the free parameters are the same for both categories of events, this amounts to assuming that the isotropic diffuse emission is the same for $Front$ and $Back$ events. This is actually not true because it contains internal background that is larger for $Back$ events (see \S~\ref{lat_and_background}).  
This effect is significant only below 400 MeV (\S~\ref{DiffuseModel}), and so the consequence is an underestimate of the low-energy flux, which results in a systematic hardening of the measured power-law spectral index but which is nearly always less than its statistical uncertainty. 
Thus in terms of the absolute change in spectral index, the effects are greatest for soft sources.

\subsection{Point Sources and Extended Sources}
\label{catalog_point_extended_sources}

Except for the diffuse emission and the 12 sources explicitly considered as spatially extended, all sources in the catalog are
assumed to be point-like.  Just as the modeling of the diffuse emission can affect the properties of point sources (as discussed in the previous section), the treatment of known or unknown extended sources can similarly influence the analysis of nearby point sources.  This influence can be felt in three ways:

\begin{enumerate}
\item  The modeling of an extended source is limited by the detailed knowledge of the $\gamma$-ray emissivity of the source as a function of position on the sky.  As noted in \S~3.4, the modeling for the catalog was done using largely geometric functions.  The true distribution can have residual excesses that the catalog analysis then treats as point sources.   Examples are the sources near the Large Magellanic Cloud:   2FGL J0451.8$-$7011, 2FGL J0455.8$-$6920, 2FGL J0532.5$-$7223, 2FGL J0533.3$-$6651,  2FGL J0601.1$-$7037  Although some of these may be unrelated to the LMC itself (two have  blazar associations), some may be residuals from the modeling.  Sources close to any of the extended sources should be treated warily in detailed analysis of such regions. 
\item  Some known or likely extended sources are not among the 12 that were modeled for the catalog analysis, having been recognized and measured only after the catalog analysis was largely complete.  In such cases, the catalog analysis finds one or more point sources at or near the possible extended source.   Two clear examples are supernova remnants.
RX J1713.7$-$3946 is represented in the catalog by 2FGL J1712.4$-$3941, but recent analysis has shown this 
SNR to be an extended GeV source \citep{LAT11_RXJ1713}.  
In Table \ref{tab:snrext},  RX J0852.0$-$4622  shows four associated 2FGL sources: 
J0848.5$-$4535,
J0851.7$-$4635,
J0853.5$-$4711, and
J0855.4$-$4625.
All of these are likely part of the spatially extended supernova remnant \citep{LAT11_RXJ0852}.  
Other clusters of sources in Table \ref{tab:snrext} may indicate yet-unresolved extended objects.  
As longer exposures with the LAT collect more of the highest-energy photons with the best angular resolution, additional spatial structure will be revealed in the data. 
\item  A spectral bias can be introduced if an extended source is analyzed as if it were a point source.  In such cases the calculated spectrum is likely to be softer than the true spectrum.  At higher energies where the LAT PSF is closer to the size of the extended source, the extension will cause such photons to be lost. 
\end{enumerate}

    \subsection{Sources Toward Local Interstellar Clouds and the Galactic Ridge}
\label{catalog_ism}
\begin{figure}[!ht]
\centering
\includegraphics[width=0.5\textwidth]{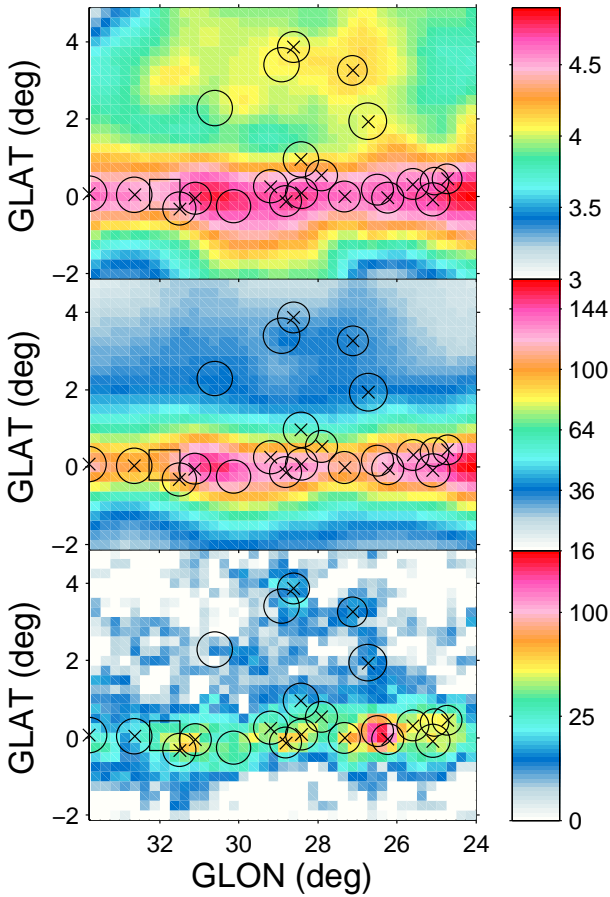}
\caption{From top to bottom: the CO contribution to the interstellar photon counts, the total interstellar photon counts, and the photon residual counts above the model for diffuse $\gamma$-ray emission, all in the 1--11 GeV energy band. The circles mark the effective 50\% containment radii of the 2FGL sources for the 1--10 GeV band. The `c' sources are crossed. The square notes an identified source. The photon residual map has been smoothed for display with a $\sigma = 0\fdg125$ Gaussian. The 2FGLc sources seen above the Galactic plane, with $TS$ ranging from 26 to 75, follow an extended and clumpy excess of interstellar emission}
\label{fig:2FGLc_gas}
\end{figure}

The interstellar part of the model for diffuse emission of the Galaxy has greatly improved since the 1FGL catalog analysis, in particular in angular resolution (\S~\ref{DiffuseModel}). However, the use of large-scale rings in the Milky Way and of a single ring in the solar neighborhood (containing most of the gas-related diffuse emission off the Galactic plane) does not allow for small-scale variations in the gas and dust properties used to derive the target mass for cosmic rays, or in the cosmic-ray spectrum itself.  The optical depth correction applied in deriving column densities from the H~{\sc I} 21-cm line observations also is necessarily based on the approximation that the gas has a uniform spin temperature across the Galaxy. As a result, extended and structured excesses of $\gamma$ radiation are present above the diffuse model. The renormalization of the diffuse model within each RoI lessens, but cannot always remove, the impact of the diffuse excesses.  Point sources detected in the regions of the excesses can be formally very significant but are necessarily suspect (see Fig.~\ref{fig:2FGLc_gas}).  

One particularly difficult region is the local arm tangent in Cygnus.  The region $75^{\circ} \leq l \leq 85^{\circ}$, $|b| \leq 15^{\circ}$ contains 18 2FGL sources, 13 without firm identifications.  Of these 13, we find that 8 sources,  with detection significances ranging up to 13 $\sigma$ have properties that are especially sensitive to the diffuse emission model (see \S~\ref{catalog_analysis_flags}, in particular Flags 1--4 in Table~\ref{tab:flags}).  The diffuse emission in Cygnus has recently been the subject of a detailed study \citep{LAT11_CygISM} and evidence suggests that some of the excess diffuse emission is due to an extended cocoon of unusually hard-spectrum cosmic rays \citep{LAT11_Cygcocoon}.

\begin{figure}[!ht]
\centering
\includegraphics[width=0.5\textwidth]{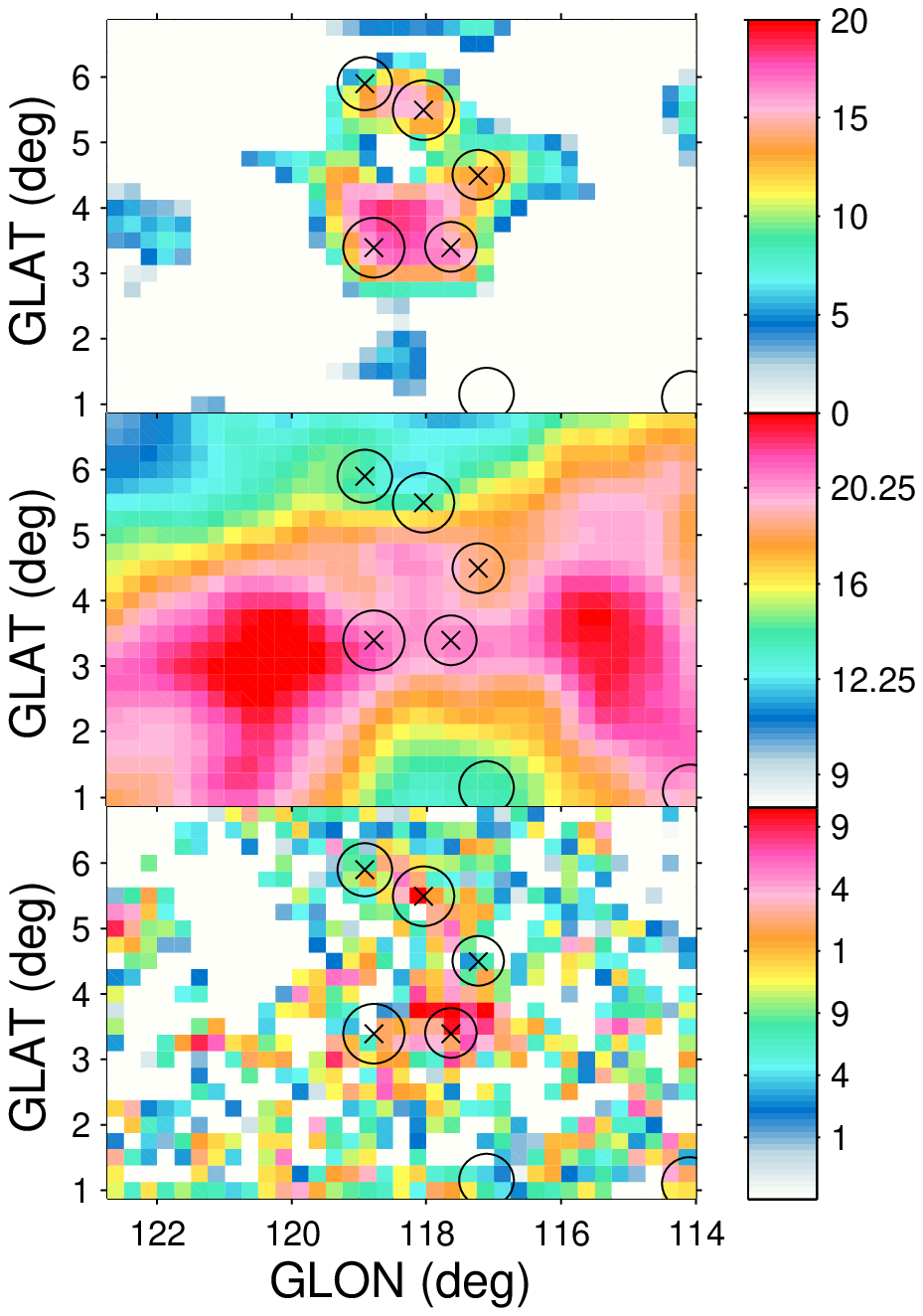}
\caption{From top to bottom: the absolute value of the dust negative residual photon counts incorporated in the diffuse model, the total interstellar photon counts, and the photon residual counts above the diffuse model, all in the 1--11 GeV energy band. The circles mark the effective 50\% containment radii of the 2FGL sources for the 1--10 GeV band. The `c' sources are crossed. The photon residual map has been smoothed for display with a $\sigma = 0\fdg125$ Gaussian. The 2FGL `c' sources are distributed along the rim of a large H~{\sc II} region where the dust temperature correction led to an overestimate of the dust column densities in the ionized gas. The negative dust residuals have artificially reduced the diffuse $\gamma$-ray intensity in these directions.}
\label{fig:2FGLc_EBVring}
\end{figure}

We have used a dust reddening map to trace substantial amounts of dark gas in addition to the atomic and molecular gas seen in H~{\sc I} and CO emission lines \citep{grenier05, planck11}. This made essential improvements over wide regions from low to medium latitudes, but inaccuracies in the infrared color corrections used to build the reddening map \citep{schlegel98} can cause diffuse residuals toward bright H~{\sc II} regions or stellar clusters by artificially lowering the gas column densities measured in their directions \citep[see Figs. 10 and 11 of][]{LAT10_1FGL}. There are fewer such artifacts in 2FGL than in 1FGL, but examples can be found in Orion, Taurus, and near the source LS\,I +61~303; see also Figure \ref{fig:2FGLc_EBVring}. Another known limitation of the diffuse model relates to the optical thickness of the CO lines and the saturation of the CO intensity toward very dense clouds. Since stellar clusters are born in the clouds, both CO saturation and dust temperature corrections may cooperate to under-predict the gas mass in dense molecular clouds. Self-absorption of the H~{\sc I} lines also leads to under-predicted column densities in the dense atomic phase. These limitations are particularly relevant at low latitudes, in the inner Galaxy or toward the tangent directions of the Galactic spiral arms. 

We have inspected all the 2FGL sources to search for potential problems with the underlying diffuse model. It is unlikely that sources with very high $TS$ can be diffuse excesses. Based on the examination of the sources toward Cygnus, Orion and other nearby clouds, as well as the 1FGL sources with the `c' designation that are not confirmed in 2FGL, we tentatively consider that sources with $TS \gtrsim 200$, 130, or 80 are unlikely to be diffuse features depending on the intensity of the diffuse background (respectively when the photon count per pixel $N_{bkgd}$, integrated from 589~MeV to 11.4~GeV in the diffuse model cube, without the isotropic contribution, is  $N_{bkgd} > 100$, $60 \leq N_{bkgd} \leq 100$, or $N_{bkgd} < 60$). 

Given the large change in the width of the PSF across the LAT energy band, we computed the effective 50\% containment radius for each source from its best-fit spectrum.  We overlaid these on predicted photon count maps from the Galactic diffuse model, both for the total emission and for the individual gas components in each phase, in seven energy bands (the five energy bands of the catalog, plus the integral 0.5--10~GeV and 1--10~GeV bands).  We also compared photon residual maps (data minus model) in the same energy bands against the predicted counts maps for the individual gas components.  We also took into account the $TS$ values reached in the five catalog energy bands and the spectral index of each source. Off the Galactic plane, we flagged (Flag 6 of Table \ref{tab:flags}) unassociated sources coinciding with dust temperature or dense CO defects, or concurrent with extended residuals that followed interstellar features (as in Figure \ref{fig:2FGLc_gas}). Sources with $TS$ larger than the background-dependent threshold quoted above or with a spectral index $\Gamma < 2$ were not flagged. In the Galactic plane (i.e. at $|b| \leq 2^{\circ}$ for $|l| \leq 70^{\circ}$, or $|b| \leq 3^{\circ}$ at higher longitudes), we flagged two types of sources: (i) unassociated sources with overlapping 50\% containment radii above 500~MeV, unless their $TS$ exceeded the background-dependent threshold or their spectral index were $< 2$; (ii) low-significance unassociated sources with $TS \leq 80$ for $N_{bkgd} \geq 160$, unless their spectral indices were $< 2$. This strategy ensured that most of the Galactic ridge sources, which are closely packed together to make up for the extended photon residuals along the plane, are flagged, but it leaves all the identified and associated, intense, and hard sources out of the systematic ridge flag we had used in 1FGL. Every source was then manually checked with the same set of maps as for the work at higher latitude.

We have added the designator `c' to the names of the flagged sources to indicate that they are to be considered as potentially confused with interstellar emission. Their position, emission characteristics, or even existence may not be reliable. The `c' designator applies to 162 sources in the 2FGL catalog. 

    \subsection{Analysis Flags}
\label{catalog_analysis_flags}

\begin{deluxetable}{cl}

\tablecaption{Definitions of the Analysis Flags \label{tab:flags}}
\tablehead{
\colhead{Flag\tablenotemark{a}} & 
\colhead{Meaning}
}

\startdata
  1  & Source with $TS > 35$ which went to $TS < 25$ when changing the diffuse model \\
     & (\S~\ref{catalog_limitations}). Note that sources with $TS < 35$ are not flagged with this bit because \\
     & normal statistical fluctuations can push them to $TS < 25$. \\
  2  & Moved beyond its 95\% error ellipse when changing the diffuse model. \\
  3  & Flux ($>$ 1~GeV) or energy flux ($>$ 100~MeV) changed by more than 3 $\sigma$ when \\
     & changing the diffuse model. Requires also that the flux change by more than \\ 
     & 35\% (to not flag strong sources). \\
  4  & Source-to-background ratio less than 20\% in highest band in which $TS > 25$. \\
     & Background is integrated over $\pi r_{68}^2$ or 1 square degree, whichever is smaller.\\
  5  & Closer than $\theta_{\rm ref}$ from a brighter neighbor.  $\theta_{\rm ref}$ is defined in highest band in \\
     & which source $TS > 25$, or the band with highest $TS$ if all are $< 25$. $\theta_{\rm ref}$ is set \\
     & to $2\fdg17$ (FWHM) below 300~MeV, $1\fdg38$ between 300~MeV and 1~GeV, $0\fdg87$ \\
     & between 1~GeV and 3~GeV, $0\fdg67$ between 3 and 10~GeV and $0\fdg45$ above \\
     & 10~GeV ($2 \, r_{68}$). \\
  6  & On top of an interstellar gas clump or small-scale defect in the model of \\
     & diffuse emission; equivalent to the `c' designator in the source name (\S~\ref{catalog_ism}). \\
  7  & Not used. \\
  8  & Inconsistent position determination (\S~\ref{catalog_localization}); best position from optimization \\
     & outside the 1 $\sigma$ (39\% in 2D) contour from the $TS$ map. \\
  9  & Elliptical quality $>$ 4 in {\it pointlike} (i.e., $TS$ contour does not look elliptical). \\
 10  & Spectral Fit Quality $> 16.3$ (Eq.\ref{eq:fitquality}). \\
 11  & Possibly due to the Sun (\S~\ref{catalog_variability}). \\
 12  & Highly curved spectrum; LogParabola $\beta$ fixed to 1 or PLExpCutoff \\
     & Spectral\_Index fixed to 0 (see \S~\ref{catalog_spectral_shapes}). \\
 \enddata
 
 \tablenotetext{a}{In the FITS version the values are encoded as individual bits in a single column, with Flag $n$ having value $2^{(n-1)}$.  For information about the FITS version of the table see Table~\ref{tab:columns} in  Appendix \ref{appendix_fits_format}.}

\end{deluxetable}

As in 1FGL we identified a number of conditions
that can shed doubt on a source. 
They are described in Table~\ref{tab:flags}.
In the FITS version of the catalog, these flags are summarized in a single integer column (\texttt{Flags}).  Each condition is indicated by one bit among the 16 bits forming \texttt{Flags}. The bit is raised (set to 1)
in the dubious case, so that sources without any warning sign have \texttt{Flags} = 0.

\begin{figure}[p]
\centering
\includegraphics[width=0.99\textwidth]{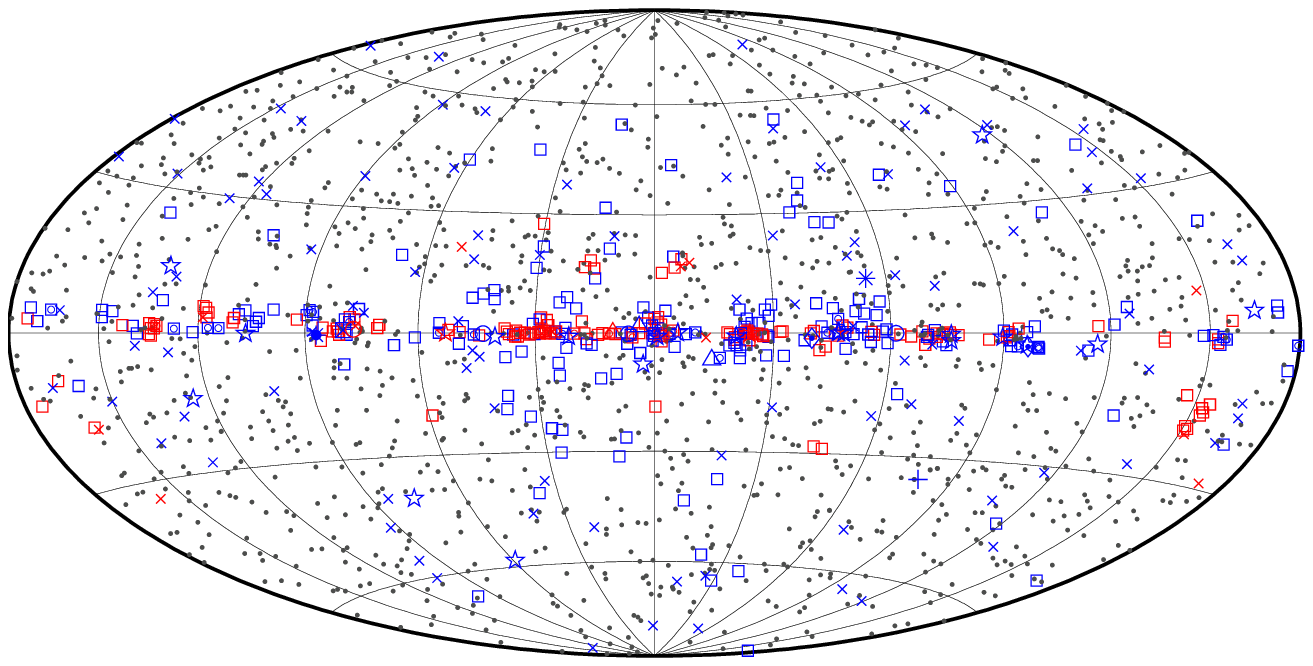}

\includegraphics[width=0.80\textwidth]{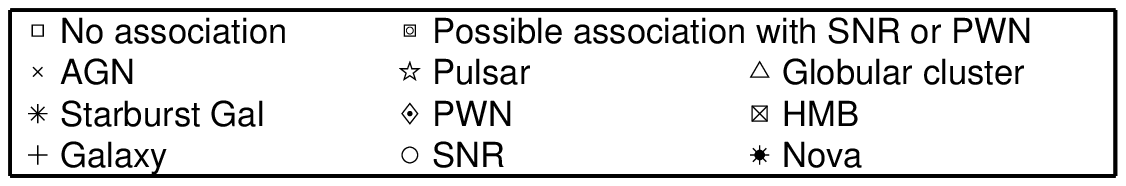}

\includegraphics[width=0.99\textwidth]{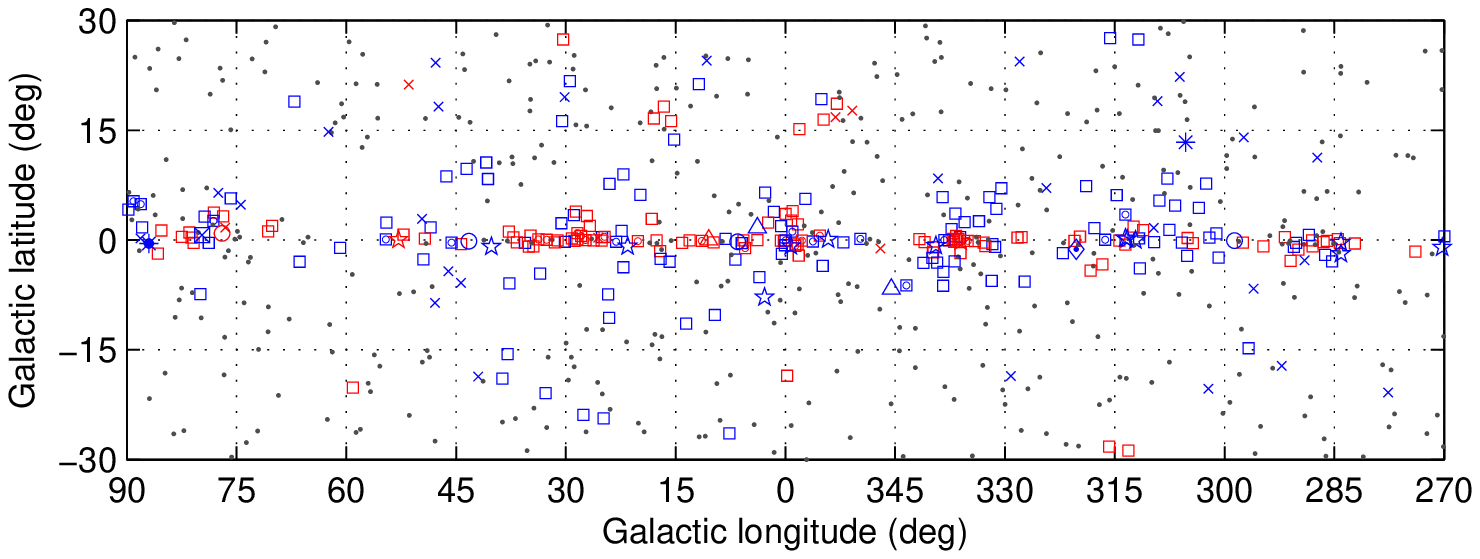}
\caption{Full sky map (top) and blow-up of the inner Galactic region
  (bottom) showing flagged sources by source class. Sources potentially confused with diffuse emission,
  i.e.,  those with a `c' designator in their names (and for which Flag
  6 is set) are shown in red; those with any other flag set are shown
  in blue. Sources with no flag set are shown as small dots.}
\label{fig:map_flags}
\end{figure}

Flags 1 to 9 have similar intent as in 1FGL, but differ in detail:
\begin{itemize}
\item In Flag 4, we reduced the threshold on source to background ratio to 20\%, because the diffuse model has improved.
\item The distances triggering Flag 5 have changed because the PSF knowledge has improved. The core of the PSF at low energy is actually better than the P6V3 estimate used in 1FGL, so the critical distance is lower at low energy. On the other hand the measured in-flight PSF at high energy is much broader than the P6V3 estimate \citep{LAT09__calib}, so the critical distance is about twice as great than for the 1FGL analysis  above 10~GeV.
\item We do not use {\it gtfindsrc} in 2FGL because it is based on unbinned likelihood. Therefore Flag 7 is not used.
\item Flag 8 compares the best position obtained from direct optimization with the contours extracted from the $TS$ maps.
\item The threshold for Flag 9 on elliptical quality was decreased to 4. The improved localization procedure allowed being a little more stringent here.
\end{itemize}
Flags 10, 11, and 12 are new.  Figure~\ref{fig:map_flags} shows the distribution on the sky of flagged
2FGL sources.

\section{The 2FGL Catalog}
\label{2fgl_description}

The basic description of the 2FGL catalog is in \S~\ref{catalog_description}, including a listing of the main table contents and some of the primary properties of the sources in the catalog.  We present a detailed comparison of the 2FGL catalog with the 1FGL catalog in \S~\ref{1fgl_comparison}.

\subsection{Catalog Description}
\label{catalog_description}

Table~\ref{tab:sources} is the catalog, with information for each of the 1873 sources; see Table~\ref{tab:desc} for descriptions of the columns.  The source designation is \texttt{2FGL JHHMM.m+DDMM} where the \texttt{2} indicates that this is the second LAT catalog, \texttt{FGL} represents {\it Fermi} Gamma-ray LAT.  Sources close to the Galactic ridge and some nearby interstellar cloud complexes are assigned names of the form \texttt{2FGL JHHMM.m+DDMMc}, where the \texttt{c} indicates that caution should be used in interpreting or analyzing these sources.  Errors in the model of interstellar diffuse emission, or an unusually high density of sources, are likely to affect the measured properties or even existence of these 162 sources (see \S~\ref{catalog_ism}).  In addition a set of analysis flags has been defined to indicate sources with unusual or potentially problematic characteristics (see \S~\ref{catalog_ism}).  The `c' designator is encoded as one of these flags.  An additional 315 sources have one or more of the other analysis flags set.  The 12 sources that were modeled as extended for 2FGL (\S~\ref{catalog_extended}) are singled out by an \texttt{e} at the ends of their names.

The designations of the classes that we use to categorize the 2FGL sources are listed in Table~\ref{tab:classes} along with the numbers of sources assigned to each class.  We distinguish between associated and identified sources, with associations depending primarily on close positional correspondence (see \S~\ref{source_assoc_automated}) and identifications requiring measurement of correlated variability at other wavelengths or characterization of the 2FGL source by its angular extent (see \S~\ref{source_assoc_firm}).  In the cases of multiple associations with a 2FGL source, we adopt the single association that is statistically most likely to be true if it is above the confidence threshold (see \S~\ref{source_assoc_automated}); the one exception is the Crab pulsar and PWN, which are listed as being associated with the same 2FGL source (see \ref{source_assoc_firm}).  Sources associated with SNRs are often also associated with PWNs and pulsars, and the SNRs themselves are often not point-like.  We do not attempt to distinguish among the possible classifications and instead list plausible associations of each class for unidentified 2FGL sources found to be positionally associated with SNRs (see \S~5.2.7).

The photon flux for 1--100~GeV ($F_{35}$; the subscript $ij$ indicates the energy range as 10$^i$ -- 10$^j$ MeV) and the energy flux for 100~MeV to 100~GeV in Table~\ref{tab:sources} are evaluated from the fit to the full band (see \S~\ref{catalog_flux_determination}), rather than sums of band fluxes as in 1FGL. We do not present the integrated photon flux for 100~MeV to 100~GeV (see \S~\ref{catalog_flux_determination}). Table~\ref{tab:fluxes} presents the fluxes in individual bands as defined in \S~\ref{catalog_flux_determination}.

\begin{deluxetable}{lrrrrrrrrrrrrrcccclcclc}
\setlength{\tabcolsep}{0.02in}
\tabletypesize{\scriptsize}
\rotate
\tablewidth{0pt}
\tablecaption{LAT 2-year Catalog\label{tab:sources}}
\tablehead{
\colhead{Name 2FGL} &
\colhead{R.A.} &
\colhead{Decl.} &
\colhead{$l$} &
\colhead{$b$} &
\colhead{$\theta_{\rm 1}$} &
\colhead{$\theta_{\rm 2}$} &
\colhead{$\phi$} &
\colhead{$\sigma$} &
\colhead{$F_{35}$} &
\colhead{$\Delta F_{35}$} &
\colhead{$S_{25}$} &
\colhead{$\Delta S_{25}$} &
\colhead{$\Gamma_{25}$} &
\colhead{$\Delta \Gamma_{25}$} &
\colhead{Mod} &
\colhead{Var} &
\colhead{Flags} &
\colhead{$\gamma$-ray Assoc.} &
\colhead{TeV} &
\colhead{Class} &
\colhead{ID or Assoc.} &
\colhead{Ref.}
}
\startdata
 J0000.9$-$0748 &   0.234 & $-$7.815 &  88.829 & $-$67.281 & 0.195 & 0.167 & 48 &      5.9 &     0.5 &   0.1 &    6.8 &    1.2 & 2.39 & 0.14 & PL & \nodata & \nodata & 1FGL J0000.9$-$0745         & \nodata & bzb & PMN J0001$-$0746  & \nodata \\
 J0001.7$-$4159 &   0.439 & $-$41.996 & 334.076 & $-$71.997 & 0.122 & 0.114 & 62 &      5.9 &     0.5 &   0.1 &    5.3 &    1.1 & 2.14 & 0.19 & PL & T & \nodata & 1FGL J0001.9$-$4158         & \nodata & agu & 1RXS J000135.5-41551 & \nodata \\
 J0002.7+6220 &   0.680 & 62.340 & 117.312 & 0.001 & 0.093 & 0.089 & 9 &     13.7 &     2.9 &   0.3 &   25.2 &    2.5 & 2.50 & 0.13 & LP & \nodata & \nodata & 1FGL J0003.1+6227         & \nodata & \nodata & \nodata & \nodata \\
 J0004.2+2208 &   1.056 & 22.137 & 108.732 & $-$39.430 & 0.194 & 0.137 & 63 &      5.4 &     0.4 &   0.1 &    6.3 &    1.2 & 2.49 & 0.15 & PL & \nodata & \nodata & 1FGL J0004.3+2207         & \nodata & \nodata & \nodata & \nodata \\
 J0004.7$-$4736 &   1.180 & $-$47.612 & 323.890 & $-$67.571 & 0.112 & 0.096 & 14 &     12.6 &     0.9 &   0.1 &   13.1 &    1.3 & 2.45 & 0.09 & PL & T & \nodata & 1FGL J0004.7$-$4737         & \nodata & bzq & PKS 0002$-$478  & \nodata \\
 J0006.1+3821 &   1.525 & 38.350 & 113.245 & $-$23.667 & 0.144 & 0.123 & 71 &     12.2 &     1.0 &   0.1 &   16.1 &    1.5 & 2.60 & 0.08 & PL & \nodata & \nodata & 1FGL J0005.7+3815         & \nodata & bzq & S4 0003+38  & \nodata \\
 J0007.0+7303 &   1.774 & 73.055 & 119.665 & 10.465 & 0.010 & 0.010 & $-$33 &    189.5 &    65.7 &   0.9 &  429.6 &    5.5 & 1.45 & 0.02 & EC & \nodata & \nodata & 1FGL J0007.0+7303         & \nodata & PSR & LAT PSR J0007+7303 & \nodata \\
  &  &  &  &  &  &  &  &  &  &  &  &  &  &  &  &  &  & 0FGL J0007.4+7303          &  &  & \\
  &  &  &  &  &  &  &  &  &  &  &  &  &  &  &  &  &  & EGR J0008+7308             &  &  & \\
  &  &  &  &  &  &  &  &  &  &  &  &  &  &  &  &  &  & 1AGL J0006+7311            &  &  & \\
 J0007.7+6825c &   1.925 & 68.423 & 118.911 & 5.894 & 0.173 & 0.170 & 64 &      6.2 &     1.0 &   0.2 &   17.5 &    2.7 & 2.61 & 0.10 & PL & \nodata & 6,10 & 1FGL J0005.1+6829         & \nodata & \nodata & \nodata & \nodata \\
 J0007.8+4713 &   1.974 & 47.230 & 115.304 & $-$14.996 & 0.062 & 0.053 & 29 &     17.6 &     2.1 &   0.2 &   23.7 &    2.1 & 2.10 & 0.06 & PL & \nodata & \nodata & \nodata & \nodata & bzb & MG4 J000800+4712  & \nodata \\
 J0008.7$-$2344 &   2.196 & $-$23.736 &  49.986 & $-$79.795 & 0.189 & 0.161 & $-$9 &      4.1 &     0.3 &   0.1 &    4.7 &    1.8 & 1.62 & 0.25 & PL & \nodata & \nodata & \nodata & \nodata & bzb & RBS 0016  & \nodata \\
 J0009.0+0632 &   2.262 & 6.542 & 104.453 & $-$54.801 & 0.129 & 0.123 & $-$10 &      5.7 &     0.5 &   0.1 &    6.7 &    1.3 & 2.40 & 0.16 & PL & \nodata & \nodata & 1FGL J0008.9+0635         & \nodata & bzb & CRATES J0009+0628  & \nodata \\
\enddata
\tablecomments{R.A. and Decl. are celestial coordinates in J2000 epoch, $l$ and $b$ are Galactic coordinates, in degrees; $\theta_1$ and $\theta_2$
are the semimajor and semiminor axes of the 95\% confidence source
location region; $\phi$ is the position angle in degrees east of north;
$F_{35}$ and $\Delta F_{35}$ are photon flux 1~GeV -- 100~GeV in
units of $10^{-9}$ cm$^{-2}$ s$^{-1}$; 
$S_{25}$ and $\Delta S_{25}$ are the energy flux 100~MeV -- 100~GeV
in units of $10^{-12}$ erg cm$^{-2}$ s$^{-1}$; $\Gamma_{25}$ and
$\Delta \Gamma_{25}$ are the photon power-law index and uncertainty
for a power-law fit; Mod is the spectral model used (PL for power-law,
 EC for exponential cutoff, and LP for log parabolic); Var is the
variability flag (see the text); Flags are the analysis flags
(see the text); $\gamma$-ray Assoc. lists associations with
other catalogs of GeV $\gamma$-ray sources; TeV indicates an association
with a point-like or small angular size TeV source (P) or extended
TeV source; Class designates the astrophysical class of the associated source (see the text);
ID or Assoc. lists the primary name of the associated source or identified counterpart; Ref. cross references LAT collaboration publications.  This table is published in its entirety in the electronic edition of the Astrophysical Journal Supplements. A portion is shown here for guidance regarding its form and content.}
\end{deluxetable}

\begin{deluxetable}{ll}
\setlength{\tabcolsep}{0.04in}
\tabletypesize{\scriptsize}
\tablecaption{LAT Second Catalog Description\label{tab:desc}}
\tablehead{
\colhead{Column} &
\colhead{Description} 
}
\startdata
Name &   \texttt{2FGL JHHMM.m+DDMM[c/e]}, constructed according to IAU Specifications for Nomenclature;  \texttt{m} is decimal\\
  &  minutes of R.A.; in the name R.A. and Decl. are truncated at 0.1 decimal minutes and 1\arcmin, respectively; \\
  &  \texttt{c} indicates that based on the region of the sky the source is considered to be potentially confused \\
  & with Galactic diffuse emission; \texttt{e} indicates a source that was modeled as spatially extended (see \S~\ref{catalog_extended}) \\
R.A. & Right Ascension, J2000, deg, 3 decimal places  \\
Decl. & Declination, J2000, deg, 3 decimal places \\
 $l$  & Galactic Longitude, deg, 3 decimal places \\
 $b$  & Galactic Latitude, deg, 3 decimal places \\
 $\theta_1$ & Semimajor radius of 95\% confidence region, deg, 3 decimal places\\
 $\theta_2$ & Semiminor radius of 95\% confidence region, deg, 3 decimal places\\
 $\phi$ & Position angle of 95\% confidence region, deg. East of North, 0 decimal places\\
$\sigma$ & Significance derived from likelihood Test Statistic for 100~MeV--100~GeV analysis, 1 decimal place \\
 $F_{35}$ &  Photon flux for 1~GeV--100~GeV, 10$^{-9}$ ph cm$^{-2}$ s$^{-1}$, summed over 3 bands, 1 decimal place \\
$\Delta F_{35}$ & 1 $\sigma$ uncertainty on  $F_{35}$ , same units and precision \\
$S_{25}$  &  Energy flux for 100~MeV--100~GeV, 10$^{-12}$ erg cm$^{-2}$ s$^{-1}$, from power-law fit, 1 decimal place \\
 $\Delta S_{25}$ & 1 $\sigma$ uncertainty on $S_{25}$, same units and precision \\
 $\Gamma$  & Photon number power-law index, 100~MeV--100~GeV, 2 decimal places \\
 $\Delta \Gamma$ & 1 $\sigma$ uncertainty of photon number power-law index, 100~MeV--100~GeV, 2 decimal places \\
Mod. &  \texttt{PL} indicates power-law fit to the energy spectrum;  \texttt{LP} indicates log-parabola fit to the energy spectrum;\\
    &     \texttt{EC} indicates power-law with exponential cutoff fit to the energy spectrum \\
 Var. &  \texttt{T} indicates $<$~1\% chance of being a steady source; see note in text  \\
Flags & See Table~\ref{tab:flags} for definitions of the flag numbers  \\
$\gamma$-ray Assoc.  & Positional associations with 0FGL, 1FGL, 3EG, EGR, or 1AGL sources \\
TeV & Positional association with a TeVCat source,  \texttt{P} for angular size $<$20$^\prime$,  \texttt{E} for extended \\
 Class & Like `ID' in 3EG catalog, but with more detail (see Table \ref{tab:classes}).  Capital letters indicate firm identifications;\\
  &  lower-case letters indicate associations. \\
 ID or Assoc.  & Designator of identified or associated source \\
 Ref. & Reference to associated paper(s) \\  
\enddata
\end{deluxetable}

\begin{deluxetable}{lcrcr}
\setlength{\tabcolsep}{0.04in}
\tablewidth{0pt}
\tabletypesize{\scriptsize}
\tablecaption{LAT 2FGL Source Classes \label{tab:classes}}
\tablehead{
\colhead{Description} & 
\multicolumn{2}{c}{Identified} &
\multicolumn{2}{c}{Associated} \\
& 
\colhead{Designator} &
\colhead{Number} &
\colhead{Designator} &
\colhead{Number}
}
\startdata
Pulsar, identified by pulsations & PSR & 83 & \nodata & \nodata \\
Pulsar, no pulsations seen in LAT yet & \nodata & \nodata & psr & 25 \\
Pulsar wind nebula & PWN & 3 & pwn & 0 \\
Supernova remnant & SNR & 6 & snr & 4 \\
Supernova remnant / Pulsar wind nebula & \nodata & \nodata & $\dagger$  & 58 \\
Globular cluster & GLC & 0 & glc & 11 \\
High-mass binary & HMB & 4 & hmb & 0 \\
Nova & NOV & 1 & nov & 0 \\
BL Lac type of blazar & BZB & 7 & bzb & 429 \\ 
FSRQ type of blazar & BZQ & 17 & bzq & 353 \\
Non-blazar active galaxy & AGN & 1 & agn & 10 \\ 
Radio galaxy & RDG & 2 & rdg &10 \\
Seyfert galaxy & SEY & 1 & sey & 5 \\
Active galaxy of uncertain type & AGU & 0 & agu & 257 \\ 
Normal galaxy (or part) & GAL & 2 & gal & 4 \\
Starburst galaxy & SBG & 0 & sbg & 4 \\
Class uncertain & \nodata & \nodata & \nodata & 1 \\
Unassociated & \nodata & \nodata & \nodata & 575 \\ 
Total & \nodata & 127 & \nodata & 1746 \\
\enddata
\tablecomments{The designation `$\dagger$' indicates potential association with SNR or PWN (see Table~\ref{tab:snrext}).  Designations shown in capital letters are firm identifications; lower case letters indicate associations. In the case of AGN, many of the associations have high confidence \citep{LAT11_2LAC}. Among the pulsars, those with names beginning with LAT were discovered with the LAT. In the FITS version of the 2FGL catalog, the $\dagger$ designator is replaced with `spp'; see Appendix \ref{appendix_fits_format}.}
\end{deluxetable}

\begin{deluxetable}{lrrrrrrrrrrrrrrr}
\setlength{\tabcolsep}{0.03in}
\tabletypesize{\scriptsize}
\tablecaption{Second LAT Catalog:  Fluxes in Bands\label{tab:fluxes}}
\tablehead{
\colhead{} & \multicolumn{3}{c}{100 MeV -- 300 MeV} & \multicolumn{3}{c}{300 MeV -- 1 GeV} & \multicolumn{3}{c}{1 GeV -- 3 GeV} & \multicolumn{3}{c}{3 GeV -- 10 GeV} & \multicolumn{3}{c}{10 GeV -- 100 GeV} \\ \cline{2-4} \cline{5-7} \cline{8-10} \cline{11-13} \cline{14-16} \\
\colhead{Name 2FGL} &
\colhead{$F_{\rm 1}$\tablenotemark{a}} &
\colhead{$\Delta F_{\rm 1}$\tablenotemark{a}} &
\colhead{$\sqrt{TS_{\rm 1}}$} &
\colhead{$F_{\rm 2}$\tablenotemark{a}} &
\colhead{$\Delta F_{\rm 2}$\tablenotemark{a}} &
\colhead{$\sqrt{TS_{\rm 2}}$} &
\colhead{$F_{\rm 3}$\tablenotemark{b}} &
\colhead{$\Delta F_{\rm 3}$\tablenotemark{b}} &
\colhead{$\sqrt{TS_{\rm 3}}$} &
\colhead{$F_{\rm 4}$\tablenotemark{c}} &
\colhead{$\Delta F_{\rm 4}$\tablenotemark{c}} &
\colhead{$\sqrt{TS_{\rm 4}}$} &
\colhead{$F_{\rm 5}$\tablenotemark{c}} &
\colhead{$\Delta F_{\rm 5}$\tablenotemark{c}} &
\colhead{$\sqrt{TS_{\rm 5}}$} 
}
\startdata
 J0000.9$-$0748  &    1.4 &    0.0 &    1.5 &    0.3 &    0.1 &    5.3 &    0.5 &    0.0 &    2.5 &    2.0 &    0.0 &    2.5 &    0.7 &    0.0 &    2.0 \\
 J0001.7$-$4159  &    1.5 &    0.0 &    2.2 &    0.2 &    0.0 &    2.8 &    0.5 &    0.0 &    2.6 &    1.6 &    0.6 &    5.2 &    0.9 &    0.0 &    0.0 \\
 J0002.7+6220  &    1.9 &    0.7 &    3.5 &    1.3 &    0.2 &    9.2 &    2.6 &    0.4 &    8.7 &    4.1 &    1.1 &    5.2 &    0.5 &    0.0 &    0.0 \\
 J0004.2+2208  &    1.6 &    0.0 &    2.0 &    0.3 &    0.1 &    4.8 &    0.4 &    0.0 &    1.6 &    1.3 &    0.6 &    3.7 &    0.5 &    0.0 &    0.0 \\
 J0004.7$-$4736  &    2.2 &    0.4 &    5.8 &    0.4 &    0.1 &    7.7 &    0.9 &    0.2 &    7.9 &    1.3 &    0.6 &    4.1 &    0.6 &    0.0 &    0.0 \\
 J0006.1+3821  &    2.7 &    0.5 &    5.9 &    0.5 &    0.1 &    7.0 &    0.9 &    0.2 &    6.2 &    2.3 &    0.0 &    2.8 &    0.6 &    0.0 &    0.0 \\
 J0007.0+7303  &   17.9 &    0.6 &   37.6 &   11.6 &    0.2 &   96.0 &   49.9 &    0.9 &  122.7 &  149.9 &    4.2 &   91.3 &   12.5 &    1.2 &   27.1 \\
 J0007.7+6825c &    2.8 &    0.0 &    0.6 &    1.1 &    0.2 &    6.1 &    0.9 &    0.3 &    3.2 &    1.3 &    0.0 &    0.0 &    1.0 &    0.0 &    1.9 \\
 J0007.8+4713  &    2.7 &    0.4 &    6.6 &    0.6 &    0.1 &    8.4 &    1.5 &    0.2 &    9.5 &    4.9 &    1.0 &    9.3 &    1.2 &    0.5 &    5.2 \\
 J0008.7$-$2344  &    0.5 &    0.0 &    0.0 &    0.1 &    0.0 &    0.1 &    0.6 &    0.0 &    3.2 &    1.4 &    0.0 &    1.6 &    1.8 &    0.0 &    3.8 \\
 J0009.0+0632  &    2.1 &    0.0 &    2.8 &    0.3 &    0.0 &    2.4 &    0.3 &    0.1 &    3.2 &    1.4 &    0.6 &    3.7 &    0.7 &    0.0 &    1.8 \\
 J0009.1+5030  &    0.9 &    0.0 &    0.3 &    0.4 &    0.1 &    6.0 &    1.5 &    0.2 &    9.6 &    5.9 &    1.1 &   10.6 &    1.9 &    0.6 &    7.0 \\
 J0009.9$-$3206  &    0.7 &    0.0 &    0.1 &    0.3 &    0.0 &    3.1 &    0.5 &    0.1 &    5.0 &    1.7 &    0.0 &    1.6 &    0.9 &    0.0 &    1.7 \\
 J0010.5+6556c &    2.7 &    0.0 &    2.8 &    1.2 &    0.2 &    7.0 &    1.8 &    0.0 &    3.0 &    3.2 &    0.0 &    1.5 &    0.7 &    0.0 &    0.5 \\
 J0011.3+0054  &    1.3 &    0.0 &    0.8 &    0.3 &    0.1 &    4.9 &    0.5 &    0.1 &    4.4 &    1.9 &    0.0 &    2.0 &    0.6 &    0.0 &    0.0 \\
\enddata
\tablecomments{This table is published in its entirety in the electronic edition of the Astrophysical Journal Supplements.  A portion is shown here for guidance regarding its form and content.}
\tablenotetext{a}{In units of $10^{-8}$ photons cm$^{-2}$ s$^{-1}$}
\tablenotetext{b}{In units of $10^{-9}$ photons cm$^{-2}$ s$^{-1}$}
\tablenotetext{c}{In units of $10^{-10}$ photons cm$^{-2}$ s$^{-1}$}
\end{deluxetable}

\begin{figure}[p]
\centering
\includegraphics[width=0.99\textwidth]{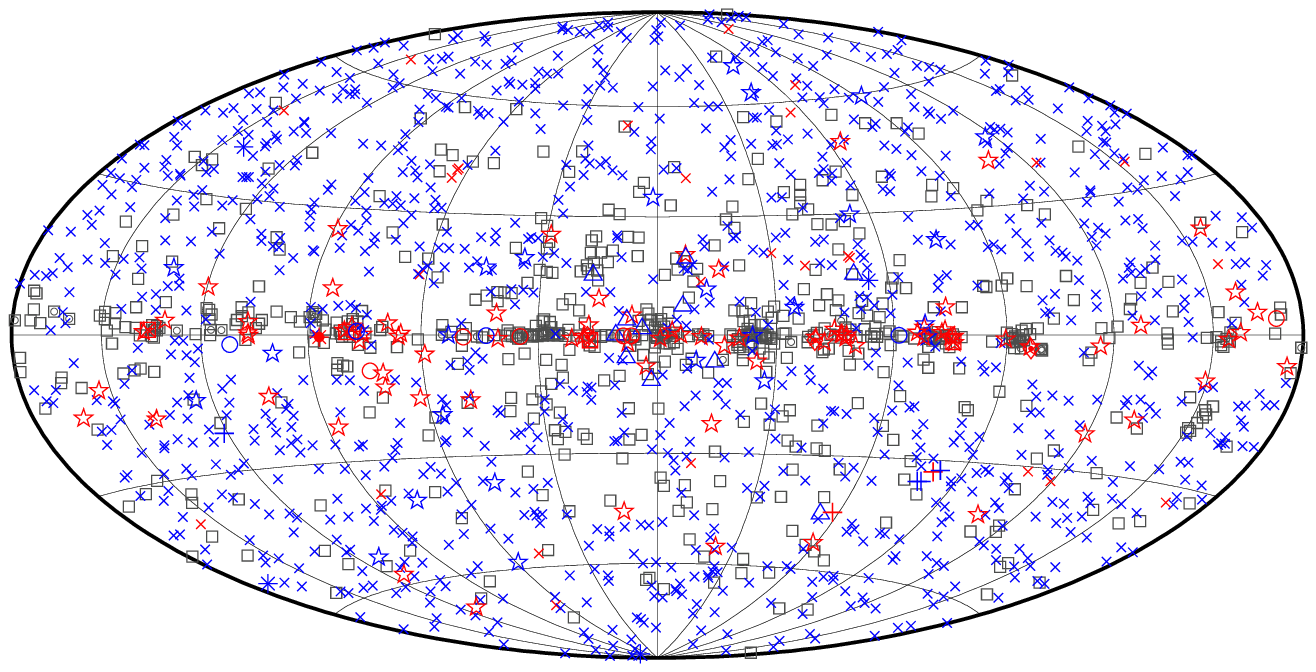}

\includegraphics[width=0.80\textwidth]{map_2fgl_legend.eps}

\includegraphics[width=0.99\textwidth]{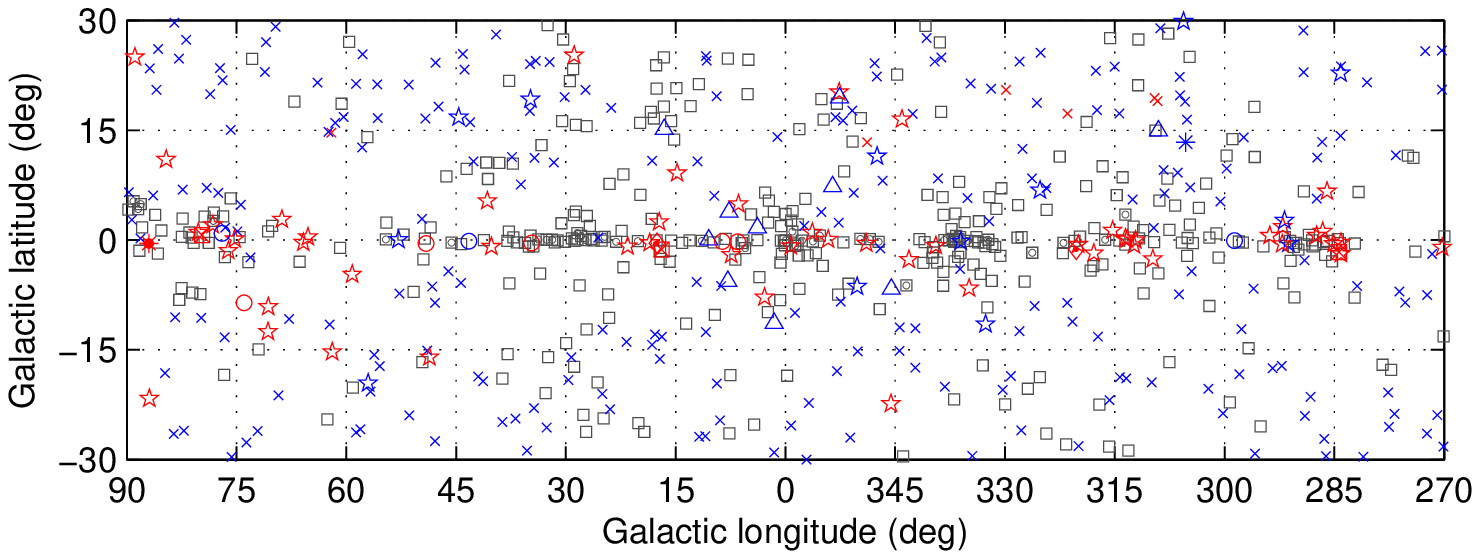}
\caption{Full sky map (top) and blow-up of the inner Galactic region
  (bottom) showing sources by source class (see
  Table~\ref{tab:classes}). Identified sources are shown with a red
  symbol, associated sources in blue.}
\label{fig:map_id_assoc}
\end{figure}

\begin{figure}
\epsscale{.80}
\plotone{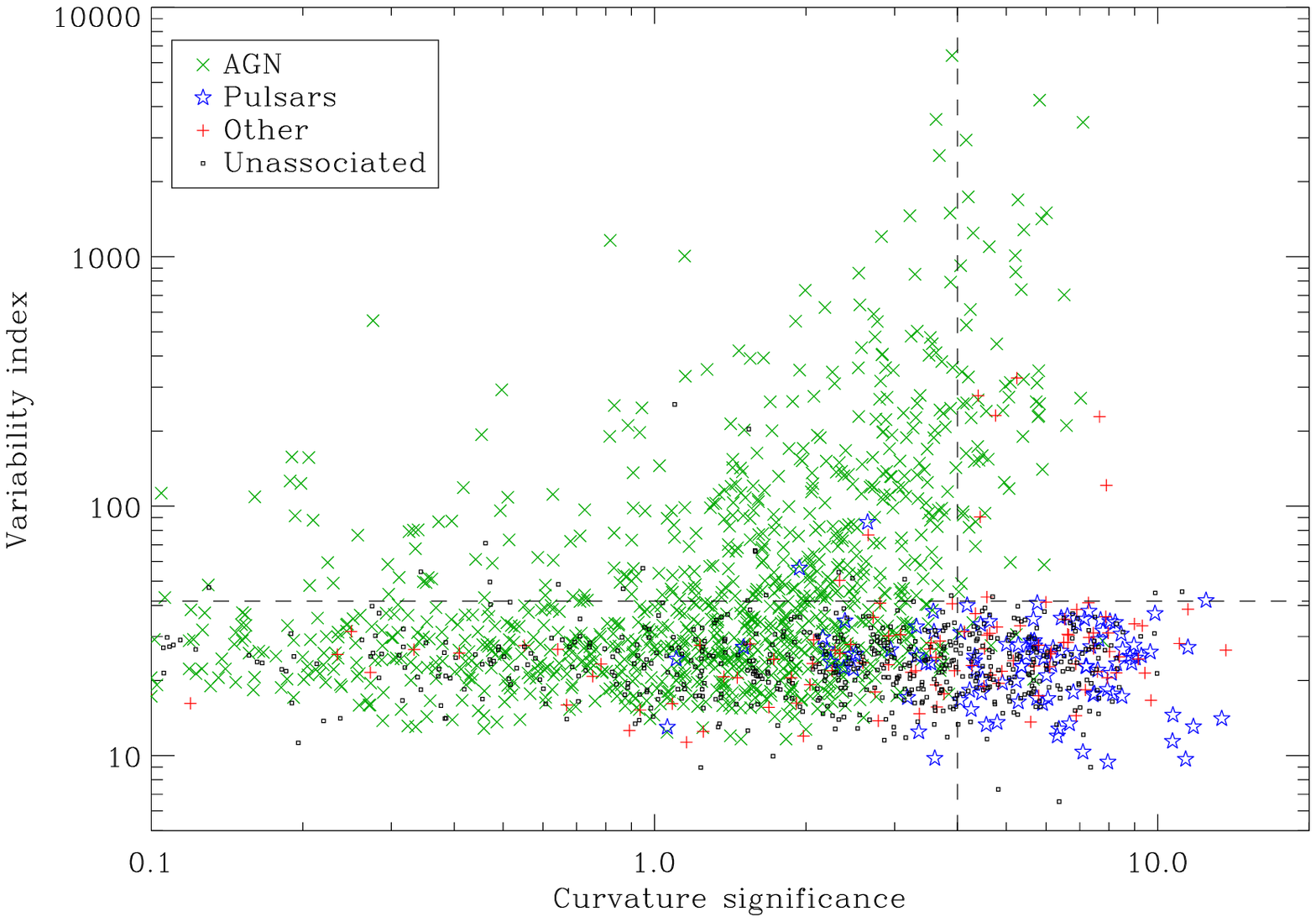}
\caption{Variability index ($TS_{var}$ in \S~\ref{catalog_variability}) plotted as a function of the curvature significance (\texttt{Signif\_Curve} in \S~\ref{catalog_flux_determination}) for different broad classes of sources. ``AGN'' here means any class starting with ``ag'' or ``bz'' in Table \ref{tab:classes}. The horizontal dashed line is set to 41.6, above which sources are likely variable. The vertical dashed line is set to 4.0, above which curved spectra are used.}
\label{fig:curvar}
\end{figure}

Figure~\ref{fig:map_id_assoc} illustrates where the different classes of sources are located in the sky.
Figure~\ref{fig:curvar} shows where the broad classes of sources appear in the curvature - variability plane. This is similar to Figure~8 of \citet{LAT10_1FGL} although the two indicators were improved. Most ``other'' curved non-variable sources are tentatively associated to SNRs. The two ``pulsars'' above the variability threshold are the Crab and PSR J1142+01. The Crab mixes the pulsar and the nebula, and we know the variability is due to the nebula \citep{LAT11_CrabFlares}. PSR J1142+01 is a newly discovered millisecond pulsar with no known LAT pulsations.

\subsection{Comparison with 1FGL}
\label{1fgl_comparison}


The 1FGL catalog \citep{LAT10_1FGL} lists 1451 sources detected during the first 11 months of operation by the LAT.  Associations between 2FGL and 1FGL sources are based on the following relation:
\be
 \Delta  \le\  d_{x} = \sqrt{ \theta_{x_{1FGL}}^2 +\theta_{x_{2FGL}}^2}
 \ee

\noindent where  ($\Delta$) is the angular distance between the sources  and $d_{x}$ is defined in terms of the semi-major axis of the  $x$\% confidence error ellipse for the position of each source, e.g., the 95\% confidence error for the automatic source association procedure (\S~\ref{source_assoc_automated}). In total, 1099 2FGL sources were automatically associated with entries in the 1FGL catalog. 
At the level of overlapping 95\% source location confidence contours the 2FGL catalog contains 774  (out of 1873) new $\gamma$-ray sources, while 352 sources previously listed in 1FGL do not have a counterpart in the 2FGL catalog. 

\begin{figure}[!ht]
\centering
\includegraphics[width=0.8\textwidth]{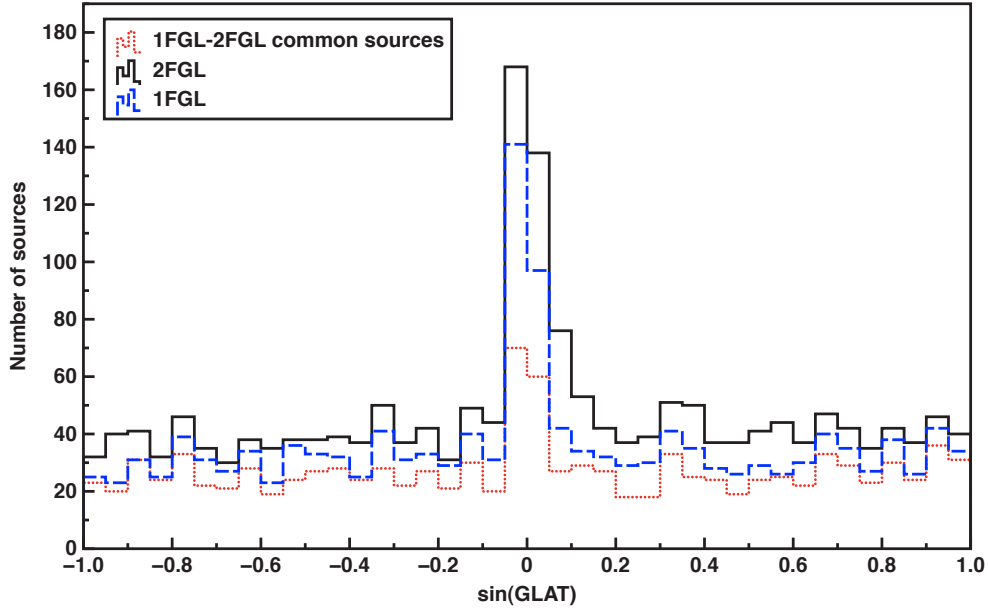}
\caption{Distributions of the Galactic latitude of the 1FGL and 2FGL sources and of the sources in common between the 1FGL and 2FGL catalogs.}
\label{fig:fig_glat}
\end{figure}

The Galactic latitude distributions of the 2FGL sources, the 1FGL sources and of the sources in common between the 1FGL and 2FGL catalogs, shown in Figure \ref{fig:fig_glat}, indicate both that most of the new 2FGL sources and  most of the missing 1FGL sources are concentrated along the Galactic plane where the Galactic diffuse emission is most intense and improvements in the model for the diffuse emission since the 1FGL analysis would be expected to have the most influence (\S~\ref{DiffuseModel}).

\begin{figure}[!ht]
\centering
\includegraphics[width=0.8\textwidth]{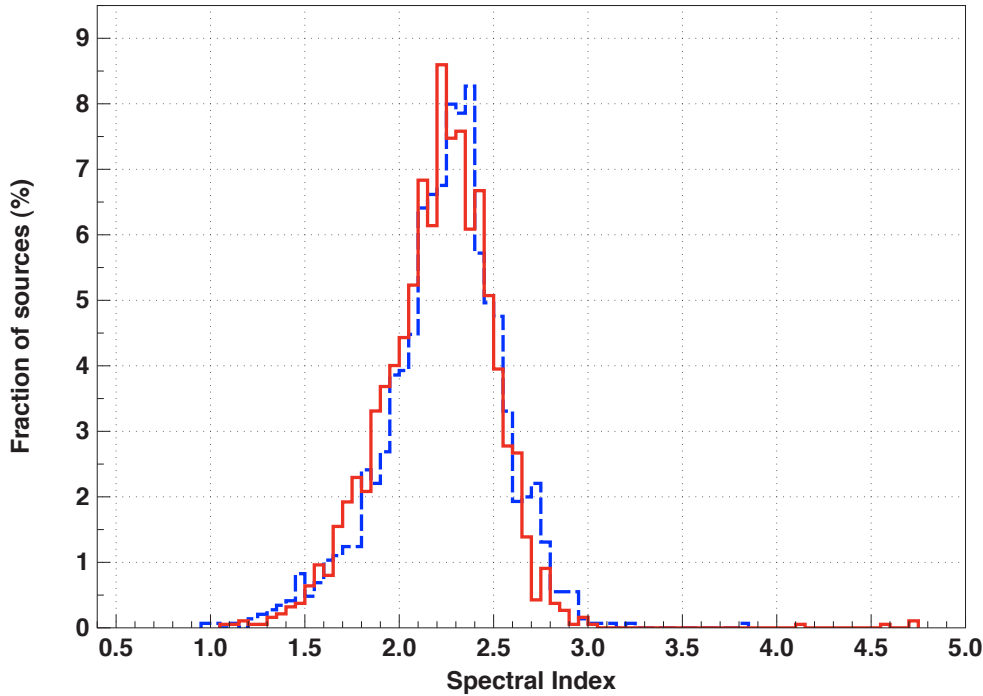}
\caption{Distributions of the  spectral index for the 1FGL (1451 sources, dashed line) and for the 2FGL (1873 sources, solid line) catalogs. }
\label{fig:fig_spec1}
\end{figure}

\begin{figure}[!ht]
\centering
\includegraphics[width=0.8\textwidth]{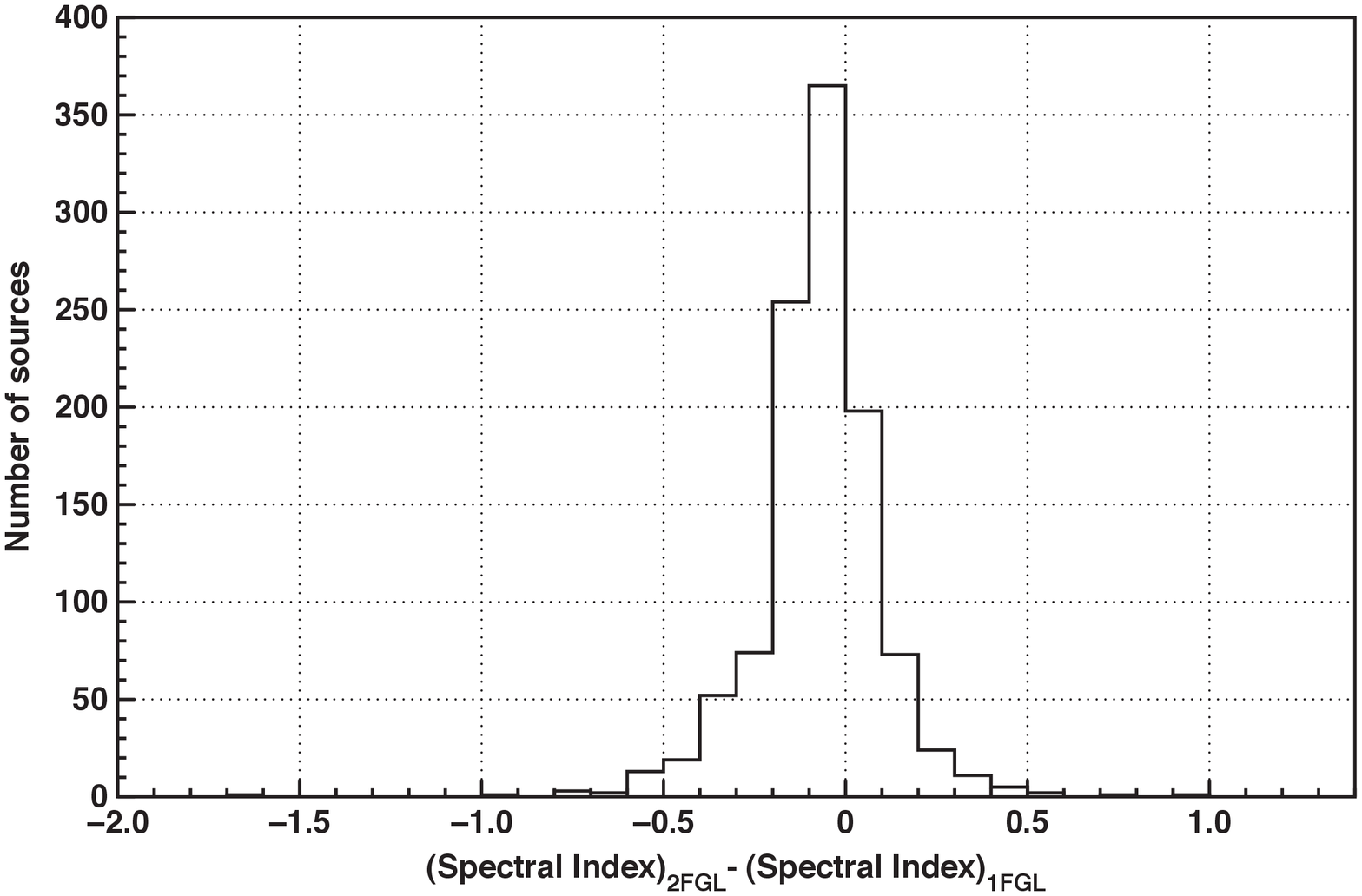}
\caption{Distribution of the   difference $\Gamma_{2FGL} - \Gamma_{1FGL}$ for the 1099 sources in common between the 1FGL and 2FGL catalogs.}
\label{fig:fig_spec}
\end{figure}

As described in \S~\ref{catalog_spectral_shapes}, in the 2FGL analysis the spectral fits are made using  power-law, power-law with an exponential cutoff, or log-parabola models.  Of the 1099 1FGL sources associated with 2FGL sources, 274 of the brightest were fitted with a curved spectral functional form.  For each 2FGL source we also evaluated the spectral index ($\Gamma$) of the best power-law fit (\S~\ref{catalog_description}) and this enables a comparison of the spectral characteristics of the 1FGL and 2FGL sources. Figure \ref{fig:fig_spec1} shows the distributions of the spectral indices of all of the sources in the 1FGL and 2FGL catalogs. The two distributions are very similar, with an average $\Gamma_{1FGL} = 2.23\pm0.01$ and an average $\Gamma_{2FGL} = 2.21\pm0.01$. However, the peaks of the two distributions are not exactly coincident; also, the skewness of the 2FGL distribution is positive while it is negative for 1FGL.  Figure \ref{fig:fig_spec} shows the distribution of the  difference $\Gamma_{2FGL}-\Gamma_{1FGL}$ for the 1099 sources in common between the catalogs. The average of the distribution is $-0.07\pm 0.01$, with the 2FGL sources  slightly harder than the 1FGL ones. This small difference in the spectral index distribution could be related to slightly difference uncertainties in the effective area between P7\_V6 and P6\_V3. 

\begin{figure}[!ht]
\centering
\includegraphics[width=0.8\textwidth]{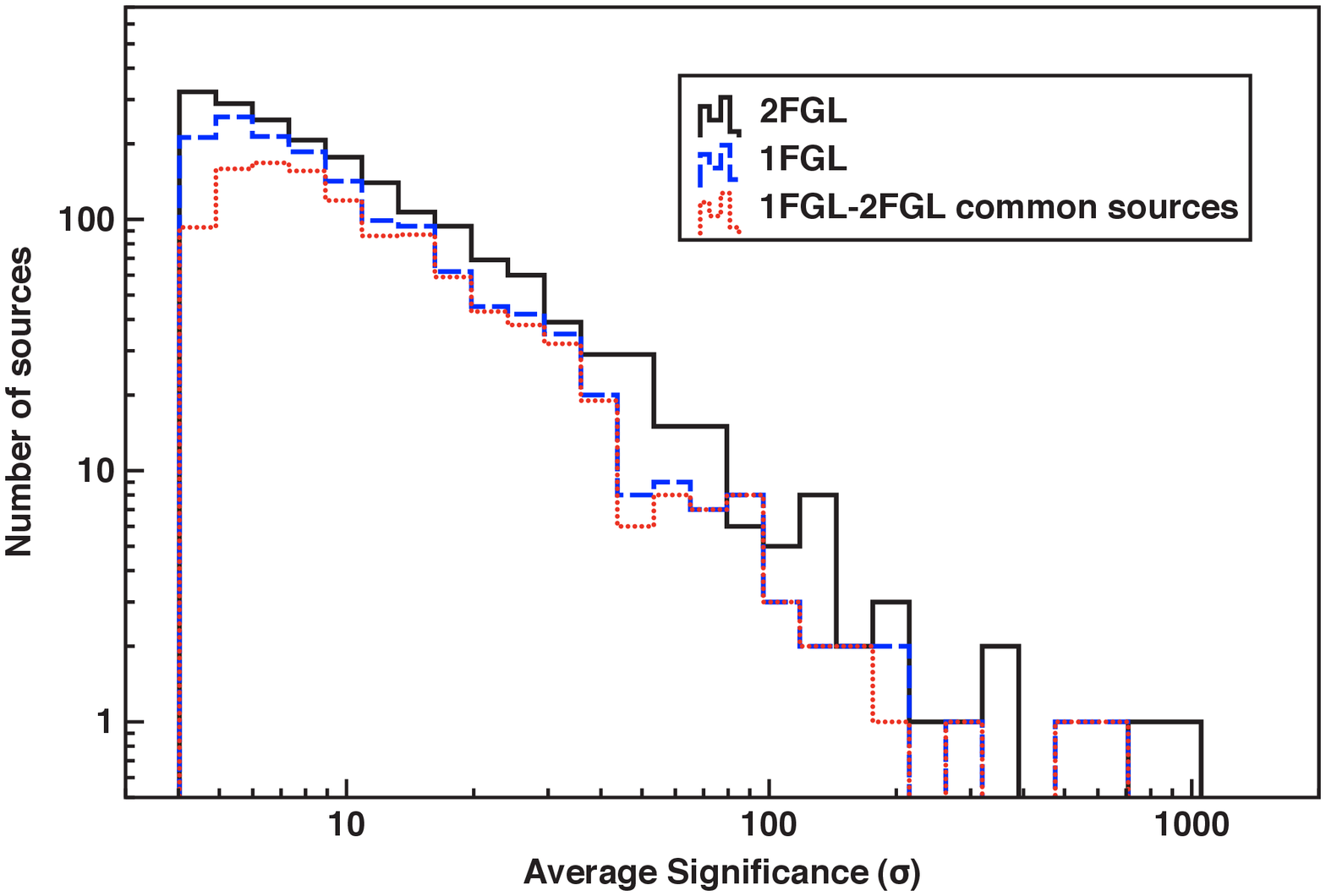}
\caption{Distributions of the significances of the sources in the 1FGL and 2FGL catalogs and of  the sources in common between the catalogs. The distribution of the significance for the 1FGL-2FGL common sources  is based on the values reported in the 1FGL catalog.}
\label{fig:fig_sign}
\end{figure}

The distributions of the source significances reported in Figure \ref{fig:fig_sign} show that for the 2FGL catalog the significance peaks between 4 $\sigma$ and 5 $\sigma$ while  for 1FGL the distribution shows a  plateau between $4\ \sigma$  and $6\ \sigma$; this indicates that 2FGL is more complete than 1FGL. Also, the distribution of the significance of the sources  that are in common between 1FGL and 2FGL shows that most of the 1FGL sources that were not recovered in the 2FGL catalog had significance less than 7 $\sigma$. In the remainder of this section we describe the variety of reasons that the additional 352 1FGL sources do not appear in the 2FGL catalog. 

Table \ref{tab:1fgl_2fgl} lists 347 of the 1FGL sources  that  do not have a corresponding source in 2FGL.  The  five other 1FGL sources that do not appear in 2FGL were not included in the table because they were already replaced by an extended source template in the 2FGL analysis.  These sources are: 1FGL\ J0523.3$-$6855 (2FGL\  J0526.6$-$6825e, LMC); 1FGL\ J1801.3$-$2322c (0FGL\ J1801.6$-$2327, 2FGL\ J1801.3$-$2326e, W28); 1FGL\ J1805.2$-$2137c (2FGL\ J1805.6$-$2136e, W30); 1FGL\ J1856.1+0122 (2FGL\ J1855.9+0121e, G034.7$-$00.4, W44); 1FGL\ J1922.9+1411 (2FGL J1923.2+1408e, G049.2$-$00.7, W51C). 

Some 1FGL sources near extended 2FGL sources remain in Table \ref{tab:1fgl_2fgl}.  An additional four 1FGL sources, 1FGL\  J0459.7$-$6921, 1FGL\  J0518.6$-$7222, 1FGL\  J0531.3$-$6716 and 1FGL\  J0538.9$-$6914, were also found in the LMC field, now replaced by an extended source in the 2FGL catalog analysis (J0526.6$-$6825e).  Furthermore, the 4 1FGL sources 1FGL\ J2046.4+3041, 1FGL\ J2049.1+3142, 1FGL\ J2055.2+3144, 1FGL\ J2057.4+3057 distributed along the Cygnus Loop (G74.0$-$8.5), one of the most famous and well-studied SNRs, were replaced by an extended source template in the 2FGL analysis (2FGL\ J2051.0+3040e), and so are not confirmed in the 2FGL catalog.  The extended source 2FGL~J1824.5$-$1351e (HESS\ J1825$-$137) replaces two 1FGL sources:  1FGL~J1821.1$-$1425c and 1FGL\ J1825.7$-$1410.  

\begin{figure}[!ht]
\centering
\includegraphics[width=0.99\textwidth]{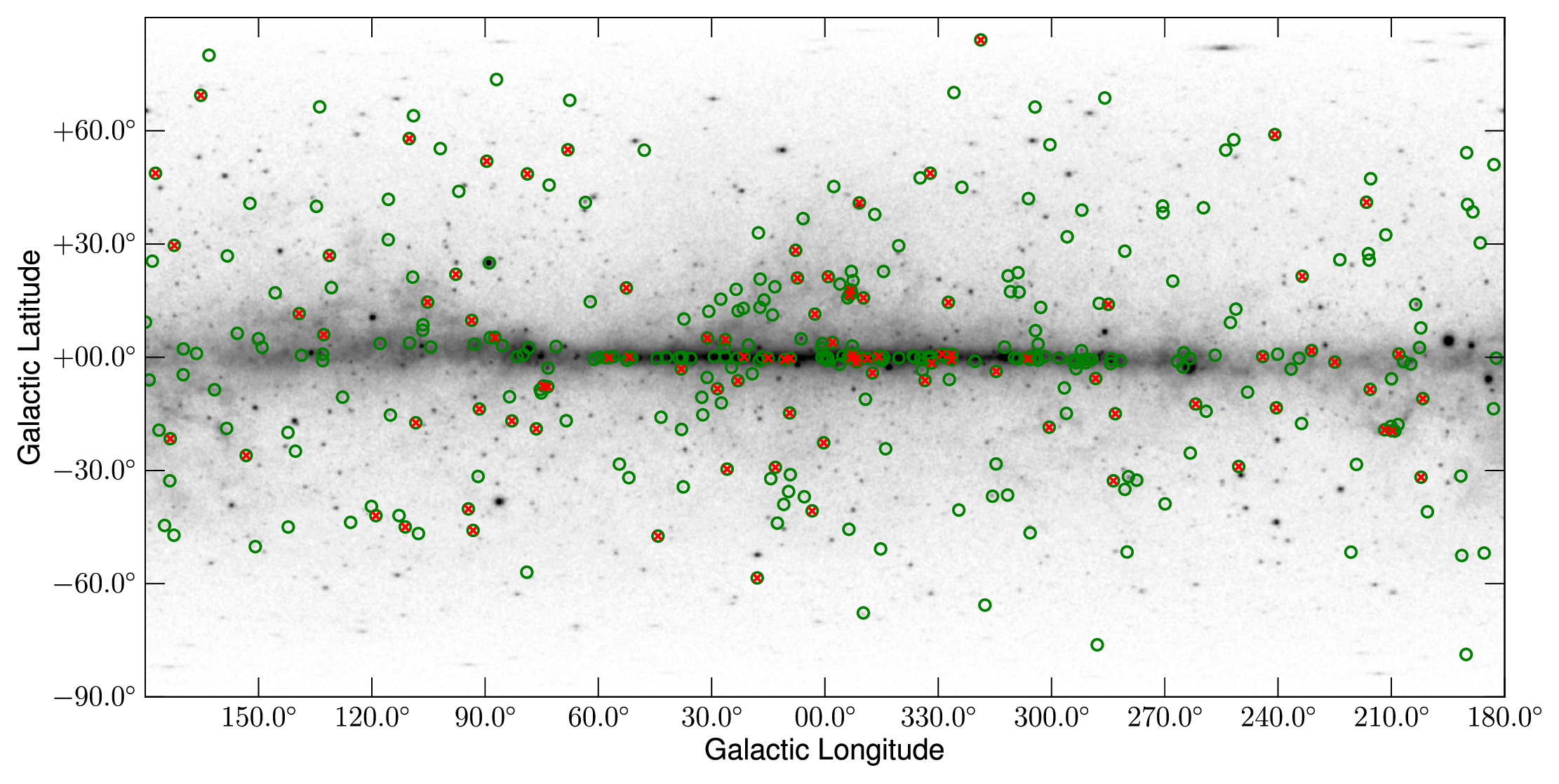}
\caption{All-sky map for energies $>$1GeV indicating the positions of the 1FGL sources that are not in the 2FGL catalog (green circles). The red crosses indicate the sources having the flag `NC' in Table \ref{tab:1fgl_2fgl})}
\label{fig:missing1fgl}
\end{figure}

About 250 of the 347 sources are located on the Galactic plane or in other regions of bright, structured diffuse emission (see Fig. \ref{fig:missing1fgl}). Of these,  88 have the `c' designation in the 1FGL name, which indicates that these sources were already recognized as possible spurious detections. Another 21 1FGL sources were flagged according to the definitions reported in Table 4 of  \citet[][]{LAT10_1FGL}. These sources were also already noted as problematic.     
In the 1FGL catalog only 67 of the 347 sources have an association with a possible counterpart, mostly AGN, while another 10 sources were associated with already known 0FGL \citep{LAT09_BSL} or 1AGL \citep{AGILEcatalog} $\gamma$-ray sources. 

In addition to the introduction of spatially extended sources in 2FGL, there are many possible causes for 1FGL sources to be absent from the 2FGL list. Among these are variability; different event selection used for the analysis (Pass 6 for 1FGL and Pass 7 for  2FGL); different IRFs; different Galactic diffuse emission models; different analysis procedures (unbinned likelihood analysis for 1FGL and  binned likelihood analysis for 2FGL); statistical threshold effects; and 1FGL sources  resolved into two  or more 2FGL sources. In the last columns of Table \ref{tab:1fgl_2fgl} we  assigned to each source one or more flags corresponding to a possible cause.  In many cases, no one reason can be singled out. 

\begin{figure}[!ht]
\centering
\includegraphics[width=0.8\textwidth]{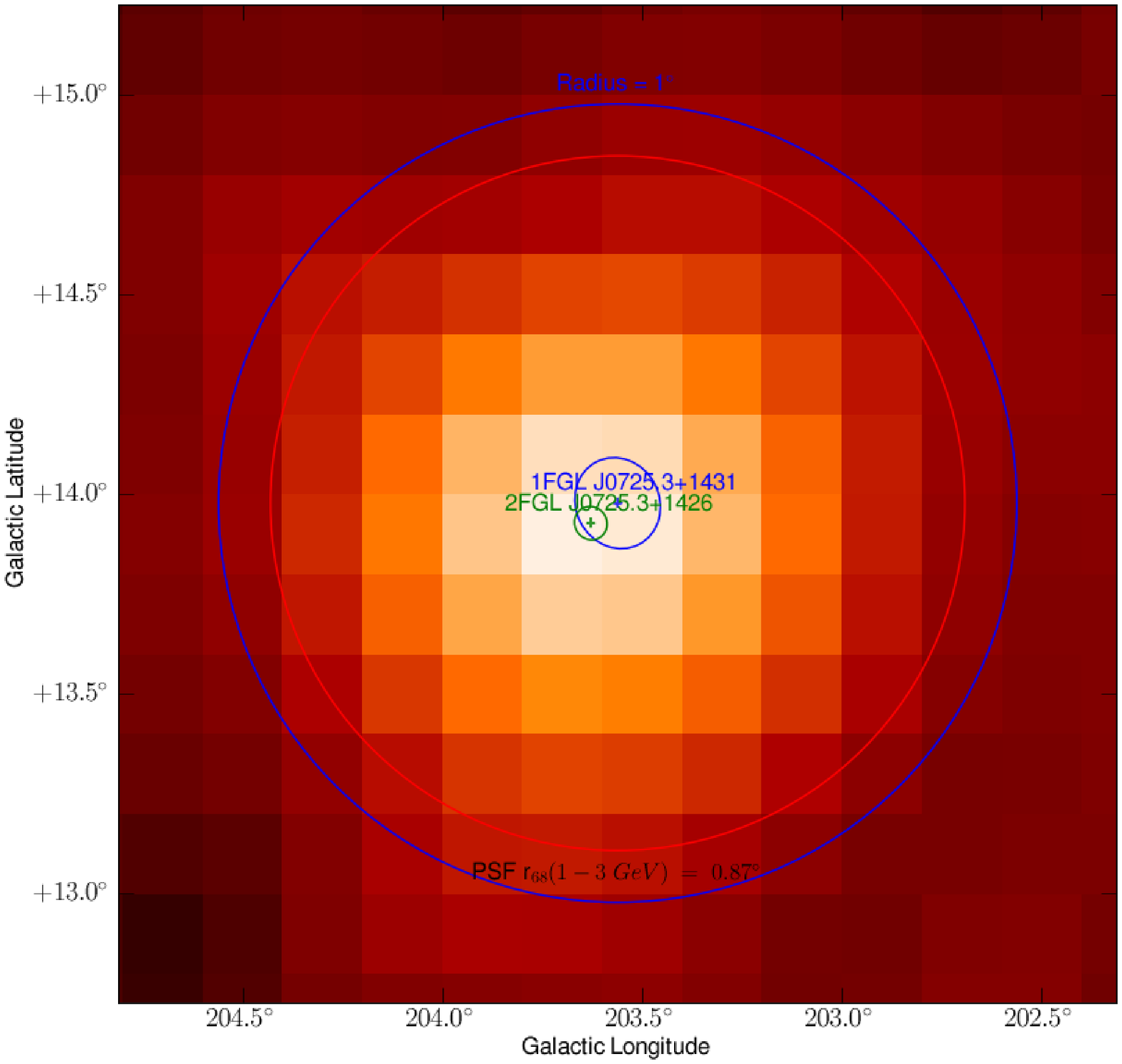}
\caption{A typical example of  a 1FGL source associated with a 2FGL source using  $d_{99.9}$. The E\ $>$\ 1\ GeV counts map (1 pixel = $0\fdg2$) was smoothed using a gaussian kernel ($\sigma$\ =\ 3 pixels). The ellipses represent the 99.9\% confidence error regions.}
\label{fig:ex_r999}
\end{figure}

The numbers of associated sources between the 1FGL and 2FGL catalogs does depend on the criterion used to define spatial coincidence (Eqn. 5).  The number of 2FGL\ -\ 1FGL associated sources increases to 1151 if we use  $\Delta < d_{99.9}$\footnote{Assuming a Rayleigh distribution for the source angular separations, $d_{99.9}$ is evaluated using $\theta_{99.9} = 1.52\ \theta_{95}$}.  The 52 additional associations (see Table \ref{tab:1fgl_2fgl} and see Figure \ref{fig:ex_r999}), represent about the 5\% of the 1451 1FGL sources,  as expected when passing from $d_{95}$ to $d_{99.9}$. Also, in 2FGL we used a better in-flight representation of the PSF that is broader than the PSF used in 1FGL at energies E$>$ 1GeV where, in general, most of the sources are detected.  Furthermore, the new and improved model of the Galactic diffuse emission used to build the 2FGL catalog together with the expected increase of the signal-to-noise ratio due to the use of 24 months data, allowed us to obtain a better localization of the sources  at positions that might be outside the 95\% confidence error regions previously reported in 1FGL. Indeed, most of the 52 additional associations concern sources located along the Galactic plane and in regions like Orion and Ophiuchus, while only about 10 were associated in regions with low diffuse emission.

\begin{figure}[!ht]
\centering
\includegraphics[width=0.8\textwidth]{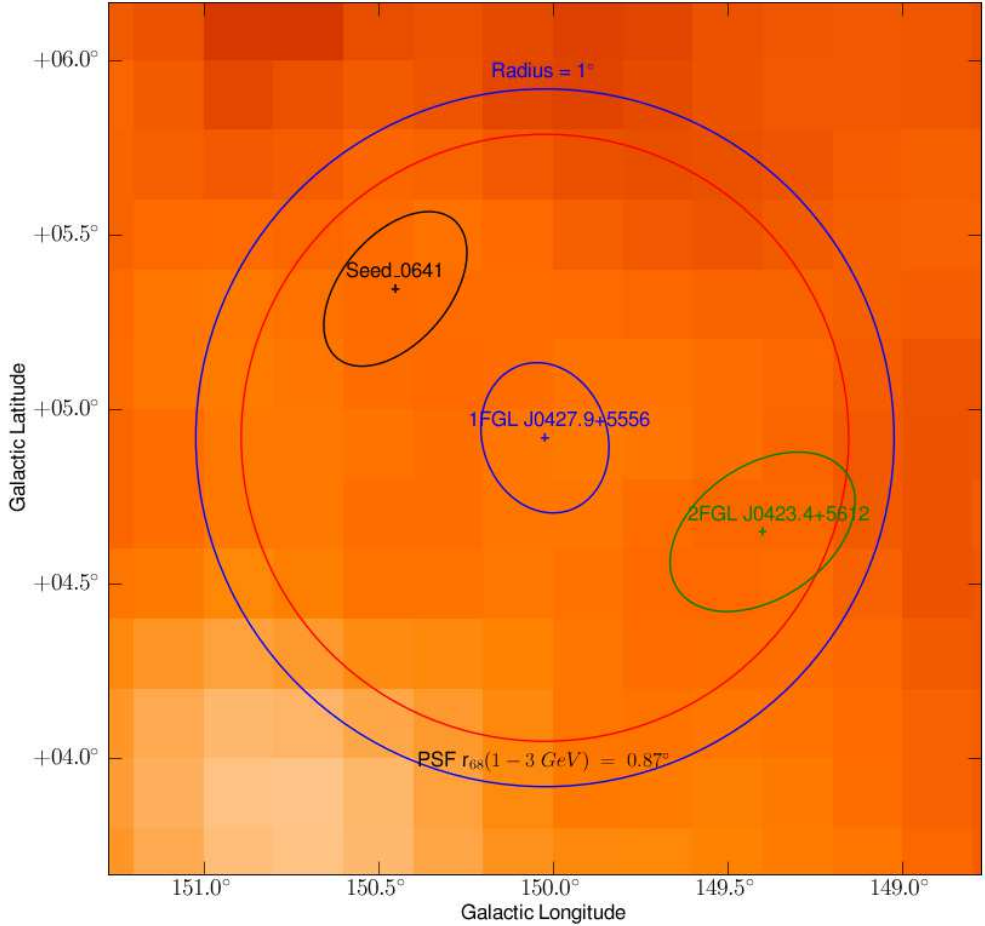}
\caption{A typical example of  a 1FGL  and a  2FGL source having an angular separation greater than  $d_{99.9}$ but less than 1$\degr$. In this particular case there is also a seed (candidate source considered in the 2FGL analysis) separated by less than 1$^\circ$ from the 1FGL source. 
The E\ $>$\ 1\ GeV counts map (1 pixel = $0\fdg2$) was smoothed using a gaussian kernel ($\sigma$\ =\ 3 pixels). The ellipses represent the 99.9\% confidence error regions. }
\label{fig:ex_close}
\end{figure}

Also, in the 1FGL catalog the positions of sources associated with the LAT--detected pulsars and X--ray binaries are the high-precision positions of the identified sources.  (These sources can be easily recognized because they have null values in the localization parameters reported in the 1FGL catalog). Not all of these associations appear in the 2FGL catalog because they cannot be associated using $d_{95}$, but some are listed in Table \ref{tab:1fgl_2fgl} because they can be associated using $d_{99.9}$. These sources are:  1FGL\ J2032.4+4057 (Cyg\ X$-$3); 1FGL\ J1836.2+5925 (LAT\ PSR J1836+5925); 1FGL\ J1124.6$-$5916 (PSR\ J1124$-$5916). 
However,  1FGL\ J1741.8$-$2101  (LAT\ PSR J1741$-$2054), 1FGL\ J1614.0$-$2230 (PSR\ J1614$-$2230) and 1FGL\ J1747.2$-$2958 (PSR\ J1747$-$2958) are still not associated, and for these sources we report the nearest 2FGL source (see, e.g., Fig. \ref{fig:ex_close}).  The last missing source in this category is  1FGL\ J1023.0$-$5746 (LAT\ PSR J1023$-$5746). It was  resolved into two 2FGL sources, 2FGL\ J1022.7$-$5741 and 2FGL\ J1023.5$-$5749.  Although both are located very close to the pulsar position, they cannot be formally associated using $d_{99.9}$.  

\begin{figure}[!ht]
\centering
\includegraphics[width=0.8\textwidth]{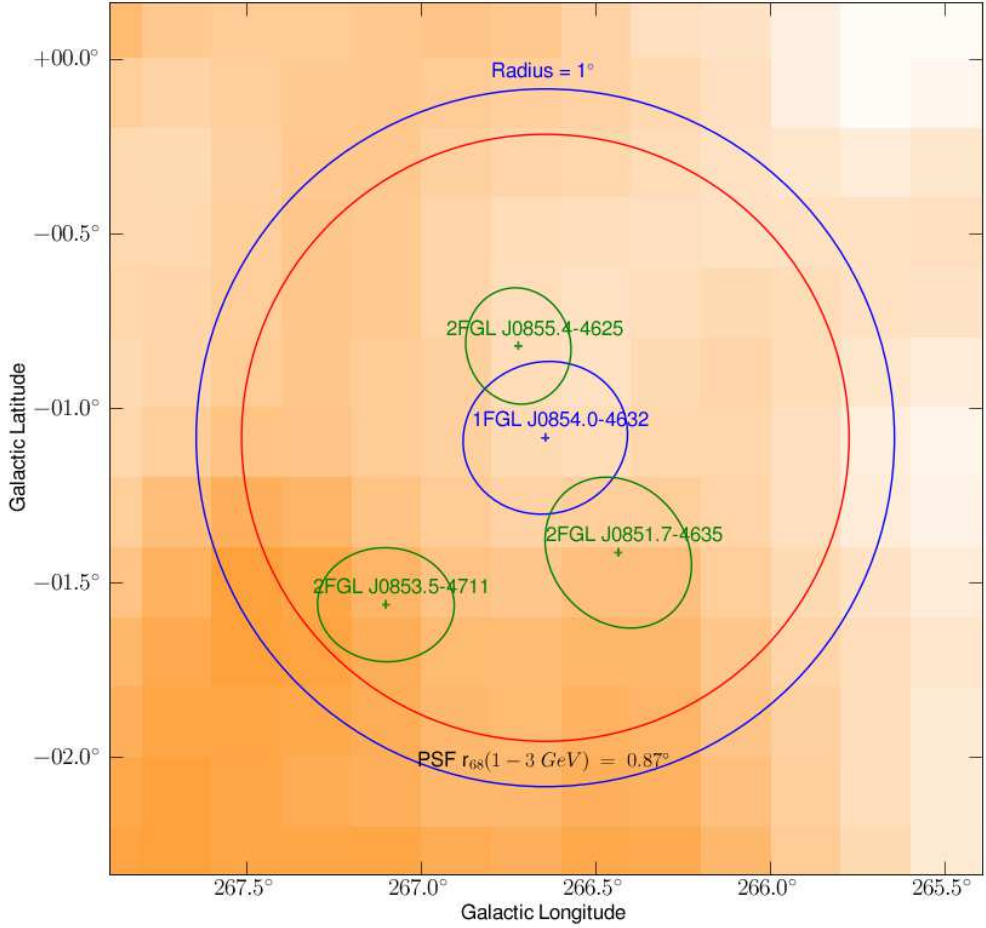}
\caption{A typical example of  a 1FGL source that was split in two 2FGL sources. The E\ $>$\ 1\ GeV counts map (1 pixel = $0\fdg2$) was smoothed using a gaussian kernel ($\sigma$\ =\ 3 pixels). The ellipses represent the 99.9\% confidence error regions. }
\label{fig:ex_split}
\end{figure}
\begin{figure}[!ht]
\centering
\includegraphics[width=0.8\textwidth]{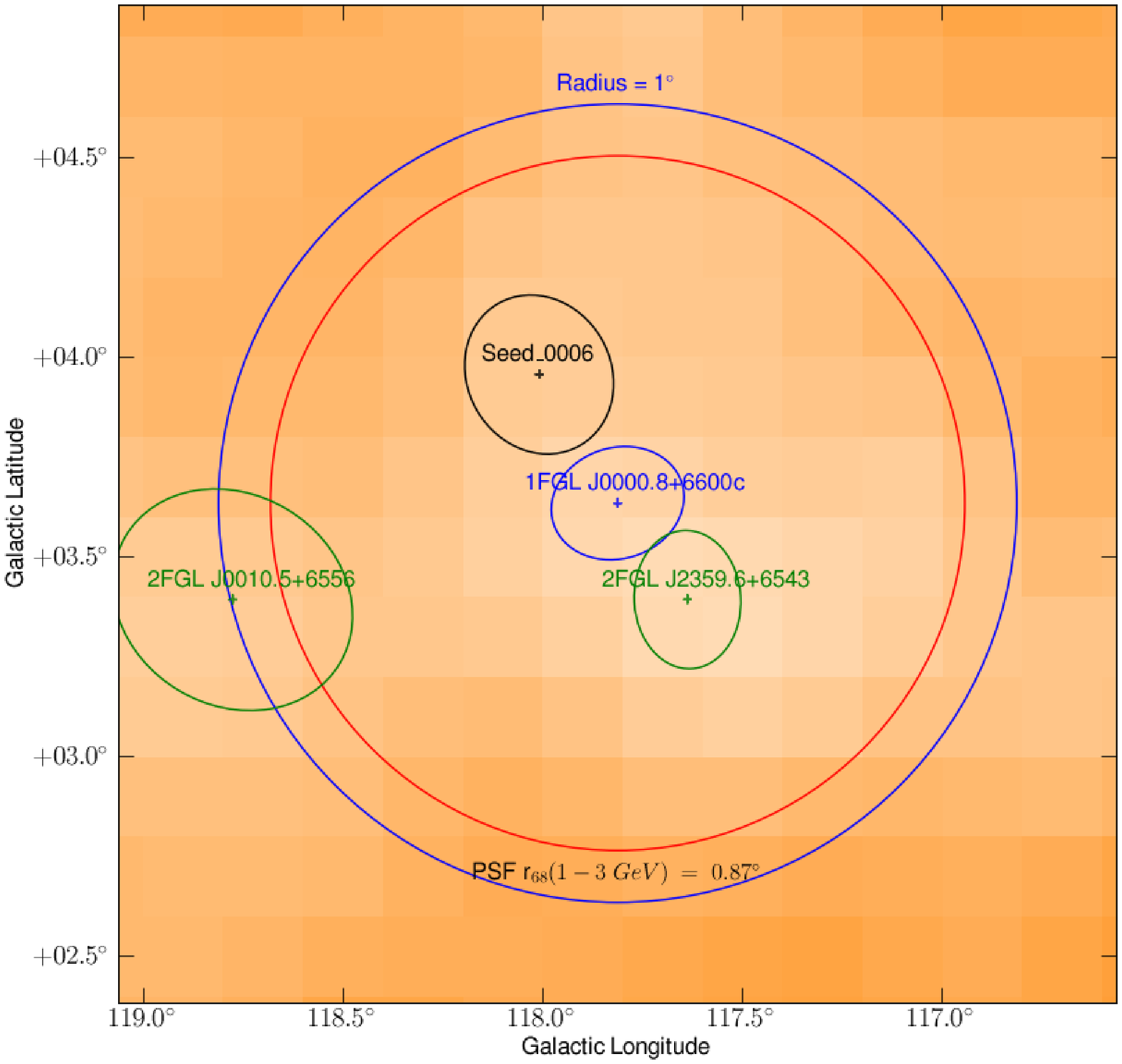}
\caption{A typical example of  a 1FGL source and a 2FGL source having overlapped 99.9\% confidence error regions. In this particular case there is also a seed (candidate source considered in the 2FGL analysis) very close to the 1FGL source and we cannot exclude the possibility that the 1FGL source was split into two seeds.
The E\ $>$\ 1\ GeV counts map (1 pixel = $0\fdg2$) was smoothed using a gaussian kernel ($\sigma$\ =\ 3 pixels). The ellipses represent the 99.9\% confidence error regions. }
\label{fig:ex_over}
\end{figure}

Several other 1FGL sources were also split into more than one candidate source seed  (`S', in  the Flags column of Table \ref{tab:1fgl_2fgl}). In some cases only one of the two seeds  reached  a TS$>$25 and so was included in the 2FGL list (see Figure \ref{fig:ex_split}).
Another example of splitting is 1FGL\ J1642.5+3947,  that was tentatively associated with the blazar 3C~345 in \citet[1LAC;][]{LAT10_AGNcat} paper.  This source has no 2FGL counterpart, because  it is now resolved into two sources:\ 2FGL J1642.9+3949  associated with  3C~345 and  2FGL J1640.7+3945 associated with NRAO~512. 
Other 1FGL sources have overlapping $\theta_{99.9}$ source location uncertainty regions with one or more 2FGL sources or seeds and have the `O' flag in Table \ref{tab:1fgl_2fgl} (see Fig. \ref{fig:ex_over}).  

Another major reason for sources to disappear between 1FGL and 2FGL is a change in the calculated significance. 
As described in the \S~\ref{residual_ts_map}, the 2FGL catalog was built starting from 3499 seeds  with  $TS > 10$ in the $pointlike$ analysis. The final $gtlike$ analysis, which did not change the positions of the seeds, resulted in the 1873 sources with  $TS > 25$ that make up the 2FGL catalog. Among the other seeds that  did not  reach the threshold, 104 can be associated with 1FGL sources (using $\Delta < d_{99.9}$).  These sources, marked with a `C' in the flags column of the table, can be considered to be confirmed sources whose significance dropped below the threshold, either as a result of time variability, change in the diffuse model, or the shift from unbinned to binned likelihood in the catalog analysis procedure. 

In order to quantify the effect of changing $gtlike$ from unbinned to binned mode, we performed a new binned analysis of the original 11-month data set, using the P6V3  Diffuse IRFs  and  the same Galactic diffuse emission model as used for the 1FGL analysis.   The analysis also started using the same 1499 seeds that were used as input to the 1FGL run  \citep[see, ][]{LAT10_1FGL}.  This analysis confirmed with TS$>$25 1138 sources of the 1451 sources that were in the 1FGL catalog. Among these confirmed sources are 168 1FGL sources that are not present in 2FGL, but were still detected at TS$ >$ 25 using the binned analysis for the 11-month data set.  In Table \ref{tab:1fgl_2fgl} the sources confirmed by the binned analysis but not included among the 3499 seeds have the flag `BC'.  In the 1FGL catalog, only 5 of  the `BC' sources were found to be variable with probabilities $p\ >$\ 90\% (see the `Var' column in Table  \ref{tab:1fgl_2fgl}).  Since the shift from unbinned to binned analysis has been excluded as a cause for these, their disappearance must be attributed to time variability or, more likely, to change in the diffuse emission model. 

The 102 sources that were no longer detected in either the binned likelihood re-analysis of the 11-month dataset  nor in any of the other all-sky analyses performed using data collected between 11 and 24 months  are considered not confirmed 1FGL sources (`NC' in  the Flags column of Table \ref{tab:1fgl_2fgl}).  
Among these sources 9 were bright during just the first months of the mission and are reported with the flag `VÕ (variable) in Table \ref{tab:1fgl_2fgl}.  They are all associated with AGN, mostly blazars.  An example is 1FGL\ J1122.9$-$6415, associated with PMN~J1123$-$6417 and included in the 0FGL list \citep{LAT09_BSL}, that, after a flare in 2008 September  was not significantly detected again until 2011 May  \citep{ATel3394}.

\begin{figure}[!ht]
\centering
\includegraphics[width=0.80\textwidth]{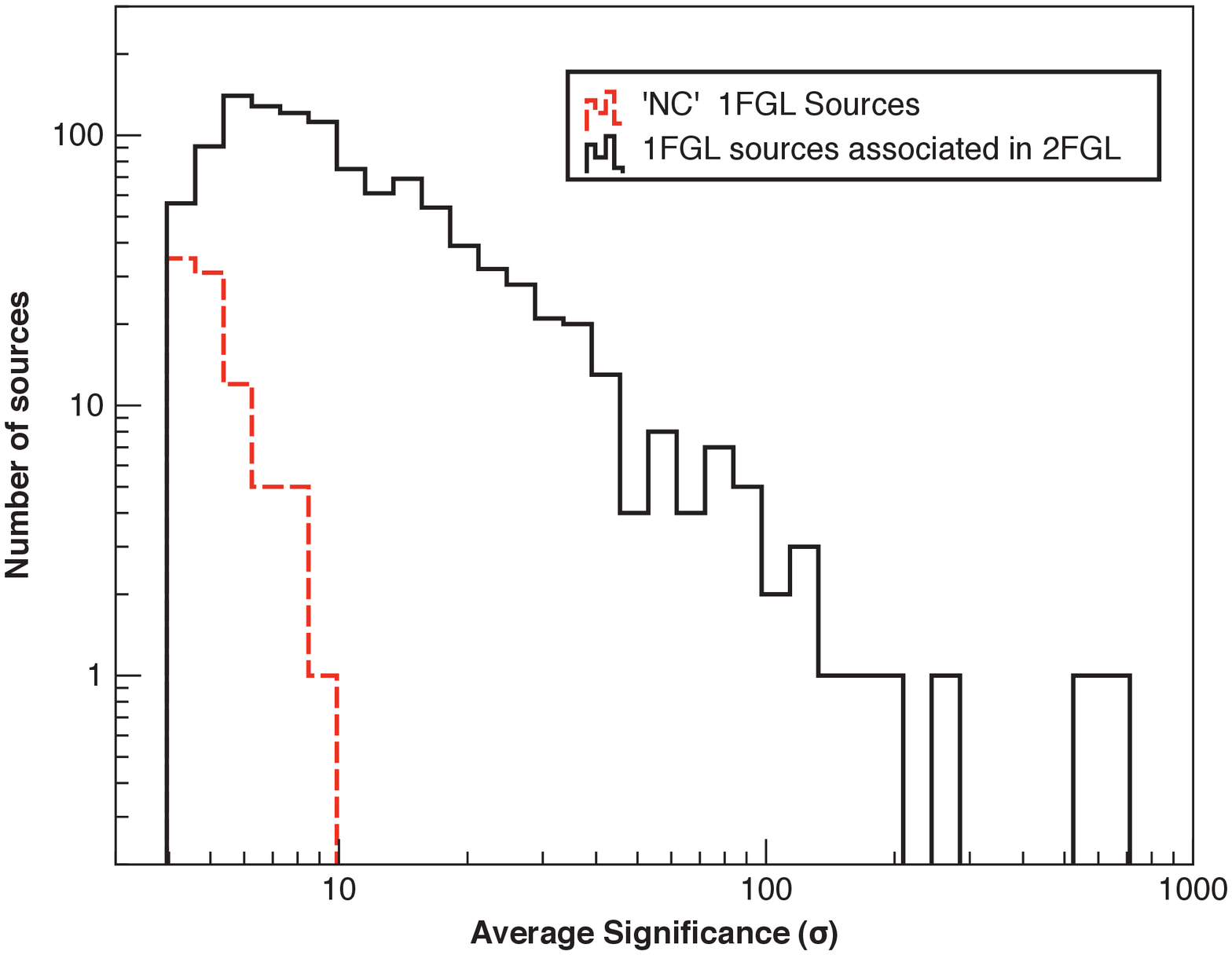}
\caption{Distribution of the significances of the unconfirmed 1FGL sources and of the 1FGL sources associated with sources in the 2FGL catalog.}
\label{fig:hist_1_1fgl}
\end{figure}

Four `NC' 1FGL sources have angular distances less than 1$^\circ$ from the ecliptic (`Sun' in the Flags column of Table \ref{tab:1fgl_2fgl}). Their light curves, which are similar to that shown in Figure \ref{fig:lc_j2124},  show significant detections of the sources only during the passage of the Sun. 
For the other 89  `NC' sources their non-associations with the 2FGL sources can be ascribed to a combination of different effects that cannot be easily disentangled. 

Most of these sources are located close to regions of enhanced diffuse emission (see Figure \ref{fig:missing1fgl}) and about 20 of them were already flagged as sources influenced by the diffuse emission in the 1FGL catalog. 
Also, the fact that these sources were not confirmed in the binned analysis of the 11-month data can be related to statistical fluctuation in the number of the sources detected close to the significance threshold. Figure \ref{fig:hist_1_1fgl} shows the distribution of the source significances, as reported in 1FGL, for the  89 non-confirmed 1FGL sources and for the 1099 1FGL sources present in the 2FGL catalog. Most of the non-confirmed sources have significances less than 6 $\sigma$, which is very close to the threshold ($\ \sim$ 4 $\sigma$) adopted in 1FGL and 2FGL catalogs. These sources are intrinsically faint and for several of them the energy flux reported in the 1FGL catalog  is just an upper limit.  Furthermore the significance values returned by the unbinned likelihood analysis by definition should be intrinsically higher than those returned by the binned analysis.  Thus, most of the 89 sources were above the threshold in the original unbinned 1FGL analysis, but not in the binned analysis.

\begin{deluxetable}{lcrrrccccccccl}
\rotate
\tablewidth{0pc} 
\setlength{\tabcolsep}{0.18cm}
\tabletypesize{\scriptsize}
\tablecaption{List of 1FGL sources not in the 2FGL  catalog\label{tab:1fgl_2fgl}}
\tablehead{ 
\colhead{1FGL} &
 \colhead{1FGL Assoc.\tablenotemark{(a)}} &
 \colhead{$l$\tablenotemark{(a)}} &
 \colhead{$b$\tablenotemark{(a)}} &
 \colhead{$\theta_{95}$\tablenotemark{(a)}} &
 \colhead{$\sigma$\tablenotemark{(a)}} &
 \colhead{$\Gamma_{25}$\tablenotemark{(a)}} &
  \colhead{Var\tablenotemark{(a)}} &
 \colhead{2FGL\tablenotemark{(b)}}&
 \colhead{2FGL\tablenotemark{(c)}}&
 \colhead{$\Delta$\tablenotemark{(d)} }&
 \colhead{$\Delta\ / d_{99.9}$} &
   \colhead{2FGL\tablenotemark{(e)}} &
  \colhead{Flags\tablenotemark{(f)}} \\
 \colhead{} &  
  \colhead{}&  
  \colhead{(deg.)} & 
  \colhead{(deg.)} & 
  \colhead{(deg.)} &  
  \colhead{} &
  \colhead{}& 
    \colhead{$p >$ 90\%}& 
  \colhead{($\Delta <d_{99.9}$)} & 
  \colhead{($d_{99.9} < \Delta< 1^\circ$) }&  
   \colhead{(deg.)} & 
   \colhead{} &
  \colhead{Seed} &   
   \colhead{} 
  }
\startdata 
J0000.8+6600c	&	\nodata	&	117.812	&	3.635	&	0.112	&	9.8	&	2.60	&	\nodata	&	\nodata	&	J2359.6+6543c	&	0.298	&	1.241	&	 \nodata	&	S	\\
J0006.9+4652	&	\nodata	&	115.082	&	$-$15.311	&	0.194	&	10.2	&	2.55	&	T	&	\nodata	&	J0007.8+4713	&	0.381	&	1.249	&	 T	&	S	\\
J0008.3+1452	&	\nodata	&	107.655	&	$-$46.708	&	0.144	&	4.7	&	2.00	&	\nodata	&	\nodata	&	\nodata	&	\nodata	&	\nodata	&	 T	&	C	\\
J0013.7$-$5022	&	BZB J0014$-$5022	&	317.624	&	$-$65.666	&	0.151	&	4.4	&	2.23	&	\nodata	&	\nodata	&	\nodata	&	\nodata	&	\nodata	&	 T	&	C	\\
J0016.6+1706	&	\nodata	&	111.135	&	$-$44.964	&	0.197	&	4.7	&	2.57	&	\nodata	&	\nodata	&	\nodata	&	\nodata	&	\nodata	&	 \nodata	&	NC	\\
J0019.3+2017	&	PKS 0017+200	&	112.787	&	$-$41.944	&	0.203	&	5.9	&	2.38	&	\nodata	&	\nodata	&	\nodata	&	\nodata	&	\nodata	&	 T	&	C	\\
J0028.9$-$7028	&	\nodata	&	305.664	&	$-$46.535	&	0.172	&	6.3	&	2.19	&	\nodata	&	J0029.2$-$7043	&	\nodata	&	0.253	&	0.704	&	 T	&	C	\\
J0038.6+2048	&	\nodata	&	118.912	&	$-$41.969	&	0.146	&	4.6	&	1.63	&	\nodata	&	\nodata	&	\nodata	&	\nodata	&	\nodata	&	 \nodata	&	NC	\\
J0041.9+2318	&	PKS 0039+230	&	120.104	&	$-$39.515	&	0.221	&	5.0	&	2.52	&	\nodata	&	\nodata	&	\nodata	&	\nodata	&	\nodata	&	 T	&	C	\\
J0059.6+1904	&	\nodata	&	125.615	&	$-$43.751	&	0.091	&	5.8	&	2.39	&	\nodata	&	\nodata	&	\nodata	&	\nodata	&	\nodata	&	 \nodata	&	BC	\\
J0110.0$-$4023	&	\nodata	&	287.889	&	$-$76.190	&	0.085	&	4.2	&	1.34	&	\nodata	&	\nodata	&	\nodata	&	\nodata	&	\nodata	&	 T	&	C	\\
J0122.2+5200	&	\nodata	&	127.740	&	$-$10.571	&	0.168	&	4.1	&	2.18	&	\nodata	&	\nodata	&	\nodata	&	\nodata	&	\nodata	&	 T	&	C	\\
J0136.3$-$2220	&	\nodata	&	190.201	&	$-$78.746	&	0.113	&	4.6	&	1.60	&	\nodata	&	\nodata	&	\nodata	&	\nodata	&	\nodata	&	 \nodata	&	BC	\\
J0147.4+1547	&	\nodata	&	142.143	&	$-$44.981	&	0.119	&	4.9	&	1.81	&	\nodata	&	\nodata	&	\nodata	&	\nodata	&	\nodata	&	 T	&	C	\\
J0202.1+0849	&	RX J0202.4+0849	&	150.851	&	$-$50.172	&	0.120	&	4.5	&	1.97	&	\nodata	&	\nodata	&	\nodata	&	\nodata	&	\nodata	&	 T	&	C	\\
\enddata 

\tablenotetext{a}{All the values reported in these columns are from the 1FGL catalog  \citep[][]{LAT10_1FGL}.}
\tablenotetext{b}{Name of the 2FGL source associated with the 1FGL one using $d_{99.9}$.}
\tablenotetext{c}{Closest 2FGL source having a distance  $d_{99.9} < \Delta < 1^\circ$ from  the position of the 1FGL source.  The 2FGL name is also reported  if the 1FGL source and one or more seeds have overlapping $\theta_{99.9}$ error regions but cannot be associated with any seed on the basis of the criterium $\Delta < d_{99.9}$.}
\tablenotetext{d}{The angular separation ($\Delta$) between the 1FGL source and the 2FGL sources associated using $d_{99.9}$ or the closest 2FGL source. }  
\tablenotetext{e}{T =  The 1FGL source and  one of the 2FGL list of initial seeds have an angular separation $< d_{99.9}$.} 
\tablenotetext{f}{C= Confirmed 1FGL  sources.  LMC, Orion, Carina and Ophiuchus  indicate that the source is in a region of the sky  with high diffuse emission and  high density of close sources; NC = not  confirmed 1FGL sources (see text); BC =  1FGL sources confirmed by the 11-m binned likelihood analysis; S = the 1FGL source was split/resolved in one or more seeds;  O = overlapping $\theta_{99.9}$ error regions with one or more seeds; V = variable source  visible only in the first 11 months; Sun = the source was detected when the Sun was at  an angular distance $<$ 1$^\circ$ and the light curve show just a flare in the time bin relative to the passage of the Sun close to the position of the source. 
This table is published in its entirety in the electronic edition of the Astrophysical Journal Supplements. A portion is shown here for guidance regarding its form and content.}
\end{deluxetable}

\section{Source Association and Identification}
\label{source_assoc_main}
    \subsection{Firm Identifications}
\label{source_assoc_firm}

As with the LAT Bright Source List \citep{LAT09_BSL} and 1FGL catalog \citep{LAT10_1FGL}, we retain the distinction between associations and firm identifications.  Although many associations, particularly those for AGN, have very high probability of being true, a firm identification, shown in the catalog by capitals in the Class column in Table~\ref{tab:classes}, is based on one of three criteria:

\begin{enumerate}
\item Periodic Variability.  Pulsars are the larger class in this category.  All PSR labels indicate that pulsed $\gamma$ rays have been seen from the source with a probability of the periodicity occurring  by chance of less than 10$^{-6}$.  A similar chance probability requirement applies to the other set of periodic sources, the high-mass binaries (HMB).  Four of these are included in the catalog:  LS\,I +61 303 \citep{LAT09_LSI}, LS 5039 \citep{LAT09_LS5039}, Cygnus X-3 \citep{LAT09_CygX3}, and 1FGL J1018.6$-$5856 \citep{ATel.3221}. 
\item Spatial Morphology.  Spatially extended sources whose morphology can be related to extent seen at other wavelengths include SNR, PWNe, and galaxies, as described in \S~\ref{catalog_extended}.  The Centaurus A  lobes and core are both marked as identified, because they are part of the same extended source, although the core itself does not show spatial extent.  As noted in \S~3.8, additional extended sources are being found but are not listed in the catalog as firm identifications, because they were analyzed as point sources for this work.  Although individual molecular clouds could in principle be included in this list, the catalog construction incorporates most known clouds into the diffuse model, and so no sources of this type are identified in the catalog. 
\item Correlated Variability.  Variable sources, primarily AGN, whose $\gamma$-ray variations can be matched to variability seen at one or more other wavelengths, are considered to be firm identifications.  Although some cases are well documented, such correlated variability is not always easily defined.   We conservatively require data in more than two energy bands for comparison.  Finding a blazar to have a high X-ray flux at the same time as a  $\gamma$-ray flare, for example, does not qualify if there is no long-term history for the X-ray emission. We include those sources whose variability properties are documented either in papers or with Astronomer's Telegrams.  This list does not represent the result of a systematic study.  Ongoing work will undoubtedly enlarge this list.  The one Galactic source identified in this way is nova V407 Cygni \citep{LAT10_V407Cyg}.
\end{enumerate}

We include one exception to these rules.  The Crab PWN is listed as a firm identification even though it does not meet any of these criteria.  The well-defined energy spectrum, distinct from the Crab pulsar spectrum and matching spectra seen at both lower and higher energies provides a unique form of identification \cite{LAT10_Crab}.

In total, we firmly identify 127 out of the 1873 2FGL sources.
Among those,
83 are pulsars,
28 are AGN,
6 are SNR,
4 are HMB,
3 are PWN,
2 are normal galaxies, and
one is a nova (Table~\ref{tab:classes}).

    \subsection{Automated Source Associations}
\label{source_assoc_automated}

Our approach for automated source association closely follows that used 
for the 1FGL catalog, and details of the method are provided in \citet{LAT10_1FGL}. 
In summary, we use a Bayesian approach that trades the positional
coincidence of possible counterparts with 2FGL sources against the expected 
number of chance coincidences to estimate the probability that a specific
counterpart association is indeed real (i.e., a physical association).
As for 1FGL, we retain counterparts as associations if they reach a posterior
probability of at least 80\%.

We apply this method to a set of counterpart catalogs for which we
calibrate the prior source association probabilities using Monte Carlo simulations of 
fake 2FGL catalogs.
In comparison to 1FGL, for which we made 100 independent simulations for
each catalog, we adapted the number of simulations (between 100 and 1000)
so that the relative accuracy in the expected false association rate is determined
to better than 5\% for each catalog.
This improved the precision of our probability computations for catalogs that
have only few associations with 2FGL sources.
The prior probabilities adopted for each catalog are listed in 
Table \ref{tab:catalogs}.

Another improvement with respect to 1FGL concerns the estimation of the
local counterpart densities $\rho_k$.
For 1FGL we estimated these densities from the number of objects in the 
counterpart catalog within a radius of $4^\circ$ around the location of the
1FGL source of interest.
 For counterpart catalogs containing strong density variations on smaller scales
(e.g., O stars, WRs and LBV stars) this choice led
to an underestimate of the actual source densities in these regions, which in turn
resulted in overestimations of the association probabilities
\citep[see discussion in][]{LAT10_1FGL}.
For 2FGL we estimate the source densities in each counterpart catalog
using an all-sky map which we implemented as a HEALPix grid with resolution 
$N_{side}=512$, corresponding to an angular resolution of about $6'$,
with the objects of each counterpart catalog binned in this grid.
We removed sparseness of the binning and attenuated the statistical fluctuations 
by applying a spherical Gaussian smoothing kernel with width adjusted adaptively
so that at least 3 sources contributed to the density estimate at each grid location.

For certain counterpart catalogs the Bayesian method could not be applied 
since either
(1) the location uncertainty of the counterpart is larger than the location uncertainty
of the 2FGL source (these catalogs are indicated by $^\ast$ in Table \ref{tab:catalogs}), or
(2) the counterpart is an extended source (these catalogs are indicated by $^\dag$
in Table \ref{tab:catalogs}).
In the first case, we consider as potential associations all objects for which the separation
from the 2FGL source is less than the quadratic sum of the 95\% confidence error
radii.  (For elliptical error regions we take the semimajor axis as the error radius.)
In the second case, we assume that the counterparts have circular extensions
and consider all objects as associations for which the extension circle overlaps 
with the 95\% confidence error radius of the 2FGL source, with the semimajor axis of the 
2FGL source location ellipse again taken as the error radius.

The list of catalogs used in the automatic association is summarized in Table \ref{tab:catalogs},
organized into four categories:
(1) catalogs of known or plausible $\gamma$-ray-emitting source classes,   
(2) catalogs of surveys at other frequencies,
(3) catalogs of GeV sources, and
(4) catalogs of identified $\gamma$-ray sources.
The first category allows us to assign 2FGL sources to object classes, while the second category reveals
multiwavelength counterparts that may suggest the possible nature of the associated
2FGL source.
The third category allows assessment of former GeV detections of 2FGL sources, and the
fourth category keeps track of all firm identifications (cf.~\S~\ref{source_assoc_firm}).
For this last category we claim associations based on the spatial overlap of the true
counterpart position with the 2FGL $99.9\%$ confidence error ellipse.

With respect to 1FGL, we updated all catalogs for which more comprehensive compilations became
available.
We now use
the 13th edition of the Veron catalog \citep{AGNcatalog},
version 20 of BZCAT\footnote{http://www.asdc.asi.it/bzcat/} \citep{BZcatalog},
version 1.40 of ATNF \citep{ATNFcatalog} that we augmented with 158 recently detected pulsars
that are not yet in the ATNF database,
the 2010 December revision of the Globular Cluster database \citep{GlobClusterCatalog}
that we augmented with 3 recently detected clusters,
version 3.1 of the Open Cluster catalog \citep{OpenClustersCatalog},
the 2010 December 5 version of the VLBA Calibrator Source List\footnote{
The VLBA Calibrator Source List can be downloaded from http://www.vlba.nrao.edu/astro/calib/vlbaCalib.txt.},
and
the most recent version of the TeVCat catalog\footnote{http://tevcat.uchicago.edu/}.
We also added new counterpart catalogs:  the Australia Telescope 20 GHz Survey \citep{AT20G} 
and the IRAS Revised Bright Galaxy Sample \citep{IRAScatalog}, from which we selected
all sources with 100~$\mu$m fluxes brighter than 50 Jy.  The latter catalog replaces
the starburst catalog used for 1FGL.

Following the philosophy for 1FGL, we split our pulsar catalog into normal pulsars and 
millisecond pulsars (MSPs) by requiring
$\log \dot{P} + 19.5 + 2.5 \times \log P < 0$
for the latter.
Because globular clusters are classified by a separate catalog and the LAT is unable to
spatially resolve individual MSPs in globular clusters, we removed all globular
cluster MSPs from the pulsar catalog.
We furthermore collect normal pulsars with
$\dot{E}/d^2 > 5 \times 10^{32}$ erg kpc$^{-2}$ s$^{-1}$
into a separate counterpart catalog to specifically select energetic and nearby pulsars
that are more likely potential $\gamma$-ray sources.
The value separating these classes corresponds to the lowest $\dot{E}/d^2$
found among all LAT identified pulsars.
We also split-off point-like supernova remnants (SNRs) from the Green catalog \citep{SNRcatalog} by selecting all 
objects with diameters $<20'$.
In parallel, we use the full Green catalog for finding matches with potentially extended SNRs.
Furthermore we divided the TeVCat catalog into point-like and extended sources by selecting 
for the latter all sources with extension radius $>0$.

We also searched for associations using
the Atlas of Radio/X-ray associations to optical objects \citep{ARXAcatalog} from which we 
selected those objects that have stellar, radio, and X-ray associations (Cl=SRX),
the Planck Early Release Catalogs \citep{Planckcatalog}, 
the 4th IBIS catalog \citep{IBIScatalog}, and
the Swift-BAT 58-Month Survey \citep{BATcatalog},
yet as these did not reveal any new reliable and plausible counterpart that has not already been 
found in one of the other catalogs, we did not include these catalogs in our final
results.

\begin{deluxetable}{lrrrrrr}
\setlength{\tabcolsep}{0.04in}
\tablewidth{0pt}
\tabletypesize{\scriptsize}
\tablecaption{Catalogs used for the automatic source association
\label{tab:catalogs}
}
\tablehead{
\colhead{Name} & 
\colhead{Objects} & 
\colhead{$P_{\rm prior}$} & 
\colhead{\nass} & 
\colhead{\nfalse} & 
\colhead{\nfalsemc} & 
\colhead{Ref.}
}

\startdata
High $\dot{E}/d^2$ pulsars & 213 & 0.037 & 29 & 0.9 & 1.0 & 2 \\
Other normal pulsars & 1657 & 0.011 & 12 & 0.6 & 0.7 & 2 \\
Millisecond pulsars & 137 & 0.014 & 45 & 0.3 & 0.4 & 2 \\
Pulsar wind nebulae & 69 & 0.009 & 25 & 0.5 & 0.6 & 1 \\
High-mass X-ray binaries & 114 & 0.003 & 2 & 0.1 & 0.2 & 3 \\
Low-mass X-ray binaries & 187 & 0.007 & 3 & 0.3 & 0.3 & 4 \\
Point-like SNR & 157 & 0.019 & 6 & 0.7 & 0.3 & 5 \\
Extended SNR$^\dag$ & 274 & n.a. & 92 & n.a. & 39.7 & 5 \\
O stars & 378 & 0.005 & 1 & 0.2 & 0.2 & 6 \\
WR stars & 226 & 0.005 & 0 & 0 & 0.2 & 7 \\
LBV stars & 35 & 0.001 & 1 & $<0.1$ & 0.2 & 8 \\
Open clusters & 2140 & 0.005 & 0 & 0 & 0.2 & 9 \\
Globular clusters & 160 & 0.028 & 11 & 0.5 & 0.6 & 10 \\
Dwarf galaxies$^\dag$ & 14 & n.a. & 7 & n.a. & 3.4 & 1 \\
Nearby galaxies & 276 & 0.014 & 5 & 0.4 & 0.4 & 11 \\
IRAS bright galaxies & 82 & 0.021 & 6 & 0.2 & 0.2 & 12 \\
BZCAT (Blazars) & 3060 & 0.341 & 691 & 7.4 & 6.9 & 13 \\
BL Lac & 1371 & 0.170 & 278 & 2.8 & 2.6 & 14 \\
AGN & 10066 & 0.009 & 8 & 0.3 & 0.4 & 14 \\
QSO & 129853 & 0.196 & 197 & 6.7 & 6.7 & 14 \\
Seyfert galaxies & 27651 & 0.028 & 29 & 2.0 & 1.9 & 14 \\
Radio loud Seyfert galaxies & 29 & 0.001 & 4 & $<0.1$ & $<0.1$ & 1 \\
\hline
CGRaBS & 1625 & 0.258 & 352 & 3.8 & 4.1 & 15 \\
CRATES & 11499 & 0.341 & 634 & 17.7 & 17.8 & 16 \\
VLBA Calibrator Source List & 5776 & 0.258 & 623 & 11.8 & 12.0 & 17 \\
ATCA 20 GHz southern sky survey & 5890 & 0.296 & 335 & 10.3 & 10.6 & 18 \\
TeV point-like source catalog$^\ast$ & 61 & n.a. & 47 & n.a. & 0.6 & 19 \\
TeV extended source catalog$^\dag$ & 57 & n.a. & 48 & n.a. & 20.1 & 19 \\
\hline
1st AGILE catalog$^\ast$ & 47 & n.a. & 57 & n.a. & 21.1 & 20 \\
3rd EGRET catalog$^\ast$ & 271 & n.a. & 116 & n.a. & 31.0 & 21 \\
EGR catalog$^\ast$ & 189 & n.a. & 69 & n.a. & 11.4 & 22 \\
0FGL list$^\ast$ & 205 & n.a. & 185 & n.a. & 5.1 & 23 \\
1FGL catalog$^\ast$ & 1451 & n.a. & 1099 & n.a. & 18.1 & 24 \\
\hline
LAT pulsars & 87 & n.a. & 80 & n.a. & 1.4 & 1 \\
LAT identified & 44 & n.a. & 43 & n.a. & 0.7 & 1 \\ 
\enddata

\tablerefs{
$^1$Collaboration internal;
$^2$\citet{ATNFcatalog};
$^3$\citet{HMXBcatalog};
$^4$\citet{LMXBcatalog};
$^5$\citet{SNRcatalog};
$^6$\citet{OStarCatalog};
$^7$\citet{vanderHucht2001};
$^8$\citet{LBVcatalog};
$^9$\citet{OpenClustersCatalog};
$^{10}$\citet{GlobClusterCatalog};
$^{11}$\citet{NearbyGalaxiesCatalog};
$^{12}$\citet{IRAScatalog};
$^{13}$\citet{BZcatalog};
$^{14}$\citet{AGNcatalog};
$^{15}$\citet{CGRaBS};
$^{16}$\citet{CRATES};
$^{17}$http://www.vlba.nrao.edu/astro/calib/vlbaCalib.txt;
$^{18}$\citet{AT20G};
$^{19}$http://tevcat.uchicago.edu/;
$^{20}$\citet{AGILEcatalog};
$^{21}$\citet{3EGcatalog};
$^{22}$\citet{EGRcatalog};
$^{23}$\citet{LAT09_BSL};
$^{24}$\citet{LAT10_1FGL}}
\end{deluxetable}

\subsubsection{Automated association summary}
\label{automated_association_summary}
The results of the automated association procedure for each of the external catalogs are summarized
in Table \ref{tab:catalogs}.
For each catalog we quote the name (Column 1), the number of objects in the catalog (Column 2),
the prior probability assigned by our calibration procedure (Column 3), and 
the number \nass\ of associations that have been found between 2FGL sources
and counterpart objects (Column 4).
Note that a given 2FGL source may have counterparts in multiple catalogs, and a given object in a
counterpart catalog may have multiple associated 2FGL sources (which may arise if the object
is spatially extended or if it has a large location uncertainty).
Consequently, the sum of the \nass\ column considerably exceeds the total number of associated
2FGL sources.
Using the posterior probabilities $P_{ik}$ that we derive by the Bayesian method for all associations 
$i$ in a counterpart catalog $k$, we compute the expected number of false associations using
$\nfalse = \sum_{P_{ik}} (1-P_{ik})$
(Column 5).
To validate that these estimates are accurate (and thus that our prior probability calibration was
precise) we alternatively estimate the number of false associations \nfalsemc\ using Monte Carlo 
simulations of 100 fake 2FGL catalogs (Column 6); we refer to \citet{LAT10_1FGL} for a detailed
description of the simulation procedure.
For all catalogs we find $\nfalse \simeq \nfalsemc$ which confirms that the posterior probabilities 
computed by the automatic association procedure are accurate.

In total we find that 1141 of the 1873 sources in the 2FGL catalog ($61\%$) have been associated 
with a least one non-GeV $\gamma$-ray counterpart by the automated procedure.
Among those, 
123 sources ($11\%$) are firmly identified objects,
790 ($69\%$) are associated with at least one object of known type, and
228 ($20\%$) have counterparts only in the multi-wavelength catalogs.
For the remaining 732 sources in the 2FGL catalog that have no non-GeV $\gamma$-ray counterpart,
322 sources ($44\%$) are associated with former GeV detections, and
410 sources ($56\%$) are new GeV sources.

Among the 2FGL sources that are not firmly identified,
940 ($92\%$) have been associated using the Bayesian method at the $80\%$ confidence 
level, while 78 ($8\%$) have been associated based on overlap of the error regions or source extents 
and have lower confidence (catalogs based on spatial overlap are indicated by
$^\dag$ in Table \ref{tab:catalogs}).
From simulations we expect that 43 of the 940 sources ($5\%$) that were associated 
with the Bayesian method are chance coincidences.
Among the 78 sources that were associated based on overlap, the expected number of chance
coincidences amounts to 55 ($71\%$), demonstrating that these associations are considerably
less reliable.
Due to this large false positive rate, we do not claim any associations based on overlap
in our final catalog.
We record, however, any spatial overlap with a TeV source in the FITS file version of the catalog, 
and use a special flag in our catalog (TEVCAT\_FLAG), distinguishing point-like (P) from extended 
(E) TeV counterparts (see Appendix C).
We furthermore list all unidentified 2FGL sources that are spatially overlapping with SNRs in
Table \ref{tab:snrext}.
Finally, 2FGL sources spatially overlapping with the LMC that are not associated with
any object in one of the other counterpart catalogs are indicated as {\em LMC field}.

\subsubsection{Active Galactic Nuclei associations}
\label{assoc:agn}

Active Galactic Nuclei (AGN), and in particular blazars, are the most prominent class of associated
sources in 2FGL.
In total, our automatic association procedure finds 917 2FGL sources that are associated with
AGN, of which 
894 are blazars, 
9 are radio galaxies, 
5 are Seyfert galaxies, and
9 are other AGN.
Among the 5 Seyfert galaxies, 4 are narrow-line Seyfert 1 galaxies that have been established as a
new class of $\gamma$-ray active AGN \citep{LAT09_Seyfert}.
The 5th object is NGC 6814, which is associated with 2FGL~J1942.5$-$1024.
Note, however, that we expect up to $\sim2$ false positives among the Seyfert galaxy associations
(cf.~Table \ref{tab:catalogs}), hence we cannot draw any firm conclusions about the possibility that 
normal Seyfert galaxies are indeed GeV $\gamma$-ray sources based on this single
association.
 
\begin{figure}
\epsscale{.80}
\plotone{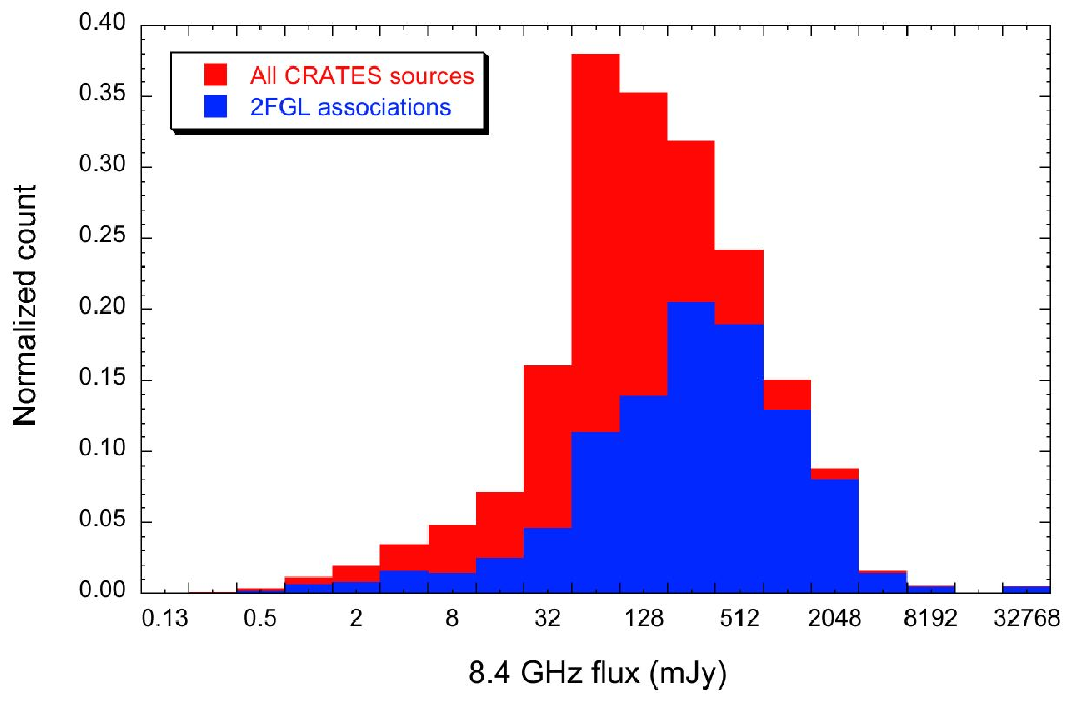}
\caption{Normalized histograms of the 8.4 GHz radio flux of CRATES sources
(red: all sources, blue: objects associated with 2FGL sources).\label{fig:crates}}
\end{figure}

AGN observed by the LAT are also sources of radio (and X-ray) emission, and we find
a clear trend that AGN associated with 2FGL sources have larger radio fluxes than the average
object in the counterpart catalogs.
This trend, which was exploited already for the association of blazars in the EGRET
catalog \citep{sowards2003}, is illustrated in Figure~\ref{fig:crates}, where we compare the
distribution of the 8.4~GHz radio fluxes of all sources in the CRATES catalog to that for objects
associated with 2FGL sources.
Obviously, the average radio flux of CRATES sources associated with 2FGL sources is about one
order of magnitude larger than the overall average for the CRATES catalog.
Similar differences are observed for other radio catalogs.

In our dedicated effort for studying the AGN population in the 2FGL catalog, which we 
publish in an accompanying paper \citep[2LAC;][]{LAT11_2LAC}, we make use of this property
to enhance the sample of associated 2FGL sources.
Briefly, instead of including all objects from the counterpart catalog in the estimation of the local
counterpart densities $\rho_k$, we count only those objects with radio (or X-ray) flux
equal or larger than the flux $S$ of the counterpart under consideration, i.e., $\rho_k(>S)$.
Using this procedure, the chance coincidence probabilities are considerably reduced, and
consequently, the posterior association probabilities are increased \citep[see also][]{sowards2003}.
We apply this procedure to a number of fairly uniform surveys of radio sources
(CRATES, NVSS, SUMSS, PMN, ATCA 20 GHz, FRBA, GAPS, CLASS and VCS)
and to the ROSAT All-Sky Survey of X-ray sources (RASS), for which we assume
the counterpart density $\rho_k(>S)$ to be position independent.  In this case
$\rho_k(>S)$ is then determined from the $\log N - \log S$ distribution of objects in
the catalog divided by the survey area, where $N$ is the total number of sources with flux $>S$.

The 2LAC association procedure increases the number of AGN associations by 173, 
resulting in a total of 1090 2FGL sources that we associate with known AGN; 
note that the 2LAC catalog lists two associated AGNs for the 27 2FGL sources for which 
more than one plausible association was found.  
The total number of 2FGL sources associated with a least one non-GeV $\gamma$-ray 
counterpart is thus 1314 ($70\%$ of all 2FGL sources).
Among the AGN associations we find 1064 blazars, of which 
432 are BL Lac (+38 with respect to the automatic association procedure),
370 are FSRQ (+24), and
262 are of unknown type (+108).
The procedure also reveals 2 additional radio galaxies
(For~A associated with 2FGL~J0322.4$-$3717 and
PKS 0943$-$76 associated with 2FGL~J0942.8$-$7558),
and one additional Seyfert galaxy
(ESO~323$-$77 associated with 2FGL~J1306.9$-$4028).
For the final AGN associations presented in the 2FGL catalog, we adopt the results
of the 2LAC procedure combined with the results of the automatic association pipeline
(see also Table \ref{tab:classes}).

Comparing to 1FGL \citep{LAT10_1FGL}, where out of 1451 sources 573 ($40\%$) were found to be
associated with blazars, 802 ($43\%$) out of 1873 sources are associated with blazars in 2FGL,
a relative increase which is readily explained by the particular effort that has been undertaken
to maximize the number of blazar associations \citep{LAT11_2LAC}.
Neglecting the 2LAC blazar associations, the fraction of 2FGL sources associated with blazars
would have been $40\%$, identical to what was found for 1FGL.
On the other hand, the proportion of active galaxies of uncertain type (designated by
`agu' in Table~\ref{tab:sources}) has increased considerably:
while 92 ($6\%$) 1FGL sources were classified `agu', 262 ($14\%$) 2FGL sources are
now in this category, more than doubling the proportion of this source class.
This increase can be explained by the extensive use of radio and X-ray surveys in the
2LAC association procedure that provides a greater number of blazar candidates
that deserve dedicated follow-up observations to assess their natures.
We also note that in the 2FGL catalog we have two new extragalactic source classes with respect to 1FGL:
radio galaxies (`rdg') and Seyfert galaxies (`sey').
Both were counted in the `non-blazar active galaxy' class (designated by `agn') in 1FGL,
and 28 ($1.9\%$) 1FGL sources were associated with that class.
Adding the `rgd' and `sey' designators to the `agn' for 2FGL amounts to
27 ($1.4\%$) associations, a number that is comparable to that found for 1FGL.

\subsubsection{Normal Galaxies}

Normal galaxies are now established as a class of high-energy $\gamma$-ray emitters
\citep{LAT10_M31}, and we associate 7 2FGL sources with such objects.
Of those, we consider
the Small Magellanic Cloud (2FGL~J0059.0$-$7242) and 
the Large Magellanic Cloud (LMC, 2FGL~J0526.6$-$6825)
as identified owing to their spatial extensions in the LAT data.
From the remaining five, 4 are classified as starburst galaxies:
M82 (2FGL~0955.9+6936), 
NGC~253 (2FGL~J0047.0$-$2516),
NGC~4945 (2FGL~J1305.8$-$4925), and
NGC~1068 (2FGL~J0242.5+0006).
The fifth is the Andromeda galaxy M31
(2FGL~J0042.5+4114).

Except for M31, all of the associated 2FGL $\gamma$-ray sources in this class were already present in 1FGL, 
yet the two starburst galaxies NGC~4945 and NGC~1068 were not associated as such as
they were not included in our very limited counterpart catalog used at that time
\citep{LAT10_1FGL}.
For 2FGL, we included a catalog of infrared bright galaxies in the automatic association procedure (see \S~\ref{source_assoc_automated})
because starburst galaxies are prominent emitters in this waveband.
Furthermore, we have found that the $\gamma$-ray fluxes of Local Group and starburst galaxies 
correlate well with star formation rates \citep{LAT10_M31}, which in turn correlate with
infrared luminosity.
Hence by selecting infrared bright galaxies from the IRAS Revised Bright Galaxy Sample 
\citep{IRAScatalog} we have added a catalog to our procedure that contains normal galaxies
that are potential $\gamma$-ray emitters.

Three 2FGL sources lie within the extended-source template for the LMC (2FGL~J0451.8$-$7011, 
2FGL~J0455.8$-$6920, and 2FGL~J0533.3$-$6651; see \S~\ref{catalog_extended}).  
Their physical association with the LMC is not certain, but they are classified here as being part of 
the LMC and as mentioned in \S~\ref{automated_association_summary} are indicated in 
Table~\ref{tab:sources} as belonging to $LMC field$.

\subsubsection{Pulsars}

As of this writing, 87 pulsars have been firmly identified by the LAT through
the detection of $\gamma$-ray pulsations.
Four of these pulsars did not pass TS~$>25$ in the catalog analysis, and therefore they were excluded
from the 2FGL catalog.
These pulsars are 
PSR~J1513$-$5908 (aka PSR~B1509$-$58),
PSR~J1531$-$5610,
PSR~J1801$-$2451, and
PSR~J1939+2134.
Of the remaining 83, 80 were formally associated by the automatic association procedure.
The remaining 3 are found to be close to 2FGL sources, but their angular separation $\Delta$ from
these sources exceeds their effective $99.9\%$ location error radius $\theta_{99.9}$.\footnote{
The effective error radius is the size of the error ellipse at the position angle toward the
counterpart. We estimate the $99.9\%$ confidence radius by multiplying the $95\%$
confidence radius by $1.52$.}
We find:
\begin{itemize}
\item PSR~J1023$-$5746 near 2FGL~J1022.7$-$5741 ($\theta_{99.9}=4.2'$, $\Delta=5.0'$).
2FGL~J1022.7$-$5741, which is in the Westerlund 2 field, lies only $10'$ from 2FGL~J1023.5$-$5749c,
so possibly the determination of its localization and/or localization uncertainty has been 
affected by this nearby source.
\item PSR~J1357$-$6429 near 2FGL~J1356.0$-$6436 ($\theta_{99.9}=9.1'$, $\Delta=9.5'$).
2FGL~J1356.0$-$6436 is a relatively isolated source, but we note a possible association
with the PWN HESS~J1356$-$645 \citep{LAT11_HESSJ1356m645}.
\item PSR~J1747$-$2958 near 2FGL~J1747.1$-$3000 ($\theta_{99.9}=2.9'$, $\Delta=3.2'$).
2FGL~J1747.1$-$3000 is located near the Galactic Center, and the localization of the source
may be affected by systematic uncertainties in the diffuse Galactic emission model.
\end{itemize}

In addition to the identified pulsars, four 2FGL sources are associated with radio pulsars:
\begin{itemize}
\item  2FGL~J1112.5$-$6105: PSR~J1112$-$6103
\item  2FGL~J1632.4$-$4820c: PSR~J1632$-$4818
\item 2FGL~J1717.5$-$5802: PSR~J1717$-$5800 (?)
\item 2FGL~J1928.8+1740c: PSR~J1928+1746 (?)
\end{itemize}
PSR~J1717$-$5800 has $\dot{E} = 2.3 \times 10^{32}$, ten times lower than for any known 
$\gamma$-ray pulsar.
The other three have $\dot{E} > 10^{34}$ erg s$^{-1}$ and the LAT team phase-folds $\gamma$
rays from their positions using radio rotation ephemerides as described by \citet{LATtiming}. 
Gamma-ray pulsations have not been detected for these pulsars.
We mark two of the associations as questionable (?) because the corresponding 2FGL
sources have spectra that are considerably softer (spectral index $\sim2.5$) than typically observed 
for $\gamma$-ray pulsars \citep{LAT10_PSRcat}.

The automatic association procedure also finds 21 2FGL sources to be associated
with MSPs.
Nineteen of those have unassociated counterparts in the 1FGL catalog, and have been
discovered in radio pulsar searches of unassociated 1FGL sources
\citep[e.g., ][]{Ransom2011_MSP_1FGL,Cognard2011_MSP_1FGL, Keith2011_MSP_1FGL, Hessels2011_MSP}.
Rotation ephemerides accurate enough to allow phase folding $\gamma$-rays from the 
directions of the newly discovered radio pulsars can require a year of radio observations 
to disentangle, e.g., binary orbital motion from annual parallax. 
As the ephemerides become available many of the unassociated 1FGL sources may reveal 
$\gamma$-ray pulsations, as has already occurred for several.
Two 2FGL sources associated with MSPs have no 1FGL counterparts:
\begin{itemize}
\item  2FGL~J1023.6$+$0040: PSR~J1023+0038
\item  2FGL~J1125.0$-$5821: PSR~J1125$-$5825 \citep{Bates2011_MSP}
\end{itemize}
\citet{Tam2010_PSRJ1023} reported the LAT detection of $\gamma$-ray
emission toward PSR~J1023+0038, the only known rotation powered MSP in a quiescent 
LMXB.
The spectrum of 2FGL~J1023.6$+$0040 is rather soft (spectral index $\sim2.5$) for an MSP,
but the system is sufficiently special that this does not necessarily rule out the
association \citep[see discussion in][]{Tam2010_PSRJ1023}.

\subsubsection{Pulsar wind nebulae}

Formally, we find 69 2FGL sources to be associated with PWNe, but except
for three, all of them are also associated with known pulsars.
Among those are three sources for which a dedicated analysis allowed us to identify both the
pulsar and the PWN; they are summarized in Table \ref{tab:psrpwn}, and the 2FGL catalog
contains both the pulsar and the PWN as separate associated sources.
For the other 63 2FGL sources, the observed pulsations firmly identify the pulsars as the primary 
source of the observed $\gamma$ rays, although some minor contribution from a PWN cannot 
be excluded.

\begin{deluxetable}{llclc}
\setlength{\tabcolsep}{0.04in}
\tablewidth{0pt}
\tabletypesize{\scriptsize}
\tablecaption{Identified PSR \& PWN
\label{tab:psrpwn}
}
\tablehead{
\multicolumn{2}{c}{Pulsar} &
\multicolumn{2}{c}{Pulsar Wind Nebula} &
\colhead{Ref} \\
\colhead{PSR} &
\colhead{2FGL} &
\colhead{PWN} &
\colhead{2FGL}
}
\startdata
J0835$-$4510 (Vela) & J0835.3$-$4510 & Vela X & J0833.1$-$4511e & 1 \\
J1509$-$5850 & J1509.6$-$5850 & MSH~15$-$52 & J1514.0$-$5915e & 2 \\
J1826$-$1256 & J1826.1$-$1256 & HESS~J1825$-$137 & J1824.5$-$1351e & 3 \\
\enddata
\tablerefs{
$^1$\citet{LAT10_VelaX};
$^2$\citet{LAT10_PSR1509};
$^3$\citet{LAT11_J1825}}
\end{deluxetable}

More interesting are the three PWN associations for which no pulsar has so far been identified.
Those are:
\begin{itemize}
\item 2FGL~J1112.1$-$6040: G291.0$-$0.1
\item 2FGL~J1640.5$-$4633: G338.3$-$0.0
\item 2FGL~J1745.6$-$2858: G359.98$-$0.05 (?)
\end{itemize}
We mark the last association as questionable because this source is located in the immediate
vicinity of the Galactic center where we know that the accuracy of our model of the diffuse Galactic emission is intrinsically limited, and because the large density of potential counterparts makes a
reliable source association difficult.

In the 1FGL catalog we reported 6 sources associated with PWNs that were not also associated with known
pulsars.
Among those, 
two are among the 3 objects mentioned above (G338.3$-$0.0 and G359.98$-$0.05),
one has turned out in fact to be a pulsar (2FGL~J1135.3$-$6054),
two are still unassociated 2FGL sources, but no longer associated with PWNs
(2FGL~J1552.8$-$5609 and 2FGL~J1635.4$-$4717c),
and one no longer has a corresponding source in 2FGL (G0.13$-$0.11, see \S~\ref{1fgl_comparison}).

\subsubsection{Globular clusters}

Eleven 2FGL sources are associated with globular clusters.
Among those, 9 have been published previously:
47~Tuc \citep{LAT09_47Tuc},
NGC~6266, NGC~6388, Terzan~5, NGC~6440, NGC~6626, NGC~6652 \citep{LAT10_1FGL},
Omega Cen \citep{LAT10_globular}, and
M~80 \citep{Tam2011_globular}.
In addition, we find two new associations:
\begin{itemize}
\item 2FGL~J1727.1$-$0704: IC~1257. With an average significance of
$4.1$ this source is near the detection threshold. It is fitted using a power law with a
spectral index of $2.2\pm0.1$, yet a 3.5 $\sigma$ curvature significance may indicate that the spectrum
is in fact curved.
\item 2FGL~J1808.6$-$1950c: 2MS-GC01. This source has already been
detected as 1FGL~J1808.5$-$1954c, but the globular cluster catalog used for the 
association of 1FGL sources did not contain 2MS-GC01, and consequently the source remained
unassociated. 2FGL~J1808.6$-$1950c has an apparently-curved spectrum (3.9 $\sigma$ significance) that is
comparable to that of other globular clusters.
\end{itemize}

\citet{Tam2011_globular} have furthermore reported the detections of
Liller~1,
NGC~6139,
NGC~6624, and
NGC~6752 using LAT data.
None of these clusters are formally associated with any of the 2FGL sources in the catalog.
NGC~6624 is near 2FGL~J1823.4$-$3014 ($\theta_{95}=7\farcm7$, $\Delta=7\farcm6$), but the formal
posterior association probability of $50\%$ is below our adopted threshold.
A source associated with NGC~6752 was in our initial list of seeds
for the catalog; however, it did not pass the detection threshold of TS~$>25$ for the 2FGL catalog.
We could not find evidence for any sources in our data that might be associated with
Liller~1 or NGC~6139.

\subsubsection{Supernova remnants}

SNRs are a special class in our association scheme because a substantial number of the known
objects are sufficiently extended to be potentially resolved with the LAT.
We thus use two separate strategies to search for SNR associations among the 2FGL
sources.
For SNRs with angular diameters $<20'$, i.e., SNRs that still should appear
point-like to the LAT, we use the Bayesian scheme to search for associations.
In total we find six 2FGL sources associated with point-like SNR, of which 2 are also associated
with firmly identified pulsars.
The remaining associations are:
\begin{itemize}
\item 2FGL~J1214.0$-$6237: G298.6$-$00.0 
\item 2FGL~J1911.0+0905: G043.3$-$00.2 (aka W49B) 
\item 2FGL~J2022.8+3843c: G076.9+01.0 
\item 2FGL~J2323.4+5849: G111.7$-$02.1 (aka Cas A) 
\end{itemize}
None of them has a concurrent association with a PWN.
Except for 2FGL~J2022.8+3843c, all of them were already present and associated in 1FGL.

In a second pass we search for all 2FGL sources for which the $95\%$ confidence error
radius overlaps with the (assumed) circular extension of the SNR.
This provides a list of 89 2FGL sources among which we estimate $\sim45\%$ chance
coincidences.
Six of the 2FGL sources correspond to SNRs that were firmly identified as $\gamma$-ray 
sources based on their spatial extensions (IC~443, W28, W30, W44, W51C, and the Cygnus Loop),
and 4 are the point-like SNRs listed above.
Twenty of the 2FGL sources are firmly identified as being either a pulsar, a PWN, or a high-mass
binary system.
This leaves 59 2FGL sources that might be associated with an
extended SNR, among which we expect $\sim26$ chance coincidences.
Due to this high chance coincidence rate, we do not claim any SNR association for this list of sources,
but we give the 2FGL names and associations in Table~\ref{tab:snrext} for reference.

Several of the SNRs have extensions that encompass multiple 2FGL sources
(G132.7+01.3, Monoceros Loop, Pup A, Vela Junior, and G089.0+04.7), in which case the 2FGL 
sources might actually correspond to local maxima of extended emission regions.
A number of the SNRs have been detected at TeV energies, which makes their possible detection
also in the LAT energy range more plausible.
Three 2FGL sources have concurrent PWN associations, which makes them also good
pulsar or PWN candidates.
We also note that one source, 2FGL~J2015.6+3709, is likely to be variable, hence a physical 
association to CTB~87 is highly improbable.

In 1FGL, 41 $\gamma$-ray sources were listed in the corresponding table of overlaps with SNRs
\citep[see Table 7 of][]{LAT10_1FGL}.
About half of SNRs that were found overlapping with 1FGL sources are still in
Table \ref{tab:snrext}, while the other half has not been found to overlap spatially
with any of the 2FGL sources.
This illustrates the relatively large uncertainty that is tied to these associations, and
should present an additional warning to treat these potential associations with great care.

\begin{deluxetable}{lcccc}
\setlength{\tabcolsep}{0.04in}
\tablewidth{0pt}
\tabletypesize{\scriptsize}
\tablecaption{Potential Associations for Sources Near SNRs
\label{tab:snrext}
}
\tablehead{
\colhead{2FGL name} &
\colhead{SNR name} &
\colhead{PWN name} &
\colhead{TeV name} &
\colhead{Common name}
}
\startdata
J0128.0+6330 & G127.1+00.5 &  &  &  \\ 
J0214.5+6251c & G132.7+01.3 &  &  &  \\ 
J0218.7+6208c & G132.7+01.3 &  &  &  \\ 
J0221.4+6257c & G132.7+01.3 &  &  &  \\ 
J0503.2+4643 & G160.9+02.6 &  &  &  \\ 
J0526.6+4308 & G166.0+04.3 &  &  &  \\ 
J0538.1+2718 & G180.0$-$01.7 &  &  &  \\ 
J0553.9+3104 & G179.0+02.6 &  &  &  \\ 
J0631.6+0640 & G205.5+00.5 &  &  & Monoceros Loop \\ 
J0636.0+0554 & G205.5+00.5 &  &  & Monoceros Loop \\ 
J0637.8+0737 & G205.5+00.5 &  &  & Monoceros Loop \\ 
J0821.0$-$4254 & G260.4$-$03.4 &  &  & Pup A \\ 
J0823.0$-$4246 & G260.4$-$03.4 &  &  & Pup A \\ 
J0823.4$-$4305 & G260.4$-$03.4 &  &  & Pup A \\ 
J0842.9$-$4721 & G263.9$-$03.3 &  &  & Vela \\ 
J0848.5$-$4535 & G266.2$-$01.2 &  & RX J0852.0$-$4622 & Vela Junior \\ 
J0851.7$-$4635 & G266.2$-$01.2 &  & RX J0852.0$-$4622 & Vela Junior \\ 
J0853.5$-$4711 & G266.2$-$01.2 &  & RX J0852.0$-$4622 & Vela Junior \\ 
J0855.4$-$4625 & G266.2$-$01.2 &  & RX J0852.0$-$4622 & Vela Junior \\ 
J1112.1$-$6040 & G291.0$-$00.1 & G291.0$-$0.1 &  &  \\ 
J1411.9$-$5744 & G315.1+02.7 &  &  &  \\ 
J1441.6$-$5956 & G316.3$-$00.0 &  &  &  \\ 
J1521.8$-$5735 & G321.9$-$00.3 &  &  &  \\ 
J1552.8$-$5609 & G326.3$-$01.8 &  &  & Kes 25 \\ 
J1615.0$-$5051 & G332.4+00.1 &  & HESS J1616$-$508 & Kes 32 \\ 
J1628.1$-$4857c & G335.2+00.1 &  &  &  \\ 
J1631.7$-$4720c & G336.7+00.5 &  &  &  \\ 
J1635.4$-$4717c & G337.2+00.1 &  & HESS J1634$-$472 &  \\ 
J1640.5$-$4633 & G338.3$-$00.0 & G338.3$-$0.0 & HESS J1640$-$465 &  \\ 
J1712.4$-$3941 & G347.3$-$00.5 &  & RX J1713.7$-$3946 &  \\ 
J1714.5$-$3829 & G348.5+00.1 &  & CTB 37A & CTB 37A \\ 
J1718.1$-$3725 & G350.1$-$00.3 &  &  &  \\ 
J1727.3$-$4611 & G343.0$-$06.0 &  &  & RCW 114 \\ 
J1731.6$-$3234c & G355.4+00.7 &  &  &  \\ 
J1737.2$-$3213 & G356.3$-$00.3 &  &  &  \\ 
J1738.9$-$2908 & G359.1+00.9 &  &  &  \\ 
J1740.4$-$3054c & G357.7$-$00.1 &  &  & Tornado Nebula \\ 
J1745.5$-$3028c & G358.5$-$00.9 &  & HESS J1745$-$303 &  \\ 
J1745.6$-$2858 & G000.0+00.0 & G359.98$-$0.05 &  & Sgr A East \\ 
J1802.3$-$2445c & G005.4$-$01.2 &  &  & Bird \\ 
J1811.1$-$1905c & G011.4$-$00.1 &  &  &  \\ 
J1828.3$-$1124c & G020.0$-$00.2 &  &  &  \\ 
J1834.3$-$0848 & G023.3$-$00.3 &  & HESS J1834$-$087 & W 41 \\ 
J1834.7$-$0705c & G024.7+00.6 &  &  &  \\ 
J1839.7$-$0334c & G028.8+01.5 &  &  &  \\ 
J1840.3$-$0413c & G027.8+00.6 &  &  &  \\ 
J1841.2$-$0459c & G027.4+00.0 &  &  & Kes 73 \\ 
J1849.3$-$0055 & G031.9+00.0 &  &  & Kes 77, 3C 391 \\ 
J1850.7$-$0014c & G032.4+00.1 &  &  &  \\ 
J1852.7+0047c & G033.6+00.1 &  &  & Kes 79 \\ 
J1916.1+1106 & G045.7$-$00.4 &  &  &  \\ 
J1932.1+1913 & G054.4$-$00.3 &  &  &  \\ 
J2015.6+3709\tablenotemark{a} & G074.9+01.2 &  &  & CTB 87 \\ 
J2019.1+4040 & G078.2+02.1 &  & VER J2019+407 & Gamma Cygni \\ 
J2041.5+5003 & G089.0+04.7 &  &  &  \\ 
J2043.3+5105 & G089.0+04.7 &  &  &  \\ 
J2046.0+4954 & G089.0+04.7 &  &  &  \\ 
J2333.3+6237 & G114.3+00.3 &  &  &  \\ 
J2358.9+6325 & G116.5+01.1 &  &  &  \\ \enddata
\tablenotetext{a}{Source is likely to be variable.}
\end{deluxetable}

\subsubsection{Binaries}

The 2FGL catalog includes four high-mass binary systems, all of which have been firmly identified
by their orbital modulation, and are described in 
separate publications:
\begin{itemize}
\item 2FGL~J0240.5+6113: LSI~+61 303 \citep{LAT09_LSI}, 
\item 2FGL~J1019.0$-$5856: 1FGL J1018.6$-$5856 \citep{ATel.3221},
\item 2FGL~J1826.3$-$1450: LS~5039 \citep{LAT09_LS5039}, and
\item 2FGL~J2032.1+4049: Cygnus X-3 \citep{LAT09_CygX3}.
\end{itemize}
No further 2FGL source is associated with a high-mass X-ray binary from Liu's catalog
\citep{HMXBcatalog}.
All four sources were already present in the 1FGL catalog, yet the orbital modulation of
1FGL~J1018.6$-$5856 was only recently discovered in a blind search using
the LAT data \citep{ATel.3221}.

Formally, the automatic association procedure associates three 2FGL sources with low-mass
X-ray binaries, but all three are located in globular clusters, and the observed emission
can be readily explained by the combined emission of MSPs \citep{LAT09_47Tuc}.
We thus conclude that no low-mass X-ray binary systems have been identified in the LAT 
data after 2 years of observations.
We came to the same conclusion for 11 months of data in our study of the 1FGL associations \citep{LAT10_1FGL}.

\subsubsection{Massive stars and open star clusters}

Among the massive star catalogs (O stars, Wolf-Rayet stars, Luminous Blue Variables) and the
open cluster catalog we find only 2 possible associations with 2FGL sources:
\begin{itemize}
\item 2FGL~J1045.0$-$5941: $\eta$ Carinae (LBV). The $\gamma$-ray emission of this well-known
peculiar binary system has been studied in detail by 
\citet{Tavani2009_EtaCarinae},
\citet{LAT10_EtaCarina}, and
\citet{Farnier2011_EtaCarinae}, yet a firm identification of the
system through periodic orbital variability in $\gamma$ rays is still missing.
\item 2FGL~J2030.7+4417: HD~195592 (O star). This O9.5Ia type star is probably a short period
(5.063 days) O+B binary system at a distance of 1.1 kpc that may have escaped from the
open cluster NGC~6913 \citep{DeBecker2010_HD195592}. We note, however, that the object is 
located in the Cygnus region where the high O star density easily could lead to false associations 
and the complex diffuse emission may render precise source localization difficult.
In addition, the spectral shape and the apparent lack of variability of 2FGL~J2030.7+4417 are 
similar to the characteristics of identified $\gamma$-ray pulsars.
Hence we caution against overinterpreting this particular O star association and we do not list
it in our final table.
\end{itemize}

\subsubsection{Multiwavelength associations}

In addition to the catalogs of classified sources, we also search for associations with catalogs of
radio and TeV sources.
Our association procedure for AGN heavily relies on associations with radio sources as most
of the $\gamma$-ray emitting AGN are bright sources of radio emission 
(see \S~\ref{assoc:agn}).
In fact, essentially all of the radio associations we find have been classified subsequently as AGN.

Eighteen 2FGL sources that have not been associated with any object in one of our
catalogs of known or plausible $\gamma$-ray-emitting source classes 
(our type 1 catalogs in \S~\ref{source_assoc_automated})
have associations with extended TeV sources.
However, due to the relatively large extents of the sources in the extended TeV catalog, we 
expect on average 20 false associations (cf.~Table \ref{tab:catalogs}), so from 
a statistical point of view, all 18 associations could be spurious.
We discuss 2FGL associations with TeV sources more deeply in \S \ref{source_assoc_tev}.

\subsubsection{Other GeV Detections}
\label{other_gev_detections}

The automated association process compares the 2FGL source locations with other catalogs 
of sources seen at GeV energies.  
Results are shown in the main table for individual sources.
From the Bright Gamma-Ray Source List \citep{LAT09_BSL} we find 185 out of 205 sources 
associated with 2FGL sources.
Comparison with the 1FGL catalog was described in detail in \S~\ref{1fgl_comparison}.
In total we find 1099 out of 1451 1FGL sources that are associated with 2FGL sources.

The only contemporaneous catalog from a different instrument is the AGILE  (1AGL) catalog 
\citep{AGILEcatalog}, which has 42 (out of 47) sources in common with the 2FGL catalog.
The five 1AGL sources that are not formally associated (1AGL~J0657+4554, 1AGL~J0714+3340,
1AGL~J1022$-$5822, 1AGL~J1803$-$2258 and 1AGL~J1823$-$1454) all lie close to 2FGL 
sources  and spatially overlap within their mutual $99\%$ confidence localization uncertainties.
Several 2FGL sources are associated with the same 1AGL source, and in total we find
57 2FGL sources associated with sources listed in 1AGL.

From the previous generation high-energy $\gamma$-ray telescope, EGRET on the 
Compton Gamma Ray Observatory, the 3EG catalog \citep{3EGcatalog} had 111 sources (out of 271)
associated by the automatic process with 2FGL sources, while the EGR catalog \citep{EGRcatalog} 
had 66 (out of 188) sources associated with 2FGL sources. 
Also here we find several 2FGL sources that are associated with the same EGRET source.
In total, 116 2FGL sources are associated with sources in 3EG, while 69 2FGL sources are associated
with sources in EGR.  
The fractions of 3EG and EGR sources with 1FGL sources were similarly low and the discussion
in the 1FGL catalog paper is still relevant \citep{LAT10_1FGL}; we also refer the reader to a study 
of unassociated 1FGL sources \citep{LAT11_1FGLUnassoc}.  An EGRET catalog based on analysis of energies above 1 GeV \citep{1997LM} found 46 high-confidence sources, of which 40 have clear 2FGL counterparts, 5 have close 2FGL sources just outside the 95\% confidence contours, and only one (GEV 2026+4124 in the confused Cygnus region) lacks a plausible 2FGL match.

Through 2011 June, 94 flaring {\it Fermi}-LAT sources were detected and promptly reported 
in more than 150 Astronomer's Telegrams. Of these, 8 are not in 2FGL. 
For 6 of these the flaring state was detected outside the time 
interval covered by 2FGL: 
SBS 0846+513 \citep[a new NLSy1 system:][]{ATel3452}, 
SHBL J001355.9$-$185406 \citep[see \S~5.3;][]{ATel3014}, 
PSR B1259$-$63 \citep[see \S~5.3;][]{ATel3085}, 
PMN~J1123$-$6417 \citep[see \S~4.2;][]{ATel3394}, 
PMN~J1913$-$3630 \citep[][]{ATel2966}, 
and the flaring source in the Galactic center region \citep[][]{ATel3162}. 
The other two sources are: 
J1057$-$6027, \citep[][]{ATEL2081} detected in 2009 June, is not included in 1FGL and 
does not have a 2FGL counterpart but could be associated with 2FGL J1056.2$-$6021 
using the 99.9\% confidence error radius; 
and PKS 1915$-$458 \citep[][]{ATel2666} a faint and high redshift blazar ($z=2.47$), detected 
in 2010 June, whose average flux between 2008 August and 2010 August is below the 2FGL catalog significance threshold.
Also, we note that two 1FGL unidentified flaring sources detected along 
the Galactic plane, 
3EG J0903$-$3531 \citep[][]{ATEL1771} and 
J0910$-$5041 \citep[2FGL J0910.4$-$5050 or 1FGL J0910.4$-$5055;][]{ATEL1788} are 
now associated with two unclassified AGN in 2FGL, PMN J0904$-$3514 and 
AT20G J0910$-$5048 respectively.
Furthermore, the 2FGL counterpart for J1512$-$3221 \citep[][]{ATel2528}, which had no 
clear association, is 2FGL J1513.6$-$3233 which is associated with blazar CRATES J1513$-$3234.

    \subsection{TeV Source Associations}
\label{source_assoc_tev}

2FGL sources that are positionally associated with sources seen by the
ground-based TeV telescopes are of particular interest because the TeV
band overlaps with the LAT energy range, suggesting the potential for
common emission mechanisms if the spectra match.  As described in
Table~\ref{tab:catalogs}, we investigated associations with the
sources in the TeVCat compilation of detections.  The compilation is 
growing with time, and information
about the sources is subject to updates and refinements, but at any
given time TeVCat represents a snapshot of current knowledge of the
TeV sky.

The association analysis was done separately for extended and
point-like TeV sources, taking into account the statistical and
systematic uncertainties in the source localization.  The `TeV' column
of Table~\ref{tab:sources} lists associations with extended sources as
`E' and point-like sources as `P'.  As the table indicates, 85 2FGL
sources are positionally consistent with TeVCat sources, although
multiple 2FGL associations are seen for some TeV sources.  In the FITS
version of the catalog, we also provide the names of the associated
TeV sources.

Of the TeV sources considered for the associations performed here,
most correspond to known objects at other wavelengths, in particular those
that lie far from the Galactic plane. A large fraction ($\sim 50\%$)
of the TeV Galactic sources, however, are still unidentified. Many of
these have plausible counterparts while others remain unassociated
despite deep searches for counterparts at other wavelengths. Among the
firm identifications in the TeV regime, there are seven different
source classes, and members of each of these source classes have been
associated with 2FGL sources. In total, 85 TeV sources have 2FGL
counterparts. Eight of these TeV sources have more than one 2FGL
association. RX\,J0852.0$-$4622  has four 2FGL associations, and the following TeV sources have 
two each:
Westerlund\,1 \citep{2010tsra.confE.257O}, Westerlund\,2, HESS\,J1632$-$478,
RX\,J1713.7$-$3946, W28 \citep{LAT10_W28}, HESS\,J1841$-$055, and MGRO\,J2019$+$37.
One  LAT source, 2FGL\,J2229.0$+$6114 is associated with two
TeV sources, Boomerang and G106.3$+$2.7. The LAT emission
from two of the TeV sources, IC\,443 and MSH\,15$-$52, is measured to be extended.

The TeV class that has the most numerous associations with the 2FGL
sources is the AGN class \citep[see][for a more detailed discussion of
the LAT AGN]{LAT11_2LAC}. There are currently 45 AGN detected at TeV
energies and all but six of these are associated with 2FGL
sources. The six that do not have 2FGL counterparts (SHBL
J001355.9$-$185406, 1ES\,0229$+$200, 1ES\,0347$-$121, PKS\,0548$-$322,
1ES\,1312$-$423, and HESS\,J1943$+$213\footnote{This source has not
  been confirmed to be a HBL but all available observations favor its
  classification as a HBL \citep{HESS_J1943p213}}) are all
high-frequency peaked BL Lacs. This is the subclass of AGN that tend to
have the lowest bolometric luminosities and their second emission peaks
at the highest energies. The six TeV AGN that did not reach the
detection threshold to be included in the 2FGL catalog are among the weakest
extragalactic TeV sources detected to date, ranging in flux from
0.4\%\,$-$\,2\% of the flux of Crab Nebula at those energies.

Both of the starburst galaxies detected at TeV energies, M\,82 and
NGC\,253, have 2FGL counterparts \citep{LAT10_starbursts}.

Four high-mass binaries (HMBs) have confirmed detections in the TeV
regime. Two of these, LS\,I\,$+$61\,303 and LS\,5039, have 2FGL
counterparts and are already the subject of LAT publications
(\citealp{LAT09_LSI, LAT09_LS5039}). We note that, although not
in 2FGL, the TeV binary PSR\,B1259$-$63 has been detected by the
LAT \citep{Tam_PSRB1259,LAT11_PSRB1259}. This system is a radio
pulsar in orbit around a Be star with an orbital period of $\sim 3.4$
years. During the time span of the 2FGL data, the system was far from periastron and
no significant GeV emission was detected, but when the system
approached periastron, variable emission, including flaring behavior,
was observed by the LAT.

The PWNe comprise the second most numerous identified TeV class that
is associated with 2FGL sources; of the 25 PWNe in TeVCat, 16 are
associated with 2FGL sources. Indeed, the association between GeV
$\gamma$-ray PSRs and the PWNe visible in the regime has been
well established already (\citealp{LAT10_PSRcat, LAT11_PSRcat}).

During the second year of LAT data taking, many more supernova remnants (SNRs) known at
TeV energies were detected at GeV energies such as Cas A
\citep{LAT10_CasA}, RX\,J1713.7$-$3946 \citep{LAT11_RXJ1713}, and Vela
Jr \citep{LAT11_RXJ0852}. Of the five SNR/Molecular Cloud associations
in TeVCat, all but one (G318.2$+$0.1) have been associated with 2FGL
sources. Ten shell-type SNR have been detected at TeV energies and
five of these now have 2FGL counterparts so, the GeV-TeV association
is established although there are still many open
questions. {\it{Fermi}}'s non-detection of RCW\,86 is surprising since
it is one of the brightest TeV SNR, with a flux of $\sim 10\%$ that of
the Crab Nebula \citep{2009ApJ...692.1500A}.

Sources of particular interest are those that are positionally
consistent between the LAT and TeV telescopes but have no obvious
associations with objects at longer wavelengths. Among the TeV sources
that have no clear identifications, 17 are associated with 2FGL
sources. In addition to these, although not formally associated with
LAT sources using the automatic pipeline
(\S~\ref{source_assoc_automated}), some other TeV sources
have possible 2FGL counterparts, for example, HESS\,J1843$-$033,
which has two potential 2FGL counterparts. Establishing a physical
connection through spectral or variability studies may help determine
the nature of these sources. In many cases, a GeV counterpart could
prove crucial for our understanding of the nature of the TeV source,
in particular for the following objects:

\begin{itemize}

\item 2FGL\,J1022.7$-$5741 and 2FGL\,J1023.5$-$5749 are spatially
  consistent with HESS\,J1023$-$\linebreak575, itself not yet firmly identified,
  but noted for its possible connection to the young stellar cluster
  Westerlund 2 in the star-forming region RCW\,49, as discussed by
  \citet{2011A&A...525A..46H}.

\item 2FGL\,J1427.6$-$6048 is associated with HESS\,J1427$-$608 which
  is, so far, without plausible counterparts \citep{2008A&A...477..353A}.

\item 2FGL\,J1503.9$-$5800 is spatially coincident with the TeV
  source HESS\,J1503$-$582, which is tentatively associated with a
  forbidden velocity region of interstellar gas \citep{2008AIPC.1085..281R}.

\item 2FGL\,J1507.0$-$6223 is spatially consistent with
  HESS\,J1507$-$622 \citep{2011A&A...525A..45H}, so far the
  only TeV unidentified source that is markedly offset from the Galactic plane
  ($\sim 3\fdg5$).

\item 2FGL\,J1615.2$-$5138 is spatially consistent with one of the
  brightest ($\sim 25\%$ of the Crab Nebula flux) TeV unidentified
  sources, HESS\,J1614$-$518 \citep{2006ApJ...636..777A}.

\item 2FGL\,J1650.6$-$4603 is spatially associated with a TeV source
  tentatively associated with the Westerlund 1 star-forming region.  (The other 2FGL source that is spatially associated with this TeV source is 2FGL J1648.4$-$4612, which is pulsar PSR J1648$-$4611.)

\item 2FGL\,J1848.2$-$0139: this source is consistent with the TeV
  source, HESS\,J1848$-$018, which is suspected to be correlated with
  the star-forming region W\,43 \citep{2008AIPC.1085..372C}.

\end{itemize}

As discussed in \S~\ref{DiffuseModel}, the Galactic Center region is particularly
complex and its study is beyond the purpose of this paper; we do,
however, find possible associations with all of the TeV $\gamma$-ray
sources detected in this region, although not all were formally
associated by the automatic pipeline analysis: the Galactic Center
source \citep{HESS09_GC}, HESS\,J1745$-$303 (2FGL\,J1745.5$-$3028c;
\citealp{HESS08_1745}), HESS\,J1741$-$302 \citep{HESS08_1741} and
HESS\,J1747$-$248 (2FGL\,J1748.0$-$2447;
\citealp{2011arXiv1106.4069H}).

    \subsection{Properties of Unassociated Sources}
\label{source_assoc_unassoc}

\begin{figure}
\epsscale{1.0}
\plotone{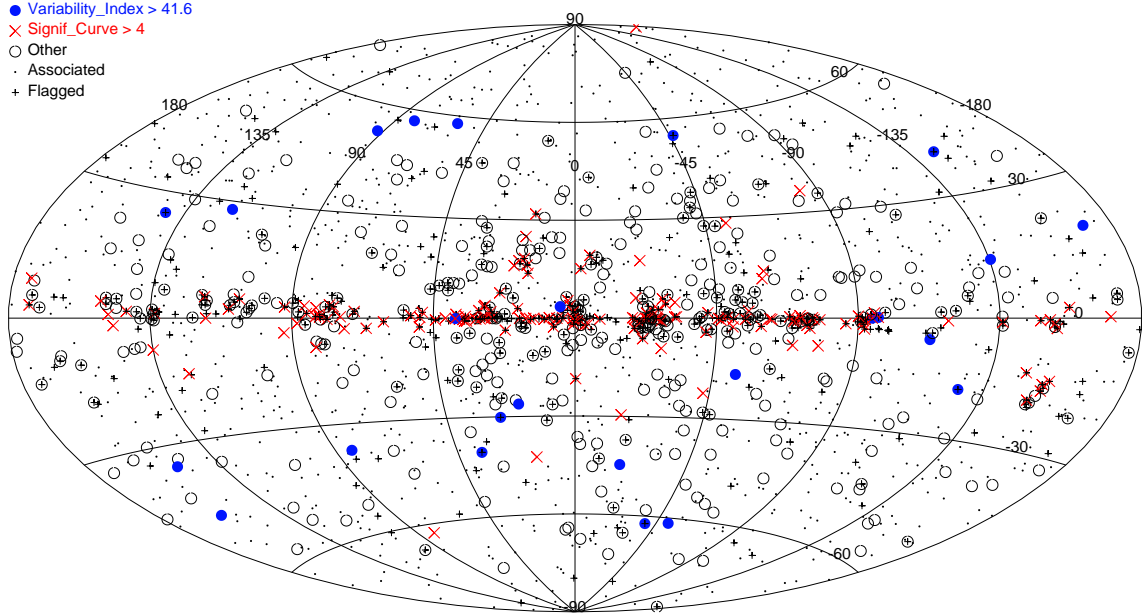}
\caption{Sky distribution of associated (dots) and unassociated sources (large symbols).
Sources that were flagged are marked by a plus.  
In particular, we mark variable unassociated sources ($TS_{var}>41.6$) using filled blue circles,
unassociated sources with a curved spectrum (\texttt{Signif\_Curve} $>4$) by red crosses,
and all other unassociated sources by open black circles. \label{fig:unassocsky}}
\end{figure}

Among the 1873 sources in the 2FGL catalog, 575 ($31\%$) remain unassociated.
Their distribution on the sky is compared in Figure~\ref{fig:unassocsky}
to the distribution of the associated sources.
We note a number of interesting features in the map that should be kept in mind
when considering unassociated 2FGL sources.

\begin{figure}
\epsscale{1.0}
\plotone{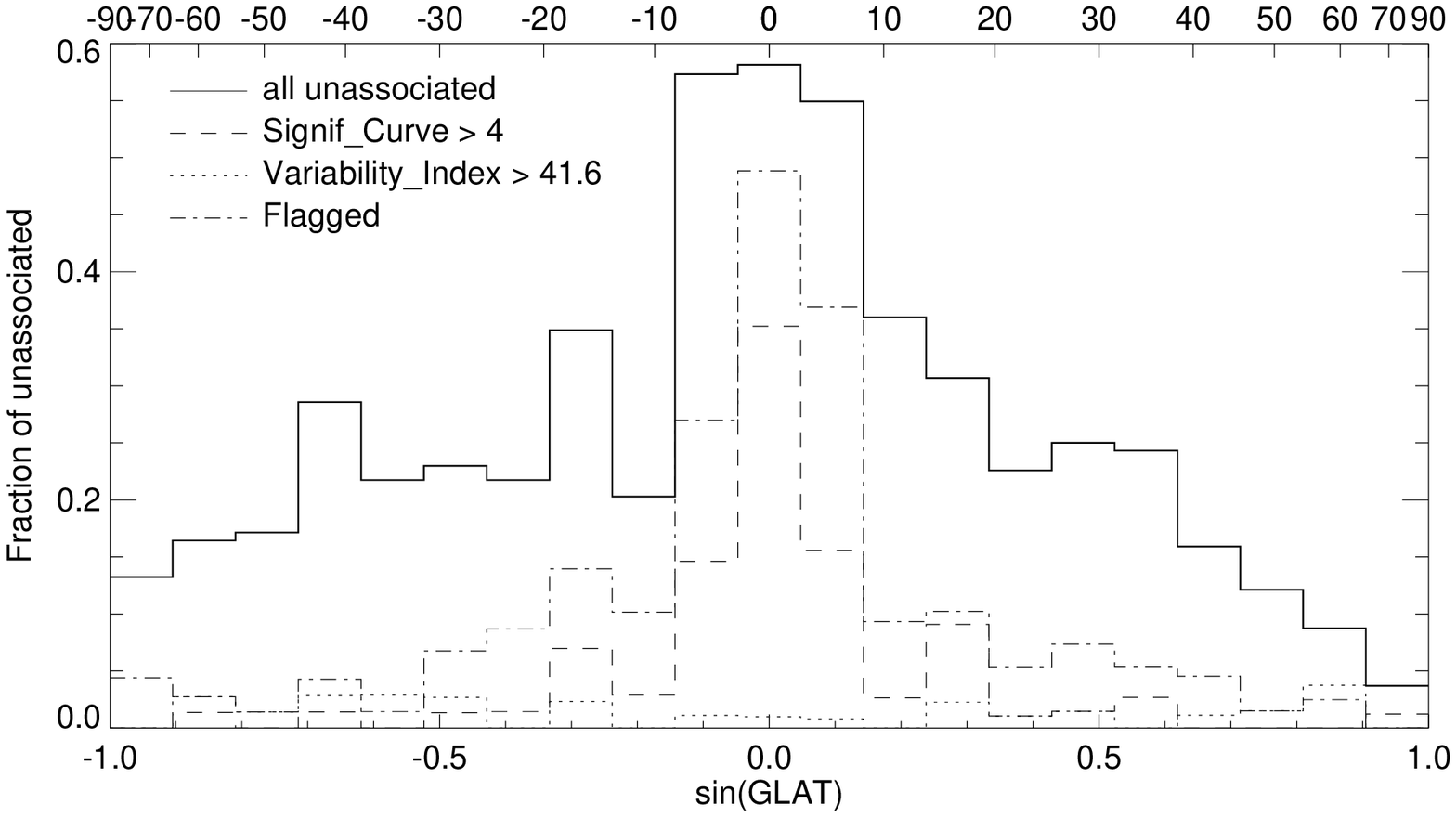}
\caption{Latitude distribution of unassociated sources. \label{fig:unassoclat}}
\end{figure}

First, the number of unassociated sources decreases with increasing Galactic latitude.
This is best illustrated by a latitude histogram of the fraction of unassociated 2FGL sources,
shown in Figure~\ref{fig:unassoclat}.
We plot here the data as function of the sine of Galactic latitude as in this representation
an isotropic distribution will appear as a flat profile.
In contrast to that, we find that the fraction of unassociated sources decreases with latitude, 
with the decrease being steeper at positive latitudes. 
This asymmetry is also present in the absolute numbers:
above Galactic latitudes $b>60\degr$ only 3 sources in 2FGL are unassociated, while
below $b<-60\degr$ we find 12 unassociated sources.  
This may be due to relative completeness in the north vs. the source of the counterpart catalogs 
used for the source association analysis (\S~\ref{source_assoc_automated}).

Second, the numbers of unassociated sources increase sharply below $|b| \approx 10\degr$.
This is attributable to the relative lack of sources below $|b|<10\degr$ in many of the
extragalactic source catalogs that we use for source association.
The Milky Way is a bright source of radio emission, limiting sensitive searches for
extragalactic sources near the Galactic plane.
Furthermore, optical identifications of radio sources are hampered by the important interstellar 
obscuration, leaving many radio sources unclassified.

Third, the numbers of 2FGL sources with curved spectra increase at low Galactic latitudes, as can be seen
in the latitude histogram (dashed line in Fig.~\ref{fig:unassoclat}) and the sky map
(red crosses in Fig.~\ref{fig:unassocsky}).
The sky map indicates that these sources tend to cluster in regions of bright Galactic
diffuse emission, such as
the inner Galactic ridge (Galactic longitudes $330\degr < l < 30\degr$), 
the Cygnus region ($l \approx 80\degr$), 
the Norma spiral arm tangent ($l \approx 330\degr$) or 
the Crux spiral arm tangent ($l \approx 300\degr$).
Whether this clustering is diagnostic of the physical natures of the sources, or whether it
indicates systematic uncertainties in the Galactic diffuse emission model that resulted in
spurious source detections remains a possibility.
We note, however, that the fraction of sources with curved spectra among the unassociated
sources is greater ($28\%$) than the fraction of curved spectra sources among the
associated sources ($16\%$).
Because the spectrum of the Galactic diffuse emission at low latitudes is itself well represented 
with a curved spectrum,  at least some fraction of the unassociated 2FGL sources at low latitudes may be local
emission maxima of diffuse Galactic emission that are not adequately modeled by our Galactic
diffuse model; see the discussion in \S~\ref{catalog_ism} and the definitions of the several analysis flags that are related to the model of the Galactic diffuse emission in \S~\ref{catalog_analysis_flags}.

Fourth, a substantial fraction of the unassociated sources have at least one analysis flag (\S \ref{catalog_analysis_flags}) set.
We find that $51\%$ of the unassociated sources have been flagged due to various issues,
while only $14\%$ of the associated sources have been flagged.
None of the flags is related to our association procedure itself, but they identify a number
of conditions that can shed doubt on the physical reality or localization quality of a source.
The fact that such a large fraction of unassociated sources are flagged may indicate that some
of these sources are indeed not real.
We emphasize that the analysis flags should be taken into consideration when using the 2FGL catalog.

Fifth, 25 unassociated sources ($4\%$) have been flagged as variable, and the spatial distribution of these
sources appears rather isotropic.
These sources are good candidates for being as-yet unassociated AGN, as this is the source class that
shows the largest flux variability in LAT data.

\section{Conclusions}
\label{conclusions}

The second {\it Fermi} LAT catalog is the product of a comprehensive analysis of the first 2 years of LAT science data.  In several ways it is an advance over the 1FGL catalog, which was based on the first 11 months of data.  The 2FGL analysis takes advantage of the new P7\_V6 Source event selection and IRFs, which in particular provide increased effective area in the range below $\sim$200~MeV.  The analysis also uses a refined model for the Galactic diffuse emission.  The source detection and localization analyses were advanced for the 2FGL analysis to iteratively optimize the definitions of the `seed' sources used for the final likelihood analysis step.  Both analysis steps allowed for non-power-law source spectra and also incorporated special models for spatially extended sources.  The source association analysis was also extensively updated for the 2FGL catalog, with updated catalogs of counterparts and local determinations of counterpart densities.  For AGN, the association analyses also included methods that took into account radio and X-ray properties of potential counterparts.  

The 2FGL catalog contains 1873 sources.  In developing the catalog analysis, we re-evaluated a number of the analysis flags used to tag sources with unusual or potentially problematic properties.  The most-prominent flag is the `c' designator, which we have appended to the names of 162 sources, that indicates potential confusion with interstellar diffuse emission or an artifact in the model for the diffuse emission.  A number of other flags are defined, and 315 sources have one or more of these other flags set.

The 2FGL catalog represents a new milestone in high-energy 
$\gamma$-ray astrophysics.  
As with any astronomical catalog, 2FGL enables a wide range of astrophysical research.  
For individual objects, the spectra and light curves offer opportunities for multiwavelength modeling 
that can lead to better physical understanding of sources.  
The catalog as a collection allows population studies for $\gamma$-ray-only sources and for comparative 
studies with other wavelengths.  In the catalog, 127 sources are considered to be identified, and  plausible associations are proposed for more than 1000 AGN.  In all identifications or associations of 2FGL sources with 15 classes of counterparts are proposed.   In addition the fact that 575 of the 2FGL sources have no plausible counterparts among known 
$\gamma$-ray-producing source classes presents discovery opportunities similar to those already 
found with the {\it Fermi} LAT Bright Source List and 1FGL catalog. 
Even the absence of 2FGL sources in predicted source classes such as clusters of galaxies 
will stimulate additional research into why these known sources of nonthermal radiation are 
not producing $\gamma$ rays at a level yet detectable with the LAT. 
We look forward to extensive use of this catalog in high-energy astrophysics.

\acknowledgments
We dedicate this paper to the memory of our colleague Patrick Nolan, who died
on 2011 November 6. His career spanned much of the history of high-energy
astronomy from space and his work on the Large Area Telescope began nearly 20
years ago when it was just a concept. Pat was a central member in the
operation of the LAT collaboration and he is greatly missed.

The {\it Fermi}-LAT Collaboration acknowledges generous ongoing support
from a number of agencies and institutes that have supported both the
development and the operation of the LAT as well as scientific data analysis.
These include the National Aeronautics and Space Administration and the
Department of Energy in the United States, the Commissariat \`a l'Energie Atomique
and the Centre National de la Recherche Scientifique / Institut National de Physique
Nucl\'eaire et de Physique des Particules in France, the Agenzia Spaziale Italiana
and the Istituto Nazionale di Fisica Nucleare in Italy, the Ministry of Education,
Culture, Sports, Science and Technology (MEXT), High Energy Accelerator Research
Organization (KEK) and Japan Aerospace Exploration Agency (JAXA) in Japan, and
the K.~A.~Wallenberg Foundation, the Swedish Research Council and the
Swedish National Space Board in Sweden.

Additional support for science analysis during the operations phase is gratefully
acknowledged from the Istituto Nazionale di Astrofisica in Italy and the Centre National d'\'Etudes Spatiales in France.

This work made extensive use of the ATNF pulsar  catalog\footnote{http://www.atnf.csiro.au/research/pulsar/psrcat}  \citep{ATNFcatalog}.  This research has made use of the NASA/IPAC Extragalactic Database (NED) which is operated by the Jet Propulsion Laboratory, California Institute of Technology, under contract with the National Aeronautics and Space Administration.

This research has made use of Aladin\footnote{http://aladin.u-strasbg.fr/}, TOPCAT\footnote{http://www.star.bristol.ac.uk/\~mbt/topcat/} and APLpy, an open-source plotting package for Python\footnote{http://aplpy.github.com}. The authors acknowledge the use of HEALPix\footnote{http://healpix.jpl.nasa.gov/} \citep{Gorski2005}.

{\it Facilities:} \facility{Fermi LAT}.


\appendix

\section{Localization Power}
\label{appendix_detection_localization}

\begin{figure}
\epsscale{.80}
\plotone{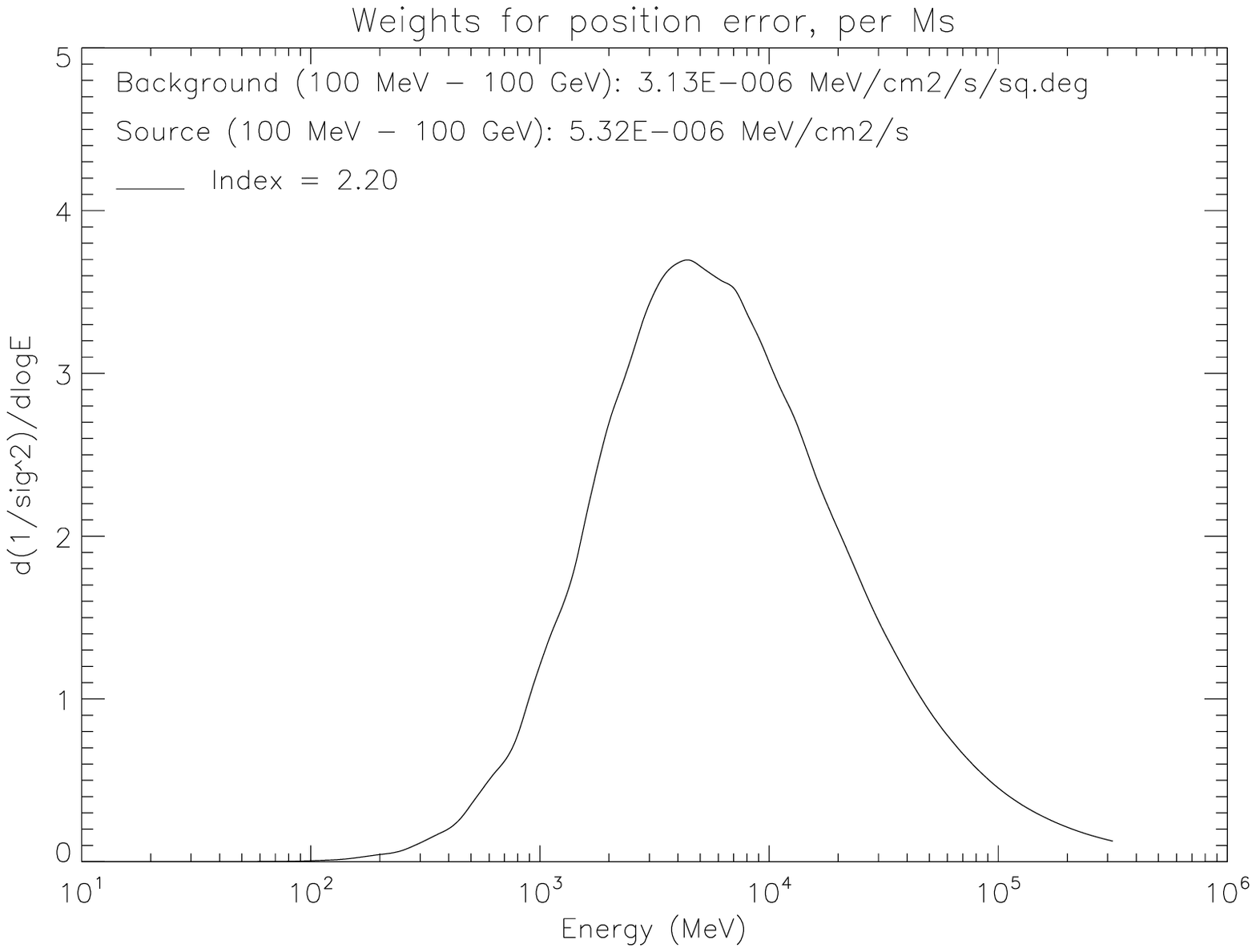}
\caption{Theoretical contribution ($W_l(E)$ of Eq.~\ref{eq:wloc})
to $\Delta \theta_0^{-2}$ per Ms and per log($E$)
interval as a function of energy for a $TS$ = 100 power-law source over the average background at $|b| > 10\degr$.
The assumed photon spectral index is 2.2.}
\label{fig:locweight}
\end{figure}

In the maximum likelihood formalism, the error ellipse (\S~\ref{catalog_localization}) is given by the covariance matrix of the position parameters after the fit.
One can obtain an approximate but reasonably accurate estimate of the
localization power of the $Fermi$-LAT for a point source,
assuming that the diffuse
background is locally uniform and considering only one source.
In that approximation the error ellipse is a circle and the 1~$\sigma$ localization precision of a source along any direction $\Delta \theta_0$ is given by
\begin{equation}
\Delta \theta_0^{-2} =  \left | \frac{\partial^2 \log \mathcal{L}}{\partial \theta_0^2} \right | = \frac{1}{2} \left | \frac{\partial^2 TS}{\partial \theta_0^2} \right |
\label{eq:covariance}
\end{equation}
and is related to the 95\% error radius by $r_{95}/\Delta \theta_0 = \sqrt{-2 \log(0.05)} = 2.45$.
Along the lines of Eq.~A1 of \citet{LAT10_1FGL}, 
denoting $S(E)$ the source spectrum, $B(E)$ the background
spectrum per unit solid angle, $T_0$ the equivalent
on-axis observing time, $A_{\rm eff}(E)$ the on-axis effective area
and the local source to background ratio
$g(\theta,E) = S(E) {\rm PSF}(\theta,E) / B(E)$ one may write
\begin{eqnarray}
\Delta \theta_0^{-2} = T_0 \int_{\log E_{\rm min}}^{\log E_{\rm max}} W_l(E) \, d\log E \\
W_l(E) = \pi E A_{\rm eff}(E) \frac{S(E)^2}{B(E)}
    \int_0^\pi \left ( \frac{\partial {\rm PSF}}{\partial \theta} \right )^2
    \frac{\sin\theta d\theta}{1 + g(\theta,E)}
\label{eq:wloc}
\end{eqnarray}
after integrating the $\cos^2\phi$ term arising from the projection along one direction.
Here $W_l(E)$ is the contribution to $\Delta \theta_0^{-2}$ per unit log(energy).
It is illustrated in Figure~\ref{fig:locweight} for a power-law source spectrum at high latitude.
Not surprisingly, the localization depends even more on high energy (where the core PSF is narrowest) than the detection itself \citep[Figure~18 of][]{LAT10_1FGL}.
For that reason, the average effect of confusion on localization is small, because it is important only at those energies when the average angular distance between sources ($2\fdg8$ at high latitude) is comparable to the PSF width.

\begin{figure}
\epsscale{.80}
\plotone{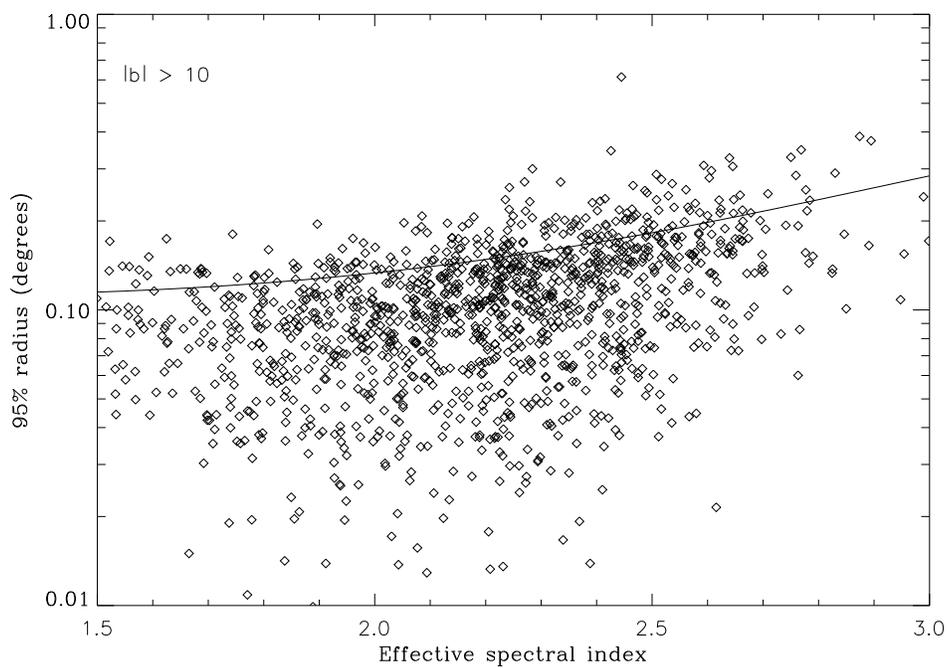}
\caption{95\% error radius of sources at $|b| > 10\degr$
as a function of spectral index.
The line shows the theoretical error radius 
for an isolated source at the detection threshold of $TS = 25$ over the average extragalactic background.}
\label{fig:locspec}
\end{figure}

For each spectral index it is possible to compute the detection threshold and then the localization precision at the detection threshold. This is normally the worst error radius one may expect in the catalog.
That prediction is compared on Figure~\ref{fig:locspec} with the actual 95\% error radius. The curve accurately predicts the dependence on spectral index. The localization is worse for softer sources, but only by a relatively small factor at a given $TS$.
A few sources are above the line. This can happen for purely statistical
reasons, because the background and exposure depend
a little on direction even after taking out the Galactic plane,
or because of another nearby source.
The highest point (worst error ellipse) is 2FGL J1952.6$-$3252 which is specifically flagged for imperfect localization (Flags 8 and 9 set).

\section{Quality of the Fit}
\label{appendix_fit_quality}

\begin{figure}
\epsscale{.80}
\plotone{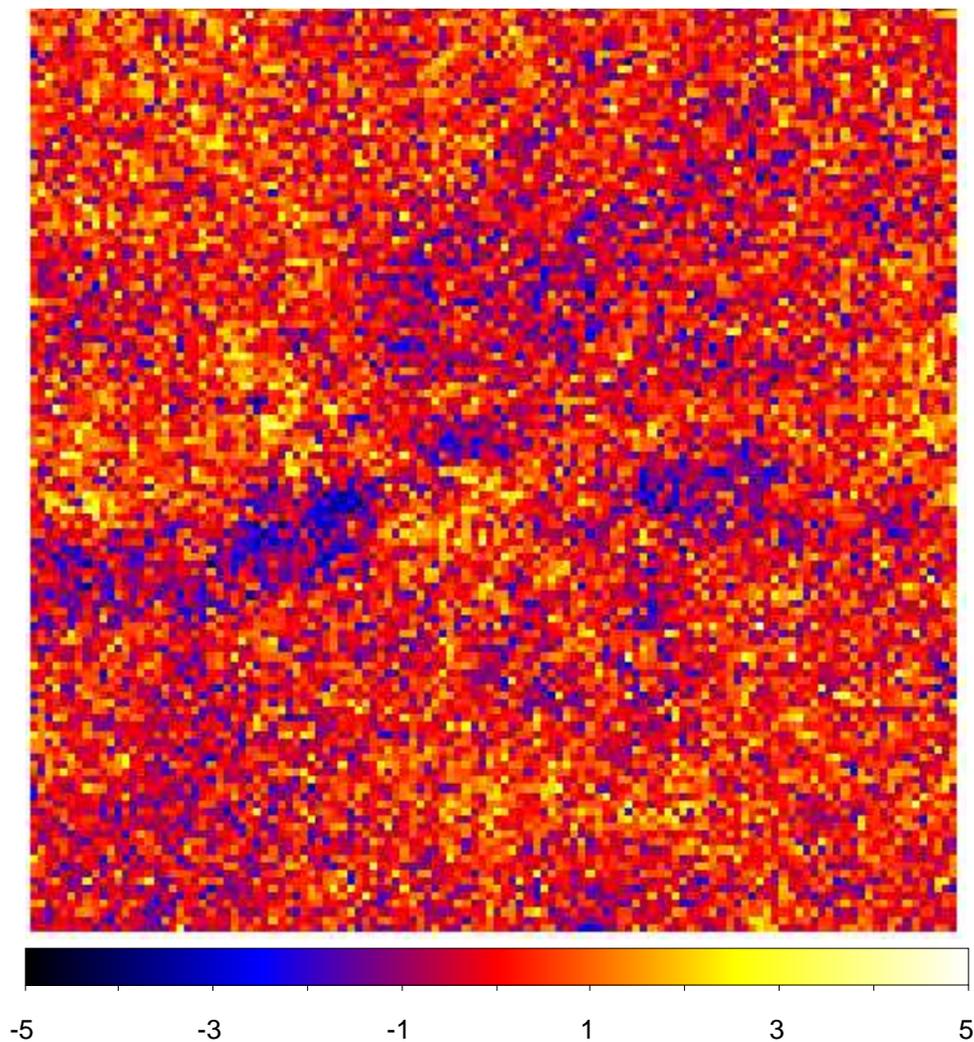}
\caption{Residuals (in $\sigma$ units) in $0\fdg5$ pixels over a $60 \times 60\degr$ area around the Galactic anticenter, summed over the full energy range (100~MeV to 100~GeV). All sources were fixed to the catalog values and the diffuse parameters were fitted as in an ordinary RoI (\S~\ref{catalog_significance}). The pixels used in the source fitting process were much smaller. The larger pixels used here allow reducing the statistical fluctuations to 5\% in the Galactic plane and 10\% at the top and bottom of the plot.}
\label{fig:ressigma_map}
\end{figure}

\begin{figure}
\epsscale{1.1}
\plottwo{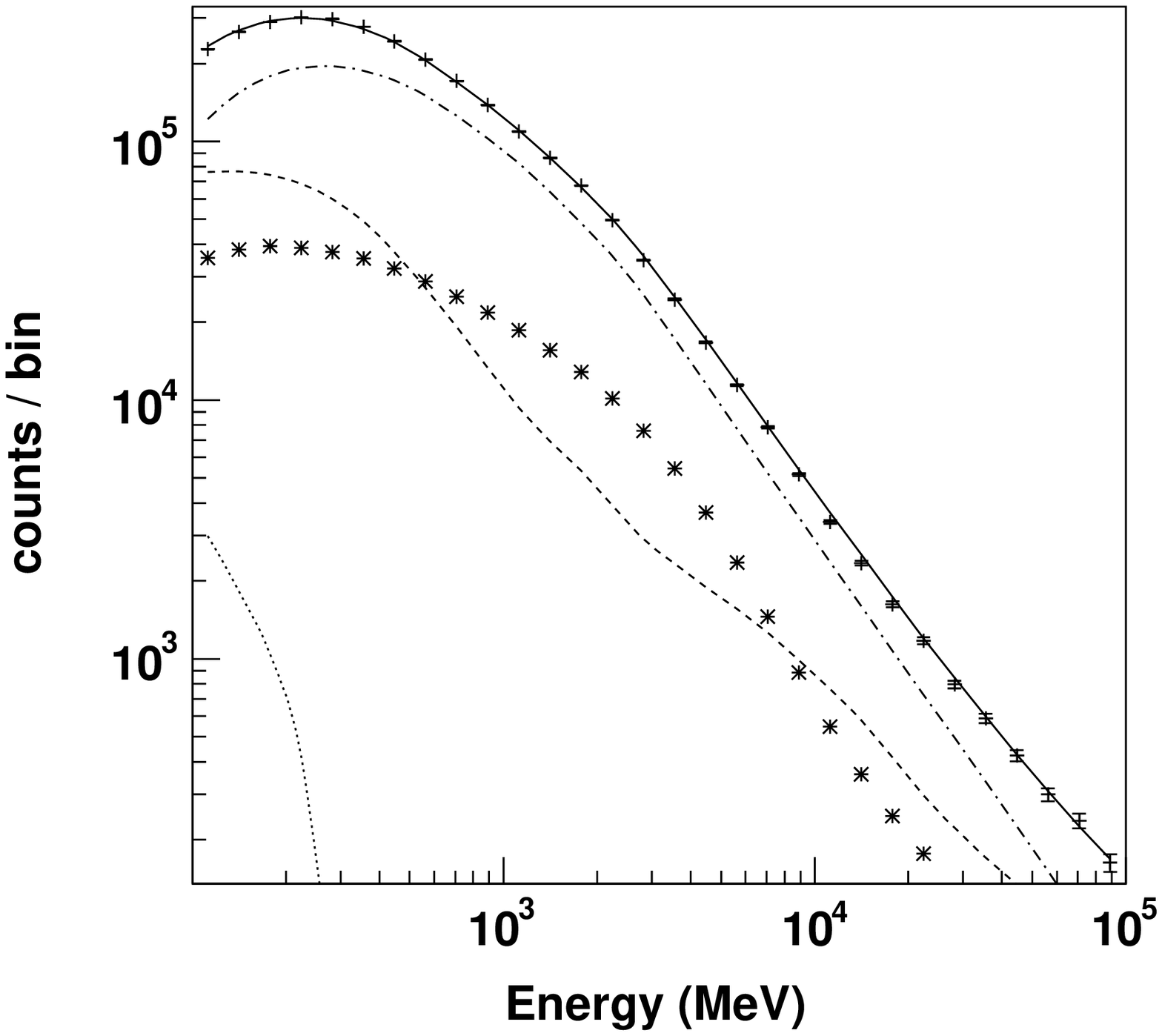}{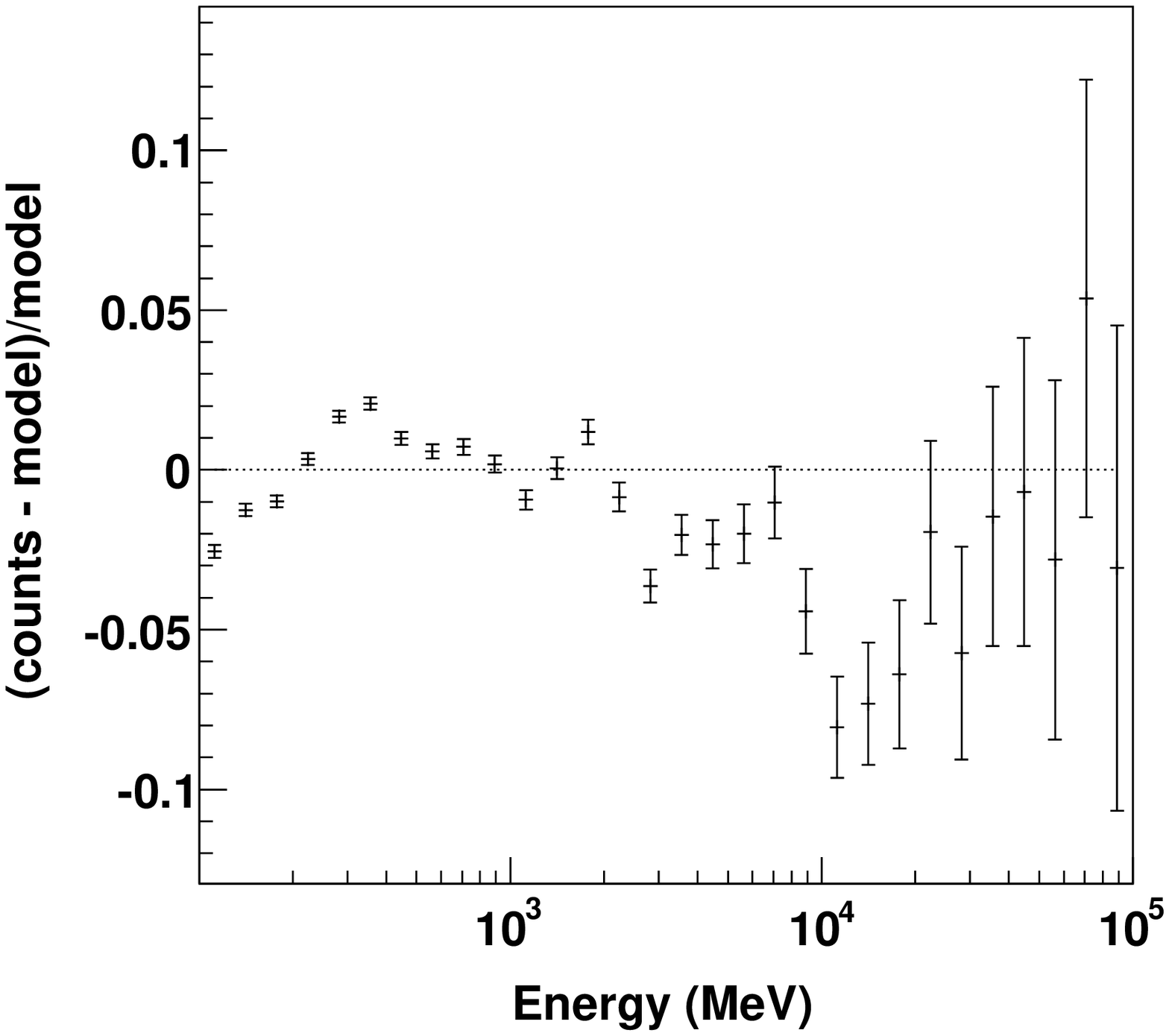}
\caption{Left: Fit to the full spectrum integrated over the same anticenter region as in Figure~\ref{fig:ressigma_map}. The spectral bins are the same as in the source fitting process. The dotted, dashed and dash-dotted lines are the Earth limb, isotropic and Galactic components, respectively. The asterisks show the total source contribution (dominated by the Geminga and Crab pulsars). The full line is the sum of all model contributions, to be compared with the data (plus signs). The statistical errors on the data are shown but barely visible except at high energy. Right: Fractional residuals (data/model$-$1) with statistical error bars. The residuals are statistically significant because of the very large number of events ($2.8 \times 10^6$ over that area) but are only a few percent.}
\label{fig:residuals_spec}
\end{figure}

In order to illustrate the global quality of the main spectral fit (\S~\ref{catalog_significance}), we show in Figures \ref{fig:ressigma_map} and \ref{fig:residuals_spec} the spatial and spectral residuals over a large sky region rather than an individual RoI which could hide cross-talk issues. We chose the Galactic anticenter which is halfway between the quiet high-latitude regions and the most difficult Galactic Ridge regions discussed in \S~\ref{catalog_ism}.

We fit the same parameters as in an ordinary RoI: normalizations of the isotropic and Galactic components $K_{\rm iso}$ and $K_{gal}$, and corrective slope of the Galactic component $\Gamma_{gal}$, such that the correction to the Galactic model is $K_{gal} (E/E_0)^{-\Gamma_{gal}}$ with $E_0$ set to 500~MeV. The fitted parameters were $K_{\rm iso} = 0.973$, $K_{gal} = 1.003$ and $\Gamma_{gal} = 0.029$.

The spatial residuals are scaled to the Poisson noise in each pixel in order to quantify whether the deviations are significant. What is shown is (data $-$ model) / $\sqrt{\rm model}$. The pixel size is large enough that there are about 100 counts per pixel outside the plane.
The distribution of spatial residuals on $0\fdg5$ pixels follows very closely a normal law. Its standard deviation is only 1.1~$\sigma$, implying that the intrinsic fluctuations are about 0.5~$\sigma$, or 5\%. They appear to be on a scale of a few degrees. The spectral residuals are a few percent and evolve slowly with energy.
Those are small imperfections of the diffuse model, which show up because of the very high statistical quality of the data.
Their impact on sources is limited because the residuals are on a larger scale than the LAT PSF except at low energy. It is quantified in \S~\ref{catalog_limitations}.

\section{Description of the FITS Version of the 2FGL Catalog}
\label{appendix_fits_format}

The FITS format version of the 2FGL catalog\footnote{The file is available  from the $Fermi$ Science Support Center, http://fermi.gsfc.nasa.gov/ssc} has four binary table extensions.  The extension {\tt LAT\_Point\_Source\_Catalog Extension} has all of the information about the sources, including the monthly light curves (Tab.~\ref{tab:columns}).  

The extension {\tt Hist\_Start} lists the Mission Elapsed Time (seconds since 00:00 UTC on 2000 January 1) of the start of each bin of the monthly light curves.  The final entry is the ending time of the last bin.  

The extension {\tt GTI} is a standard Good-Time Interval listing the precise time intervals (start and stop in MET) included in the data analysis.  The number of intervals is fairly large because on most orbits ($\sim$95~min) $Fermi$ passes through the South Atlantic Anomaly (SAA), and science data taking is stopped during these times.  In addition, data taking is briefly interrupted on each non-SAA-crossing orbit, as $Fermi$ crosses the ascending node.  Filtering of time intervals with large rocking angles, other data gaps, or operation in non-standard configurations introduces some more entries.  The GTI is provided for reference and would be useful, e.g., for reconstructing the precise data set that was used for the 1FGL analysis.

The extension {\tt ExtendedSources} contains information about the 12 spatially extended sources that are modeled in the 2FGL catalog, including locations and shapes (Tab.~\ref{tab:ExtendedSourcesColumns}).

\begin{deluxetable}{llll}
\setlength{\tabcolsep}{0.04in}
\tablewidth{0pt}
\tabletypesize{\scriptsize}
\tablecaption{LAT 2FGL FITS format:  LAT\_Point\_Source\_Catalog Extension\label{tab:columns}}
\tablehead{
\colhead{Column} &
\colhead{Format} &
\colhead{Unit} &
\colhead{Description}
}
\startdata
Source\_Name & 18A & \nodata & \nodata  \\
RAJ2000 & E & deg & Right Ascension \\
DEJ2000 & E & deg & Declination \\
GLON & E & deg & Galactic Longitude \\
GLAT & E & deg & Galactic Latitude \\
Conf\_68\_SemiMajor & E & deg & Long radius of error ellipse at 68\% confidence \\
Conf\_68\_SemiMinor & E & deg & Short radius of error ellipse at 68\% confidence \\
Conf\_68\_PosAng & E & deg & Position angle of the 68\% long axis from celestial North, \\
 & \nodata & & positive toward increasing RA (eastward) \\
Conf\_95\_SemiMajor & E & deg & Long radius of error ellipse at 95\% confidence \\
Conf\_95\_SemiMinor & E & deg & Short radius of error ellipse at 95\% confidence \\
Conf\_95\_PosAng & E & deg & Position angle of the 95\% long axis from celestial North, \\
 & & & positive toward increasing RA (eastward) \\
Signif\_Avg & E & \nodata & Source significance in $\sigma$ units (derived from Test Statistic) \\
Pivot\_Energy & E & MeV & Energy at which error on differential flux is minimal \\
Flux\_Density & E & cm$^{-2}$ MeV$^{-1}$ s$^{-1}$ & Differential flux at Pivot\_Energy \\
Unc\_Flux\_Density & E & cm$^{-2}$ MeV$^{-1}$ s$^{-1}$ & 1 $\sigma$  error on differential flux at Pivot\_Energy \\
Spectral\_Index & E & \nodata & Best fit photon number power-law index.  For LogParabola spectra, \\
& \nodata & & index at Pivot\_Energy; for PLExpCutoff spectra, low energy index.\\
Unc\_Spectral\_Index & E & \nodata &  1 $\sigma$ error on Spectral\_Index \\
Flux1000 & E & cm$^{-2}$ s$^{-1}$ & Integral flux from 1 to 100 GeV \\
Unc\_Flux1000 & E & cm$^{-2}$ s$^{-1}$ & 1 $\sigma$  error on integral flux from 1 to 100 GeV \\
Energy\_Flux100 & E & erg cm$^{-2}$ s$^{-1}$ & Energy flux from 100 MeV to 100 GeV obtained by spectral fitting \\
Unc\_Energy\_Flux100 & E & erg cm$^{-2}$ s$^{-1}$ & 1 $\sigma$  error on energy flux from 100 MeV to 100 GeV \\
Signif\_Curve & E & \nodata & Significance (in $\sigma$ units) of the fit improvement between power-law \\
& & & and either LogParabola (for ordinary sources) or PLExpCutoff (for pulsars). \\
& & & A value greater than 4 indicates significant curvature. \\
SpectrumType & 18A & \nodata & Spectral type (PowerLaw, LogParabola, PLExpCutoff). \\
beta & E & \nodata & Curvature parameter ($\beta$) for LogParabola. NULL for other spectral types \\
Unc\_beta & E & \nodata & 1 $\sigma$ error on $\beta$ for LogParabola. NULL for other spectral types\\
Cutoff & E & MeV & Cutoff energy as exp(-E/Cutoff) for PLExpCutoff. NULL for other spectral types \\
Unc\_Cutoff & E & MeV & 1 $\sigma$ error on cutoff energy for PLExpCutoff. NULL for other spectral types\\
PowerLaw\_Index & E & \nodata & Best fit power-law index. Equal to Spectral\_Index if SpecrumType is PowerLaw.\\
Flux30\_100 & E & cm$^{-2}$ s$^{-1}$ & Integral flux from 30 to 100~MeV (not filled) \\
Unc\_Flux30\_100 & E & cm$^{-2}$ s$^{-1}$ & 1 $\sigma$  error on integral flux from 30 to 100~MeV (not filled) \\
Sqrt\_TS30\_100 & E & \nodata & Square root of the Test Statistic between 30 and 100~MeV (not filled) \\
Flux100\_300 & E & cm$^{-2}$ s$^{-1}$ & Integral flux from 100 to 300~MeV \\
Unc\_Flux100\_300 & E & cm$^{-2}$ s$^{-1}$ & 1 $\sigma$  error on integral flux from 100 to 300~MeV\tablenotemark{a}\\
Sqrt\_TS100\_300 & E & \nodata & Square root of the Test Statistic between 100 and 300~MeV \\
Flux300\_1000 & E & cm$^{-2}$ s$^{-1}$ & Integral flux from 300~MeV to 1~GeV \\
Unc\_Flux300\_1000 & E & cm$^{-2}$ s$^{-1}$ & 1 $\sigma$  error on integral flux from 300~MeV to 1~GeV\tablenotemark{a} \\
Sqrt\_TS300\_1000 & E & \nodata & Square root of the Test Statistic between 300~MeV and 1~GeV \\
Flux1000\_3000 & E & cm$^{-2}$ s$^{-1}$ & Integral flux from 1 to 3~GeV \\
Unc\_Flux1000\_3000 & E & cm$^{-2}$ s$^{-1}$ & 1 $\sigma$  error on integral flux from 1 to 3~GeV\tablenotemark{a} \\
Sqrt\_TS1000\_3000 & E & \nodata & Square root of the Test Statistic between 1 and 3~GeV \\
Flux3000\_10000 & E & cm$^{-2}$ s$^{-1}$ & Integral flux from 3 to 10~GeV \\
Unc\_Flux3000\_10000 & E & cm$^{-2}$ s$^{-1}$ & 1 $\sigma$  error on integral flux from 3 to 10~GeV\tablenotemark{a} \\
Sqrt\_TS3000\_10000 & E & \nodata & Square root of the Test Statistic between 3 and 10~GeV \\
Flux10000\_100000 & E & cm$^{-2}$ s$^{-1}$ & Integral flux from 10 to 100~GeV \\
Unc\_Flux10000\_100000 & E & cm$^{-2}$ s$^{-1}$ & 1 $\sigma$ error on integral flux from 10 to 100~GeV\tablenotemark{a} \\
Sqrt\_TS10000\_100000 & E & \nodata & Square root of the Test Statistic between 10 and 100~GeV \\
Variability\_Index & E & \nodata & Sum of 2$\times$Log(Likelihood) comparison between the flux fitted in 24 time\\
& & & segments and a flat lightcurve over the full 2-year catalog interval. \\
& & & A value greater than 41.64 indicates $< $1\% chance of being a steady source.  \\
Signif\_Peak & E & \nodata & Source significance in peak interval in $\sigma$ units \\
Flux\_Peak & E & cm$^{-2}$ s$^{-1}$ & Peak integral flux from 100~MeV to 100~GeV \\
Unc\_Flux\_Peak & E & cm$^{-2}$ s$^{-1}$ &  1 $\sigma$  error on peak integral flux \\
Time\_Peak & D & s (MET) & Time of center of interval in which peak flux was measured \\
Peak\_Interval & E & s & Length of interval in which peak flux was measured \\
Flux\_History & 11E & cm$^{-2}$ s$^{-1}$ & Integral flux from 100~MeV to 100~GeV in each interval (best fit from \\
& & &  likelihood analysis with spectral shape fixed to that obtained over 2 years).\\
Unc\_Flux\_History & 11E & cm$^{-2}$ s$^{-1}$ &  Error on integral flux in each interval using method \\
&  & &  indicated in Unc\_Flag\_History column and added in quadrature \\
&  & &  with 3\% systematic component. \\
Unc\_Flag\_History & 11B &  &  1 if it is half of the difference between the 2 $\sigma$ upper limit \\
& & &  and the maximum-likelihood value given in Flux\_History, 0 if it is the \\
& & &  1 $\sigma$ uncertainty derived from a significant detection in the interval\\ 
Extended\_Source\_Name & 18A & \nodata & Cross-reference to the ExtendedSources extension for extended sources, if any \\
0FGL\_Name & 18A & \nodata & Name of corresponding 0FGL source, if any \\
1FGL\_Name & 18A & \nodata & Name of corresponding 1FGL source, if any \\
ASSOC\_GAM1 & 18A & \nodata & Name of likely corresponding 1AGL source \\
ASSOC\_GAM2 & 18A & \nodata & Name of likely corresponding 3EG source \\
ASSOC\_GAM3 & 18A & \nodata & Name of likely corresponding EGR source \\
TEVCAT\_FLAG & A & \nodata & P if positional association with non-extended source in TeVCat \\
   & \nodata & & E if associated with a more extended source in TeVCat, N if no TeV association \\
ASSOC\_TEV & 24A & \nodata & Name of likely corresponding TeV source from TeVCat \\
CLASS1 & 3A & \nodata & Class designation for associated source; see Table~\ref{tab:classes} \\
CLASS2 & 3A & \nodata & Second class designation for associated source \\
ASSOC1 & 24A & \nodata & Name of identified or likely associated source \\
ASSOC2 & 24A & \nodata & Alternate name of identified or likely associated source \\
Flags & I & \nodata & Source flags (binary coding as in Table~\ref{tab:flags}) \\
\enddata
\tablenotetext{a} {The upper limit is set equal to 0 if the flux in the corresponding energy band is an upper limit ($TS < 10$ in that band).  The upper limits are 2 $\sigma$.}
\end{deluxetable}

\begin{deluxetable}{llll}
\setlength{\tabcolsep}{0.04in}
\tablewidth{0pt}
\tabletypesize{\scriptsize}
\tablecaption{LAT 2FGL FITS format:  ExtendedSources Extension\label{tab:ExtendedSourcesColumns}}
\tablehead{
\colhead{Column} &
\colhead{Format} &
\colhead{Unit} &
\colhead{Description}
}
\startdata
Source\_Name & 18A & \nodata & \nodata  \\
1FGL\_Name & 18A & \nodata & \nodata  \\
RAJ2000 & E & deg & Right Ascension of centroid\\
DECJ2000 & E & deg & Declination of centroid\\
GLON & E & deg & Galactic Longitude of centroid \\
GLAT & E & deg & Galactic Latitude of centroid\\
Model\_Form & 24A & \nodata & Spatial shape (2D Gaussian, Disk, Ring, Template, ...) \\
Model\_SemiMajor & E & deg & Long radius of source. Full size for bounded shapes (disk, ring). \\
 & & & 68\% containment for unbounded shapes (Gaussian) \\
Model\_SemiMinor & E & deg & Short radius of source \\
Model\_PosAng & E & deg & Position angle of the long axis from celestial North, \\
 & & & positive toward increasing RA (eastward) \\
Spatial\_Filename & 68A & \nodata & Name of spatial template file\tablenotemark{a}  \\
\enddata
\tablenotetext{a} {Spatial\_Filename refers to external files that should be included with the catalog distribution.}
\end{deluxetable}

\end{document}